\documentclass[reprint,pra,onecolumn,nobibnotes,preprintnumbers,superscriptaddress,amsmath,amssymb,floatfix]{revtex4-2}

\usepackage[labelformat=simple]{subcaption}

\usepackage{lipsum}
\usepackage{amsthm}
\usepackage{amssymb}
\usepackage{graphicx}
\usepackage{subcaption}

\begin{document}

\title{Percolation Theories for Quantum Networks}

\author{Xiangyi Meng}
\affiliation{Network Science Institute, Northeastern University, Boston, Massachusetts
02115, USA}
\affiliation{Department of Physics and Astronomy,
Northwestern University, Evanston, Illinois 60208, USA}
\author{Xinqi Hu}
\affiliation{School of Mathematical Sciences, Jiangsu University, Zhenjiang, Jiangsu 212013, China}
\author{Yu Tian}
\affiliation{Nordita, KTH Royal Institute of Technology and Stockholm University,  SE-106 91 Stockholm, Sweden}
\author{Gaogao Dong}
\affiliation{School of Mathematical Sciences, Jiangsu University, Zhenjiang, Jiangsu 212013, China}
\author{Renaud Lambiotte}
\affiliation{Mathematical Institute, University of Oxford,  Oxford OX2 6GG, UK}
\affiliation{Turing Institute, London NW1 2DB, UK}
\author{Jianxi Gao}
\affiliation{Department of Computer Science, Rensselaer Polytechnic Institute, Troy, New York 12180, USA}
\affiliation{Network Science and Technology Center, Rensselaer Polytechnic Institute, Troy, New York 12180, USA}
\author{Shlomo Havlin}
\email{havlins@gmail.com}
\affiliation{Department of Physics, Bar-Ilan University, Ramat Gan 52900, Israel}

\date{\today}

\begin{abstract}
Quantum networks have experienced rapid advancements in both theoretical and experimental domains over the last decade, making it increasingly important to understand their large-scale features from the viewpoint of statistical physics. This review paper discusses a fundamental question: how can entanglement be effectively and indirectly (e.g.,~through intermediate nodes) distributed between distant nodes in an imperfect quantum network, where the connections are only partially entangled and subject to quantum noise? We survey recent studies addressing this issue by drawing exact or approximate mappings to percolation theory, a branch of statistical physics centered on network connectivity. Notably, we show that the classical percolation frameworks do not uniquely define the network's indirect connectivity. This realization leads to the emergence of an alternative theory called ``concurrence percolation,'' which uncovers a previously unrecognized quantum advantage that emerges at large scales, suggesting that quantum networks are more resilient than initially assumed within classical percolation contexts, offering refreshing insights into future quantum network design.
\end{abstract}

\maketitle


\section{Introduction}

Quantum information~\cite{q-comput-q-inf} is a fast-developing field that has transcended its roots originally in quantum mechanics and information theory to other areas like condensed matter physics~\cite{fundam-condens-matter-phys}, statistical physics~\cite{equilib-stat-phys,stat-field-theor-1,stat-field-theor-2}, and network science~\cite{stat-of-netw_ab02,struct-dyn-netw}. At the core of quantum information lies the quantum bit, or {qubit}, the basic quantum information carrier. Two qubits can be designed into a relationship, called entanglement, which is an essential quantum resource~\cite{q-resour_cg19} for quantum computing. Yet, entanglement is notoriously fragile, especially when qubits are spatially distant. Fortunately, by path routing and adding in-between sites for replaying, entanglement between remote qubits may eventually be established in an indirect way. Such an action, called \emph{entanglement distribution}~\cite{entangle-distrib_czkm97}, is a fundamental benefit of quantum networks (QN)~\cite{QEP_acl07,q-internet_k08,QEP-mix-state_bdj09,QEP-lattice_lwl09,QEP-mix-state_bdj10,albert2000error,QEP-complex-netw_wz11,q-netw-simul_lck12,q-netw-summ_pjcla13,QEP-gossip_s17,q-netw_casbcs17,q-netw_dkd18,q-netw-route_p19,q-netw-summ_bfd19,q-netw_umfydmk19,q-netw_cbclf19,q-netw_kbdgl19,q-netw_rwbbgb19,multipartite-q-netw_hpe19,Zhuang2021,q-netw_sshd23}.

In general, a QN is a network representation of different parties (nodes) that share entanglement (links) as connections. A significant part of our interest lies in distributing entanglement between \emph{two} arbitrary nodes in the network, a process we refer to as ``entanglement transmission.''  Entanglement across different parties is essentially transmitted through quantum communication protocols. Successful demonstrations of quantum communication protocols have already been made on small-scale QN using diamond nitrogen-vacancy centers~\cite{q-netw_krhbknbtmh17,q-netw_hkmsvtmh18,q-netw_phbbhsvtmdwh21} and ion traps~\cite{entangle-swap_rmkvschb08,q-netw-ion-trap_ahprshde09}.
However, the big question that looms is how to scale this to much larger networks. A large-scale, practical QN would offer significant advantages for many industrial and scientific applications. For example, financial institutions and governments would benefit from quantum cybersecurity  providing an unprecedented level of secure communication. Researchers could also use networked quantum computers to dramatically increase the simulation speed of the physical and chemical processes of many interacting particles. Yet, if the individual channels (links) along the routed path are too noisy,  the entanglement transmission may fail. Study of such dependence of ``indirect'' transmission ability on the noise level of individual links requires tools from statistical physics and complex network theories.

One of such theory that has proven to be useful is percolation theory~\cite{percolation-theor_e80, stauffer92,diffus-react-fractal-disorder-syst,bunde2012}. Percolation theory offers a mathematical framework for understanding how networks behave when subjected to random processes (can be treated as a form of noise), such as how water percolates through soil or how diseases spread through populations. In the context of QN, percolation could provide valuable insights into the robustness and efficiency of entanglement distribution. By applying percolation theory, we can model and analyze the network structure directly and identify the most effective ways to maintain and distribute quantum entanglement across it. This lays the groundwork for examining QNs through the lens of \emph{statistical physics} and opens up new avenues for understanding the upper limits of entanglement distribution in these networks.

In this work, we will explore and summarize the developments of the QN framework and how a mapping to percolation offers unique tools for dissecting the problem of entanglement transmission. Specifically, we will show that the mapping to percolation theories---and a definition of how a combination of pairwise edges combines into indirect connectivity---are, indeed, \emph{not} unique. A new, alternative percolation-like theory termed \emph{concurrence percolation}~\cite{conpt_mgh21} emerges, and it underlies an unexpected ``quantum advantage,'' revealing that QNs are more robust than we initially thought within the  classical percolation framework. Moreover, the finding is scalable with network size and adaptable to different network topologies, suggesting a macroscopic improvement over classical considerations from a statistical physics perspective.

This paper focuses on the comparison between classical percolation and concurrence percolation when mapped based on QN. It is structured as follows: In Section~\ref{sec_qn}, we give a definition of the QN theoretical framework as well as its possible generalizations to other QN-based structures {(e.g.,~hypergraphs)}. In Section~\ref{sec_percolation}, we briefly review the concept and definition of percolation theory and, in particular, how it relates to network connectivity at large scales. In Sections~\ref{sec_classical}~and~\ref{sec_quantum}, we will focus on the discoveries that the new concurrence percolation theory surpasses the traditional percolation theory (which we refer to as ``classical percolation'' for comparison). In Section~\ref{sec_algorithm}, we will delve into the algorithms developed for calculating concurrence percolation. Finally, in Section~\ref{sec_discussion}, 
we will discuss the open questions and practical implications of the findings, both theoretically and for real-world communications.

\begin{figure}[t!]
	\centering
	\includegraphics[width=397pt]{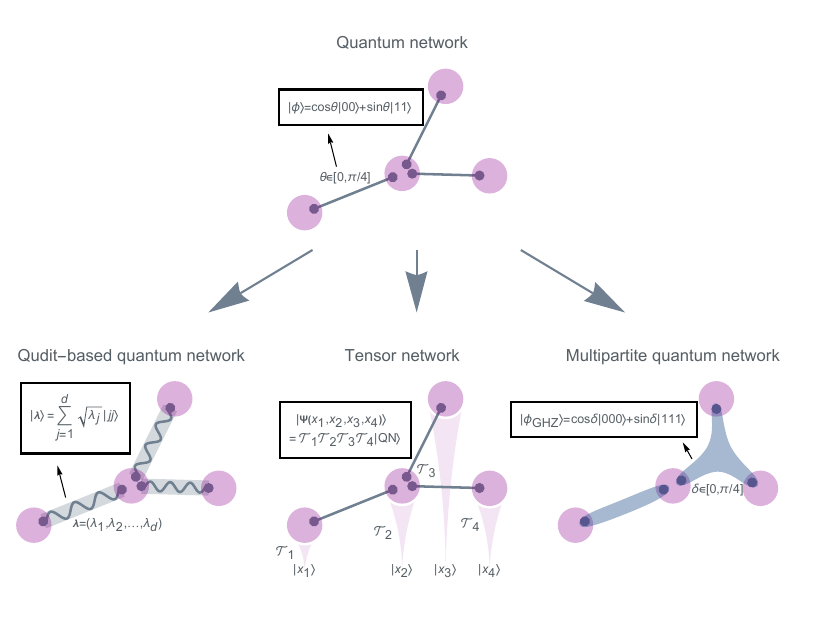}
	\caption{\label{fig_qn}{A pure-state quantum network (QN) consists of nodes (purple) and links (gray line).} Each node comprises a collection of qubits (gray dots) that are entangled with qubits belonging to other nodes, and each link represents a bipartite entangled pure state $\left|\phi\right\rangle$ connecting the two qubits at its endpoints. This QN model can be extended to $d$-dimensional qudits {(bottom left)}, which allow higher bandwidth for transmitting information, or~to tensor networks {(bottom middle)} by employing linear transformations $\mathcal{T}_i$ at each node $i$. Moreover, the~QN can be adapted to higher-order graphs {(bottom right)}, where each link manifests as a hyperedge, denoting a multipartite entangled pure state.\hfill\hfill
}
\end{figure}

\section{Quantum networks (QN)}
\label{sec_qn}

As in traditional network theories, a QN resembles a topological graph or a graph-like structure, comprising nodes and links. This paper primarily focuses on a pure-state version of QN (Fig.~\ref{fig_qn})~\cite{QEP_acl07,QEP_p10}. 
The QN is defined based on the following three principles:
\begin{enumerate}
    \item Each \emph{node} (purple) comprises a collection of qubits (gray dots) that are entangled with qubits belonging to other nodes.
    \item Each \emph{link} (gray line) represents a bipartite entangled pure state $\left|\phi\right\rangle$ connecting the two qubits at its endpoints.
    \item A \emph{weight} $\theta$ is assigned to each link to characterize the degree of the link's entanglement.
\end{enumerate}
Using the Dirac notation, a link, which corresponds to a bipartite entangled pure state connecting two nodes (e.g.,~Alice and Bob), can be written as $\left|\phi\right\rangle=\cos\theta\left|00\right\rangle+\sin\theta\left|11\right\rangle$. Here, w.l.o.g.,~the weight parameter $\theta$ is constrained within the range $0\le \theta \le \pi/4$, ensuring that $\cos\theta\ge\sin\theta$ is satisfied.
In this notation, the first ``$0$'' in $\left|00\right\rangle$ and the first ``$1$'' in $\left|11\right\rangle$ represent the two possible states of Alice's qubit, $\left|0\right\rangle_{\text{Alice}}$ and $\left|1\right\rangle_{\text{Alice}}$, respectively. Similarly, the second ``$0$'' and ``$1$'' represent the two possible states of Bob's qubit, $\left|0\right\rangle_{\text{Bob}}$ and $\left|1\right\rangle_{\text{Bob}}$. The entanglement between Alice's and Bob's qubits is evident by the presence of only two terms $\left|00\right\rangle$ and $\left|11\right\rangle$ in $\left|\phi\right\rangle$, while $\left|01\right\rangle$ and $\left|10\right\rangle$ are absent. This implies that upon measuring the state $\left|{\phi}\right\rangle$ in the ${\left|0\right\rangle, \left|1\right\rangle}$ basis from either Alice's or Bob's side, the state will randomly collapse to either $\left|00\right\rangle$ or $\left|11\right\rangle$. Consequently, if Alice's (or Bob's) measurement yields ``$0$,'' it guarantees that the other party's measurement result will also be ``$0$.''
This highlights a correlation feature that can be harnessed for communication in the quantum realm. 
Similar to how correlation in classical communication is measured using mutual information, we can quantify this quantum correlation using quantum mutual information~\cite{q-comput-q-inf}, which is  given by $-2\cos^2\theta \ln\left(\cos^2\theta\right)-2\sin^2\theta \ln\left(\sin^2\theta\right)$. The quantum mutual information reaches its maximum value when $\theta=\pi/4$, which corresponds to a maximally entangled state,  $\left|\phi_{\perp}\right\rangle=\sqrt{1/2}\left|00\right\rangle+\sqrt{1/2}\left|11\right\rangle$, commonly referred to as a Bell state or a singlet~\cite{QEP_acl07}.

The entire QN, comprising many links, can be regarded as a huge pure state $\left|\text{QN}\right\rangle$---the tensor product of all the individual bipartite pure states associated with each link. Consequently, the QN solely focuses on ``quantum noise,'' which comes from the fact that when $\theta<\pi/4$, the link exhibits only partial entanglement.
This partial entanglement, when employed in quantum communication tasks such as quantum teleportation, leads to errors in the teleported qubits, 
affecting the QN's overall communication capacity. 
Yet, as a pure state, the QN does not involve any classical noise (i.e.,~mixed states). This makes $\left|\text{QN}\right\rangle$ an excellent medium for examining quantum phenomena without the interference of classical noise. Thus, this ``minimalist'' construction of the QN can serve as an ideal framework for investigating quantum theories and concepts on large scales.

At present, the choice to define nodes as collections of qubits rather than individual qubits may appear arbitrary. What is the physical meaning of a node as a collection of qubits? And what about the qubits that belong to the same node---are they also entangled?

To answer these questions, it is crucial to comprehend the concept of \emph{locality}. While the theoretical framework of quantum mechanics is inherently nonlocal,
practical implementations of quantum information technologies often necessitate considering a ``distant lab'' paradigm~\cite{locc_clmow14}. In such scenarios, when a quantum system is distributed among multiple spatially distant parties or laboratories, it  becomes unrealistic to assume the feasibility of executing global quantum operations. Instead, the parties are typically constrained to apply quantum operations exclusively to their respective subsystems (in their own ``labs''), rather than collectively to the global system. This subset of quantum operations is known as local operations (LO).

For example, given the entangled state $\left|\phi\right\rangle=\cos\theta\left|00\right\rangle+\sin\theta\left|11\right\rangle$ between Alice and Bob, Alice may apply local unitary transformation on her qubit (e.g.,~a rotation $\{\left|0\right\rangle,\left|1\right\rangle\}\to\{\frac{\left(\left|0\right\rangle+\left|1\right\rangle\right)}{\sqrt{2}}, \frac{\left(\left|0\right\rangle-\left|1\right\rangle\right)}{\sqrt{2}}$), and Bob may apply the same transformation as well. This yields a new state $\left|\phi\right\rangle \to \left|\phi'\right\rangle=\cos\theta\frac{\left(\left|0\right\rangle+\left|1\right\rangle\right)\left(\left|0\right\rangle+\left|1\right\rangle\right)}{2}+\sin\theta\frac{\left(\left|0\right\rangle-\left|1\right\rangle\right)\left(\left|0\right\rangle-\left|1\right\rangle\right)}{2}$. 
Furthermore, LO also allows Alice or Bob to locally measure their qubits as well, resulting in the random collapse of $\left|\phi\right\rangle$ to one of its eigenstates.
However, Alice and Bob cannot transform their state globally and obtain a singlet, $\left|\phi\right\rangle \to \left|\phi_{\perp}\right\rangle=\sqrt{1/2}\left|00\right\rangle+\sqrt{1/2}\left|11\right\rangle$. This is not counted as LO.

On top of LO, Alice and Bob are also free to communicate classical information (CC), sharing their results of quantum measurements. Together, this set of operations is called the local operations and classical communication (LOCC). The LOCC defines a set of strategies to share and manipulate quantum information under the locality constraint. One of the most powerful theorems in quantum information states that the average entanglement between two parties can never be increased if only LOCC is allowed for the two parties. This establishes the role of entanglement as a quantum {resource}, given that LOCC is the ``free operations'' of the system~\cite{q-resour_cg19}. 

Therefore, qubits belonging to the same node are not constrained by LOCC. They can be freely entangled or disentangled  as needed, but the entanglement is not viewed as a resource for communication. Only the entanglement of qubits belonging to different nodes matters. In other words, quantum networks are an effective representation of the \emph{fundamental constraints of locality}, manifested by assigning qubits to different local compartments and entanglement to inter-compartment connections.

\subsection{Qudit-based quantum networks}
A natural extension of QN is to use more general $d$-dimensional ``qudits'' (qutrits, ququarts, etc.) instead of qubits (Fig.~\ref{fig_qn}). 
Each link, as a bipartite pure state of qudits, can be written as
\begin{equation}
\label{eq_qudits}
    \left|\boldsymbol{\lambda}\right\rangle=\sum_{j=1}^{d}\sqrt{\lambda_{j}}\left|jj\right\rangle.
\end{equation}
Here, $\lambda_{1}\ge\lambda_{2}\ge\cdots\ge\lambda_{d}\ge0$ and $\sum_{j=1}^{d}\lambda_{j}= 1$. In this generalization, the weight of each link is no longer a single number but a list of non-negative numbers known as Schmidt numbers, denoted as $\boldsymbol{\lambda}=\left(\lambda_1, \lambda_2, \cdots, \lambda_d\right)$.
When $d=2$, the bipartite pure state reduces to qubits, where $\lambda_1=\cos^2\theta$ and $\lambda_2=\sin^2\theta$. 

The consideration of qudit-based QN offers both theoretical and practical advantages. Theoretically, a $d$-dimensional qudit inherently carries $\log_2 d$ times more information than a qubit. Therefore, as the value of $d$ increases, a single carrier can transmit more information, increasing the bandwidth. This enhanced capability is also evident in the robustness of entanglement for entangled states of qudits. Indeed, even when some coefficients are erased ($\lambda_{j}\to 0$) from $\boldsymbol{\lambda}$ in the presence of noise, the pure state $\left|\boldsymbol{\lambda}\right\rangle$ can still remain entangled, as long as the two largest Schmidt numbers $\lambda_{1}$ and $\lambda_{2}$ remain positive.

In experiments, qubit systems are commonly realized using two-level atoms or superconducting states. However, isolating these two levels from other nearby levels can be challenging. By including nearby levels and increasing the potential dimension $d$, the experimental design may become more feasible. In fact, several experiments have employed qudits to achieve better performance, including applications in quantum scrambling~\cite{qutrit-scramble_brsobkdmyys21} and superdense coding~\cite{ququart-superdense_hglhlg18}.

\subsection{Quantum networks are the basis of tensor networks}
There is also an interesting and deep connection between QN and tensor networks (Fig.~\ref{fig_qn}), the latter being a familiar and powerful tool in condensed matter physics, mostly used for the purpose of facilitating computations and simulations in quantum physics and materials science. To be specific, tensor networks are designed to \emph{efficiently} represent many-body quantum states~\cite{tensor-netw_o14}. These quantum states, which are essentially large, high-dimensional tensors in mathematical terms, can be factorized into smaller tensors using tensor networks.
In particular, tensor networks are useful for representing the ground state of quantum systems, which typically exhibit strong ordering compared to excited states. This strong ordering often means that entanglement does not grow very fast with the length scale, which, in turn, allows for easier and more efficient factorization of the corresponding ground state.

To delve deeper into the concept, note that a general $N$-body quantum state reads
\begin{equation}
\label{eq_tensor_network_0}
    \left|\Psi \right\rangle= \sum^D_{x_1,x_2,\cdots,x_N=1} T^{x_1 x_2 \cdots x_N} \left|x_1\right\rangle \left|x_1\right\rangle \cdots \left|x_N\right\rangle,
\end{equation}
which lives in a $D^N$-dimensional Hilbert space that is the tensor product of $N$ ``single-body'' Hilbert spaces ($i=1,2,\cdots,N$). Each space is spanned by a basis of $D$ vectors, $\left|x_i\right\rangle$, where $x_i=1,2,\cdots, D$. The complex tensor $T$ that stores the coefficients is exponentially large ($\sim D^N$), effectively preventing direct computations of the quantum state $\left|\Psi \right\rangle$'s characteristics. However, there may be a significant level of redundancy in the coefficients present stored in the tensor.
Consider an example where every entry in $T$ can be fully factorized, such that $T=a\otimes b\otimes c\otimes \cdots$ where $a,b,c,\cdots$ are $D$-dimensional vectors. In this case, it becomes unnecessary to store the entire tensor or perform calculations on it. Rather, it suffices to simply store the vectors $a,b,c,\cdots$ of which the total size is $DN$.

This, indeed, is how a tensor network works---by leveraging different ways of factorization that can be depicted through different graphical network structures~\cite{lstm_my21}.
Among various tensor networks, the matrix product state (MPS) is among the most researched~\cite{mps_vc06}. In computer science, it is often called the tensor-train network~\cite{tensor-train-spectr_bekm16}. The MPS is commonly utilized to represent one-dimensional many-body quantum states. When extended to higher dimensions, this becomes what is known as the projected entangled pair state (PEPS)~\cite{mps_cpgsv21}.
More involved tensor network structures, such as the multiscale entanglement renormalization ansatz (MERA)~\cite{mera_v08}, are also routinely used to study critical quantum systems.

For example, a MPS representation of Eq.~\eqref{eq_tensor_network_0} can be written as
\begin{equation}
\label{eq_tensor_network_mps}
    \left|\Psi \right\rangle= \sum^D_{x_1,x_2,\cdots,x_N} \text{tr}\left\{A_1^{x_1}A_2^{x_2}\cdots A_N^{x_N}\right\} \left|x_1\right\rangle \left|x_1\right\rangle \cdots \left|x_N\right\rangle,
\end{equation}
where for each single-body $i$, a set of $D$ different matrices, $A_i^1,A_i^2,\cdots,A_i^D$, are introduced.  Each matrix is of size $d\times d$, where $d$ is called the bond dimension. Thus, the total number of parameters is $NDd^2$, which is linear in $N$.
For a sufficiently large $d$, the MPS has enough degrees of freedom to exactly represent any tensor $T$. However, it is frequently observed that a small $d$ can approximately, if not perfectly, reproduce $T$. This occurs when the information stored in $T$ scales linearly with $N$, a condition often found in the ground states of one-dimensional noncritical quantum systems.

Intriguingly, the MPS offers a new physical perspective---the valence-bond picture~\cite{mps_cpgsv21}. To be specific, we map each single-body Hilbert space (spanned by $\left|x_i\right\rangle$) to a physical site and assume that there are two $d$-dimensional qudits located at each site. For every two neighboring sites ($1\leftrightarrow 2, 2\leftrightarrow3,\cdots, N\leftrightarrow 1$), two qudits from each site are fully entangled, forming a ``valence bond'' that can be written as an unnormalized maximally entangled state,
\begin{equation}
\label{eq_tensor_network_valence_bond}
    \left|\psi \right\rangle= \sum^d_{j=1} \left|j\right\rangle \left|j\right\rangle \rightarrowtail \psi=\begin{pmatrix}
        1&0&\cdots\\
        0&1&\cdots\\
        \vdots &\vdots & \ddots
    \end{pmatrix}.
\end{equation}
Here, the state is also represented (matricized) into the matrix form $\psi$. Combining this with Eq.~\eqref{eq_tensor_network_mps}, we obtain
\begin{eqnarray}
    \left|\Psi \right\rangle &=& \sum^D_{x_1,x_2,\cdots,x_N} \text{tr}\left\{A_1^{x_1} \psi A_2^{x_2} \psi \cdots A_N^{x_N} \psi\right\} \left|x_1\right\rangle \left|x_1\right\rangle \cdots \left|x_N\right\rangle\nonumber\\
    &=& \left(\mathcal{T}_1 \otimes \mathcal{T}_2 \otimes \cdots  \otimes \mathcal{T}_N \right) \left|\psi \right\rangle^{\otimes N},
\end{eqnarray}
where $\mathcal{T}_i=\sum_{x_i=1}^{D} \sum_{j,j'=1}^{d} \left(A_i^{x_i}\right)_{j j'} \left|x_i \right\rangle \left\langle j,j'\right| $ represents a linear transformation acting on the two qudits (labeled by $j$ and $j'$) on-site $i$. The valence-bond picture is now evident: In this picture, the many-body state is not the primary entity. Instead, it is built upon something more fundamental---a network of qudits and ``valence bonds.'' The linear transformations $\mathcal{T}_i$ are then employed on top of it to form the tensor network.

Note that this fundamental network $\left|\psi \right\rangle^{\otimes N}$ is a one-dimensional (periodic) quantum network consisting of maximally entangled states, making it remarkably suitable to be \emph{generalized} to partially entangled states [Eq.~\eqref{eq_qudits}]. This can be achieved by replacing $\psi$ in Eq.~\eqref{eq_tensor_network_valence_bond} by
\begin{equation}
    \psi=\begin{pmatrix}
        \lambda_1&0&\cdots\\
        0&\lambda_2&\cdots\\
        \vdots &\vdots & \ddots
    \end{pmatrix}.
\end{equation}
The physical meaning of inserting such a partially entangled state is that since LO cannot increase the entanglement, the entanglement between neighboring sites will be upper bounded by the amount of entanglement in $\psi$.

The valence-bond picture is not limited to MPS but can be generalized to arbitrary tensor networks. Indeed, suppose $A_i$ at site $i$ does not denote a set of matrices but a set of tensors, $A_i^1,A_i^2,\cdots,A_i^D$, each having entries $\left(A^{x_i}_i\right)_{j j' j''\cdots j^{(k)}}$ labeled by $k$ subscripts $j,j',j'',\cdots,j^{(k)}$. Each subscript denotes a qudit on-site $i$. The site has $k$ qudits in total, indicating that the corresponding node $i$ has degree $k$ (i.e.,~$k$ incident links) in the QN.
The linear transformation then becomes
\begin{equation}
\label{eq_tensor_network_linear_transformation}
    \mathcal{T}_i=\sum_{x_i=1}^{D} \sum_{j,j',\cdots,j^{(k)}=1}^{d} \left(A^{x_i}_i\right)_{j j' \cdots j^{(k)}} \left|x_i \right\rangle \left\langle j,j',\cdots,j^{(k)}\right|,
\end{equation}
and the many-body state is expressed as 
\begin{eqnarray}
    \left|\Psi \right\rangle = \left(\mathcal{T}_1 \otimes \mathcal{T}_2 \otimes \cdots  \otimes \mathcal{T}_N \right) \left|\text{QN} \right\rangle,
\end{eqnarray}
where $\left|\text{QN} \right\rangle$ represents the entire QN considered as a huge pure state (Fig.~\ref{fig_qn}).

\subsection{Multipartite quantum networks}
Our attention has been narrowed to bipartite entanglement. However, a complete QN framework should take \emph{multipartite} entangled states into account. This is since multipartite entangled states have a specialized and unique role in certain quantum communication applications, such as secret sharing~\cite{multipartite-secret-share_fmfp07}.
Although multipartite entanglement has been widely explored, we still lack a unified, clear method to precisely detect, measure, and define it.  For example, even with an entangled state of just three qubits, there exist two non-equivalent forms of genuine tripartite entanglement. The first is known as the GHZ class, characterized by five real parameters, $\alpha$, $\beta$, $\gamma$, $\delta$, and $\theta$, and can be expressed as~\cite{tripartite_dvc00}
\begin{eqnarray}
\label{eq_ghz}
    \left|\phi_\text{GHZ} \right\rangle \propto  \cos\delta \left|000 \right\rangle+\sin\delta e^{i \theta}\left(\cos\alpha \left|0 \right\rangle+\sin\alpha \left|1 \right\rangle \right)\left(\cos\beta \left|0 \right\rangle+\sin\beta \left|1 \right\rangle \right)\left(\cos\gamma \left|0 \right\rangle+\sin\gamma \left|1 \right\rangle \right).\nonumber\\
\end{eqnarray}
The second form, called a W class, has a general representation as~\cite{tripartite_dvc00}
\begin{eqnarray}
\label{eq_w}
    \left|\phi_\text{W} \right\rangle =  \sqrt{a}\left|001 \right\rangle+\sqrt{b}\left|010 \right\rangle+\sqrt{c}\left|100 \right\rangle+\sqrt{d}\left|000 \right\rangle,
\end{eqnarray}
with the real parameters $a,b,c>0$ and $d=1-a-b-c\ge 0$.
Both the GHZ and W classes represent a level of correlation that goes beyond just pairwise interactions, meaning that a measurement on any single qubit among the three will instantaneously affect the outcomes of the other two. Despite this, states within one class cannot be converted to those in the other class using LOCC. As a result, we cannot directly compare the degree of entanglement of states belonging to different classes. This represents a fundamentally challenging quantum ``three-body problem'' that complicates the practical applications of multipartite entanglement. For example, a W state may outperform in certain applications, while in others, a GHZ state may be more effective. Note that states belonging to the W class are characterized by only three real d.o.f., whereas the GHZ class requires five. Hence, a generic tripartite state typically belongs to the GHZ class.

Traditionally, each link in a network is also ``bipartite,'' connecting exactly two nodes. As a result, to study a  QN consisting of multipartite entangled states, it is essential to go beyond ``bipartite'' network theory and consider multipartite entangled states as higher-order interactions~\cite{high-ord-netw_babbfafiklmmpvp21,lambiotte2019networks}. These can be mathematically represented as ``hyperedges'' of \emph{hypergraphs}~\cite{hypergr-theor-introd}. Here, Fig.~\ref{fig_qn} shows an example of a hypergraph-based QN consisting of hyperedges in the form of
\begin{eqnarray*}
    \left|\phi_\text{GHZ} \right\rangle =  \cos\delta \left|000\cdots \right\rangle+\sin\delta \left|111\cdots \right\rangle,
\end{eqnarray*}
which represents a special case of the GHZ class [Eq.~\eqref{eq_ghz}]. 
Of course, this is only \emph{one} specific form of multipartite entangled states, characterized by the sole parameter $\delta$. Yet it illustrates the necessity of representing these as hypergraphs and studying them through higher-order network theories, such as higher-order percolation theories (see Section~\ref{sec_percolation_hypergraph} for a brief review).

\section{Percolation of complex network}
\label{sec_percolation}

Percolation theory, serving as a foundational model for investigating disordered systems~\cite{percolation-theor_e80, kesten1982percolation, stauffer92, turcotte1997fractals}, is mainly concerned with understanding the geometric connectivity of random media. 
Constructing a percolation model is straightforward: Take, for example, a square lattice (or a lattice of any shape) in which each link is randomly either present with a probability $p$ or absent with a probability $1-p$.
In a real-world application, one could consider the present links as electrical conductors and the absent ones as insulators~\cite{bunde2012}. The electrical current would then flow solely through the conductor links.
When $p$ is small, almost no paths exist that connect the lattice's two distant boundaries (e.g.,~the left and right boundaries in the square lattice). However, as $p$ grows, various conduction pathways begin to emerge. A phase transition~\cite{stanley1973introduction} is eventually triggered when $p$ crosses a critical threshold, labeled as $p_\text{th}$, effectively changing the composite material from an insulating to a conducting state. At this point, the probability of a path connecting the two distant boundaries becomes greater than zero. (This specific probability of connecting distant boundaries is termed the ``sponge-crossing probability,'' about which we will delve into more details in Section~\ref{sec_conpt}).

This phenomenon of a phase transition between two phases of different \emph{connectivity} is prevalent in real-life scenarios. An illustrative example from biology is the spread of epidemics~\cite{grassberger1986spreading,newman2002spread,pastor2002immunization,parshani2010epidemic,lindquist2011effective,vespignani2012modelling, pastor2015epidemic,croccolo2020spreading}. In its most basic manifestation, an epidemic commences with an infected individual who, with a probability denoted as $p$, can transmit the infection to its nearest neighbors over time until it propagates extensively. A comparable methodology can be applied to model forest fires, where the probability of a burning tree igniting its nearest neighbor tree in the subsequent time step replaces the infection probability~\cite{mackay1984forest, ritzenberg1984first}.
Another notable application of this concept can be found in polymerization processes within chemistry, where the activation of bonds between small branched molecules leads to the formation of larger molecules~\cite{de1979scaling, herrmann1986geometrical}. This transformation is known as a gelation transition. An illustrative example of this gelation process can be observed when boiling eggs.
Percolation theory has a wide range of other applications as well, spanning fields such as quantum systems~\cite{wave-localis_sah82,wave-localis_slg92,strong-disord_mmhf00,wave-localis_cb14,contin-percolation_fbcb14,meas-induc_srn19,percol_pteg19}, material science~\cite{gardner1992percolation,vigolo2005experimental,anekal2006dynamic}, geophysics~\cite{berkowitz1993percolation,berkowitz1998percolation,gaonac2003percolating,king2019percolation,ghanbarian2022soil}, social dynamics~\cite{kacperski2000phase,watts2002simple,centola2007cascade,iniguez2009opinion}, and infrastructures~\cite{echenique2005dynamics,vespignani2010fragility,li2015percolation,del2016spreading,baggag2018resilience}.

\subsection{Percolation of single-layer networks}

Percolation theory is closely associated with a wide range of concepts of critical phenomena, including scaling laws, fractals, self-organization criticality, and renormalization, holding significance across diverse statistical physics disciplines~\cite{stauffer92}.
The traditional characterization of  phase transition  in percolation hinges on the statistical properties of \emph{clusters} near $p_\text{th}$.  For $p < p_\text{th}$, only finite clusters exist. As $p > p_\text{th}$, a unique, infinite cluster emerges. A crucial parameter is $P_{\infty}$, signifying the relative size of the infinite cluster, which exhibits a power-law near $p_{\text{th}}$~\cite{bunde2012}:
\begin{equation}
    P_{\infty} \sim\left|p-p_\text{th}\right|^\beta.
\label{EqBeta}
\end{equation}
The parameter $P_{\infty}$ serves as a measure of order within the percolation system and can be identified as the order parameter. 

If we exclude the infinite cluster (if it exists), then the rest of the finite clusters follow a distribution:
\begin{equation}
    n_s \sim s^{-\tau} e^{-s / s^*}.
\end{equation}
Here, $s$ is the cluster size, and $n_s$ is the number of clusters of size $s$. At criticality, the characteristic size $s^*$ diverges:
\begin{equation}
 s^* \sim\left|p-p_\text{th}\right|^{-1 / \sigma}.
\end{equation}
Consequently, the tail of the distribution $n_s$ becomes a power law, $n_s \sim s^{-\tau}$.

The mean cluster size, i.e.,~how large a finite cluster is on average, also diverges:
\begin{equation}
    \langle s\rangle \sim \sum_s s^2 n_s \sim \left|p-p_\text{th}\right|^{-\gamma},
    \label{EqGamma}
\end{equation}
with the same exponent $\gamma$ above and below $p_\text{th}$. 

Finally, the correlation length $\xi$, defined as the average distance between two sites on the same finite cluster, also diverges:
\begin{equation}
    \xi \sim\left|p-p_\text{th}\right|^{-\nu},
    \label{EqNu}
\end{equation}
again, with the same exponent $\nu$ above and below $p_\text{th}$.

These exponents, namely $\beta$, $\tau$, $\sigma$, $\gamma$, and $\nu$, encapsulate the critical behavior of key quantities associated with the percolation transition and are collectively referred to as the critical exponents. Notably, they satisfy the \emph{scaling relations}:
\begin{equation}
    \beta=\frac{\tau-2}{\sigma} \text{ and } \gamma=\frac{3-\tau}{\sigma}.
\end{equation}

It is worth emphasizing that these exponents exhibit universality, meaning they remain invariant irrespective of the specific structural attributes of the lattice (e.g., square or triangular) or the type of percolation (site or bond). Instead, their values are solely determined by the dimensionality of the lattice. 
Besides, at the critical point,  $\xi$ and $s^*$ also follow a relation,
\begin{equation}
    s^* \sim \xi^{d_f} .
\end{equation}
The exponent $d_f$ is often called the fractal dimension~\cite{bunde2012}, characterizing the structure of the infinite cluster at the critical
point. Assuming the dimension of the system is $d$, there is another relation between critical exponents, called the \emph{hyperscaling relation},
\begin{equation}
    d_f=d-\frac{\beta}{\nu}.
\end{equation}
Thus, the fractal dimension of the infinite cluster at $p_\text{th}$ is not a new independent
exponent but depends on $\beta$, $\nu$ and $d$. 

Finally, for complex network structures, similar critical exponents following Eqs.~\eqref{EqBeta}-\eqref{EqNu} can also be identified.
For example, in scale-free networks~\cite{Barabasi1999,stat-of-netw_ab02,albert2000error,degree-degree-distance_zms20,scale-free_mz23}, which are characterized by a power-law distribution $P(k)\sim k^{-\lambda}$ of its degree $k$, the values of critical exponents depend on the power-law exponent $\lambda$, as outlined in Table~\ref{TabCriticalExp}. As an essential process inherently associated with the notion of connectivity in networked systems, percolation has been generalized to models that go beyond undirected networks, with studies dedicated to directed networks~\cite{boguna2005generalized}, temporal networks~\cite{badie2022directed}, and, as we discuss in more detail in the next sections, networks of networks and hypergraphs.

\begin{table}
    \centering
    \begin{tabular}{lllllll}
\hline 
$\lambda$ & $\beta$ & $\gamma$ & $\nu$ & $\sigma$ & $\tau$ & $d_f$\\
\hline
(2, 3) & 1/(3-$\lambda$) & -1 & ($\lambda$-1)/(3-$\lambda$) & (3-$\lambda$)/($\lambda$-2) & (2$\lambda$-3)/($\lambda$-2) \\
(3, 4) & 1/($\lambda$-3) & 1 & ($\lambda$-1)/($\lambda$-3) & ($\lambda$-3)/($\lambda$-2) & (2$\lambda$-3)/($\lambda$-2) &
2($\lambda$-2)/($\lambda$-3)\\
(4, $\infty$) & 1 & 1 & 3 & 1/2 & 5/2 & 4\\
\hline
    \end{tabular}
    \caption{The critical exponents for the classical percolation transition in scale-free networks~\cite{cohen2002percolation}.\hfill\hfill}
    \label{TabCriticalExp}
\end{table}

\subsection{Percolation of networks of networks}
\label{sec_non}

\begin{figure}[h!]
\centering \includegraphics[width=0.44\textwidth]{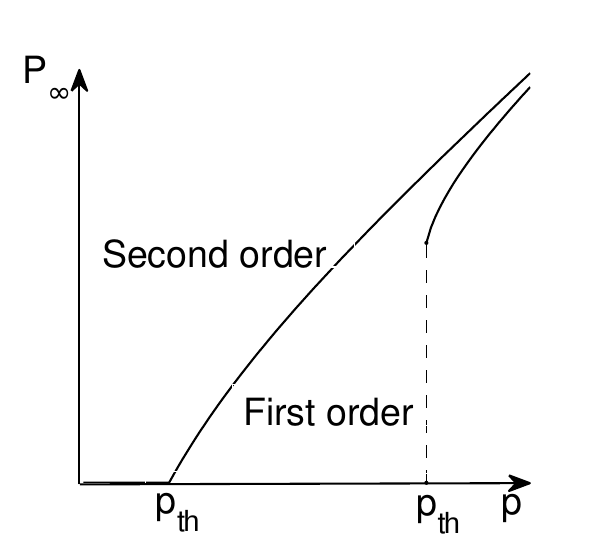}
\caption{\label{fig1_interdependent}{Schematic demonstration of first and second order
  percolation transitions.} In the second order case, the giant component is
  continuously approaching zero at the percolation threshold $p=p_{\text{th}}$. In
  the first order case the giant component approaches zero
  discontinuously.\hfill\hfill}
\end{figure}

In many real-world systems, an individual network is one component within a much larger complex network of interdependent networks~\cite{gao2011robustness,gao2012networks,kivela2014multilayer,gao2022introduction}. In interdependent networks, the failure of nodes in one network leads to the failure of dependent nodes in other networks, which may cause further damage to the first network, leading to cascading failures and possibly catastrophic consequences. In 2010, Buldyrev et al.~studied the percolation of two fully interdependent networks subject to cascading failures based on a generating function formalism. They found a surprising first-order discontinuous phase transition, dramatically different from the second-order continuous phase transition in single-layer networks~\cite{buldyrev2010catastrophic}, as shown in Fig.~\ref{fig1_interdependent}. Later, Parshani et al.~studied two partially interdependent networks and found that the percolation transition changes from first to second order as the coupling strength decreases~\cite{parshani2010interdependent}. Considering a malicious attack, Huang et al.~developed a mathematical framework for understanding the percolation of two interdependent networks under targeted attack, later extended to targeted attack on partially interdependent networks~\cite{huang2011robustness}. Each node in one network may depend on multiple nodes in another network. Therefore, Shao et al.~proposed a theoretical framework for understanding the percolation of interdependent networks with various support and dependence relationships~\cite{shao2011cascade}. The study of interdependence between networks also led researchers to realize that other types of interactions are important. One example closely related to interdependence is antagonistic interactions~\cite{zhao2013percolation}. Here, for a node to be active, the antagonistic node in another network has to be active, as can happen if each pair of nodes competes for some limited resource. Considering that more than two networks may depend on one another, Gao et al.~developed the analytical framework to study the percolation of a network formed by $n$ interdependent networks~\cite{gao2012robustness}, which was later extended to the study of targeted attacks on high-degree nodes~\cite{dong2012percolation,dong2013robustness}. Baxter et al.~studied the percolation of multiplex networks, which can be considered as the percolation of tree-like network of networks in Ref.~\cite{baxter2012avalanche}. Liu et al.~developed a theoretical framework based on generating functions and percolation theory to understand the percolation of interdependent directed networks~\cite{liu2020robustness}. In the past decade, we have witnessed fruitful results and discoveries related to the percolation of networks of networks~\cite{d2014networks,reis2014avoiding,boccaletti2014structure,bianconi2018multilayer,gomez2013diffusion,liu2022network,bashan2013extreme,lu2014synchronization,ouyang2017mathematical,duan2019universal,cellai2013percolation,kenett2015networks,de2023more}, as well as multilayer networks and interconnected networks.

In general, the percolation of networks of networks extends that of  single-layer networks. For example, when the $n$ interdependent Erd\H{o}s--R\'{e}nyi (ER) networks form a tree-like topology and have the same average degree $\bar{k}$, $\bar{k}_i=\bar{k}$ ($i=1,2,...,n$),  the giant connected component in each layer, $P_\infty$, as a function of $\bar{k},p,$ and $n$ follows~\cite{gao2011robustness}:
\begin{equation}\label{chap4:eq22}
P_{\infty}=p[1-\exp(-\bar{k}P_{\infty})]^n.
\end{equation}
Note that for Eq.~(\ref{chap4:eq22}), the particular case $n=1$ is the known ER second order percolation law for a single-layer network~\cite{erdos-renyi_er59,ER1960,Bollob1985}. When $n\ge 2$, the system shows a first-order phase transition. Using the generating function, we obtain that $p_{\text{th}}$ and $P_{\infty}|_{p\to p_{\text{th}}^{+}}$ satisfy
\begin{equation}\label{chap4:eq19}
p_{\text{th}} = -\frac{w}{\bar{k}[1+1/(nw)]^{n-1}},
\end{equation}
and
\begin{equation}\label{chap4:eq20}
P_{\infty}|_{p\to p_{\text{th}}^{+}} = -(w+1/n)/ \bar{k}.
\end{equation}
where $w$ is given by $w = W_-(-1/n\exp(-1/n))$,
and $W_-(x)$ is the smallest of the two real roots of the Lambert equation $\exp(W_-)W_-=x$. For $n=1$ we obtain the known ER results $p_{\text{th}}=1/\bar{k}$, and $P_{\infty}|_{p\to p_{\text{th}}^{+}}=0$ at $p=p_{\text{th}}$. Substituting $n=2$ in Eqs.~(\ref{chap4:eq19}) and (\ref{chap4:eq20}), we obtain the exact results derived by Buldyrev et al.~\cite{buldyrev2010catastrophic}. 

\subsection{Percolation of hypergraphs}
\label{sec_percolation_hypergraph}
Hypergraphs generalize graphs by allowing that interactions, the hyperedges,  connect an arbitrary number of vertices~\cite{bretto2013hypergraph}. Hypergraphs, and, up to a certain extent, simplicial complexes, offer more flexibility to model interacting systems,  and they have become popular models of many real-world networks over recent years~\cite{bianconi2021hyper,bick2023hyper}. For example, more than two molecules can participate in some reactions~\cite{klamt2009biology,jost2019hyperL}, and group interactions also frequently occur for collaborations of scientific papers~\cite{taramasco2010collab,krumov2011collab}. It has been shown that higher-order interaction may significantly change the physical properties of dynamical processes from those on ordinary networks with only pairwise connections~\cite{neuhauser2020multibody,bianconi2021hyper,high-ord-netw_babbfafiklmmpvp21,majhi2022higher}. However, there are only a few works exploring the robustness or the percolation of hypergraphs~\cite{coutinho2020percolhyper,sun2021percolhyper,sun2023percolhyper,lee2023kqcore,peng2022hyperattack,peng2023hypermp}. Specifically, Coutinho et al.~introduced two generalizations of core percolation to hypergraphs, and offered analytical solutions to certain types of random hypergraphs accordingly~\cite{coutinho2020percolhyper}. Sun and Bianconi later proposed a general framework for accessing hypergraph robustness, and further characterized the critical properties of simple and higher-order percolation processes~\cite{sun2021percolhyper}. Sun et al.~also considered a paradigmatic type of higher-order interactions, triadic interactions, where a node regulates the interaction between two other nodes, and provided a general theory, accurately predicting the full phase diagram on random graphs~\cite{sun2023percolhyper}. More recently, Bianconi and Dorogovtsev have further developed a theory for hyperedge and node percolation on hypergraphs, and showed that, in contrast to ordinary networks, the node and hyperedge percolation problems for hypergraphs strongly differ from each other~\cite{bianconi2023percoltheory}.    

\section{Classical percolation in quantum networks}
\label{sec_classical}

Why is percolation theory useful in the study of QN? 
The roots of this interest can be traced back to a 2007 paper~\cite{QEP_acl07}. In the seminal work, the authors first proposed a mapping between percolation theory and a particular entanglement transmission scheme, which they discovered and accordingly termed as the classical entanglement percolation (CEP) scheme. Within this context, an entanglement transmission \emph{scheme} refers to a (possibly infinite) series of quantum communication protocols that may be applied collectively to a QN for distributing entanglement between two nodes. This pioneering discovery has opened up a new approach to studying QN from a \emph{statistical physics} perspective, with a focus on understanding the large-scale, collective characteristics of the entanglement transmission task and how they are influenced by the topology of the QN.

\subsection{Classical entanglement percolation (CEP)}

As previously noted, the LOCC cannot increase the average entanglement. However, it does not mean that one cannot use LOCC as a form of ``gambling"---to enhance the entanglement with a certain probability $p$, even though it might reduce the entanglement with probability $1-p$. This principle forms the foundation of the CEP scheme.
To be specific, the CEP scheme involves two steps~\cite{QEP_acl07}. First, we ``gamble'' to enhance the entanglement of each link, aiming to obtain a singlet (maximally entangled state) with a probability of $p$. The optimal probability for this is referred to as the singlet conversion probability, given by $p=2\sin^2\theta$. Second, if a path of links connecting the source ($s$) and target ($t$) has all been converted to singlets, then a specific protocol known as entanglement swapping can be applied. This protocol converts every two singlet links sharing a common node (Relay, R) into a single singlet linking the two end nodes. For example, if there is a singlet between Alice and the Relay, and another between the Relay and Bob, the entanglement swapping protocol can merge the two into one singlet between Alice and Bob. By applying this protocol recursively along the singlet path connecting $s$ and $t$, we arrive at a final singlet between $s$ and $t$, fulfilling the transmission task.

Equipped with these concepts, the mapping between CEP and (classical) percolation theory is straightforward. The probability $p=2\sin^2\theta$ represents the probability for links to be present or absent.  The CEP scheme succeeds if $s$ and $t$ are connected after the random percolation process is applied. Furthermore, this connection implies a nontrivial critical threshold for the CEP scheme on infinitely large QN. Specifically, when $2\sin^2\theta$ falls below the percolation threshold $p_\text{th}$, $s$ and $t$ are almost certainly disconnected if they are infinitely apart. Hence, $p_\text{th}$, which solely depends on the network topology, serves as a metric of the overall capacity of the QN in the context of CEP.

The CEP scheme represents a great simplification of the QN entanglement transmission task to a pure percolation problem. Nevertheless, the CEP is not necessarily the optimal scheme. Indeed, even when $2\sin^2\theta \le p_\text{th}$, there might still be other schemes that can fulfill the transmission task, as we will explore in the following sections.

\subsection{Quantum entanglement percolation (QEP)}

It is expensive to obtain a singlet from a partially entangled state given its ``gambling'' nature. Even worse, the swapping protocol spends all the singlets along a path and converts them into just one singlet. This process leads to a waste of singlets and causes the inefficiency of the CEP scheme. Naturally, this leads to the question: Is it necessary to convert \emph{every} link into a singlet? As we will see, the answer is negative, paving the way for the QEP scheme~\cite{QEP_acl07}.

The QEP scheme is based on the discovery that given two partially entangled states between three parties, Alice--Relay--Bob, there exists a LOCC protocol that can yield a \emph{higher} probability of obtaining a singlet between Alice and Bob. This probability is higher than obtaining two singlets (Alice--Relay, Relay--Bob) individually and followed by a swapping protocol. Indeed, the optimal probability is found to be $\min\{2\sin^2\theta_\text{AR},2\sin^2\theta_\text{RB}\}$~\cite{QEP-series-rule_bvk99}, outperforming the probability $\left(2\sin^2\theta_\text{AR}\right)\left(2\sin^2\theta_\text{RB}\right)$ achieved by CEP.

What about three partially entangled states between four parties (Alice--Relay1--Relay2--Bob)? Unfortunately, the optimal conversion probability does not intuitively simplify to $\min\{2\sin^2\theta_{\text{A}\text{R}_1},2\sin^2\theta_{\text{R}_1\text{R}_2}, 2\sin^2\theta_{\text{R}_2\text{B}}\}$, but takes on a much more complicated form. This forbids us to generalize the optimal result to larger scale. 
For readers who wish to delve deeper, further details can be found in Ref.~\cite{QEP-detail_pcalw08}.

Even though we cannot generalize the optimal result, we can still extend the \emph{improvement}, by bypassing one relay every other step. This gives rise to the QEP scheme, which avoids the need to create singlets for every link, bypassing (half of) the Relays. But this approach is not without its trade-offs, especially on a large scale. Since the Relays are bypassed, the QN misses out on the potential connectivity to other paths through the Relays. Thus, it still remains a question whether QEP can achieve a lower critical threshold than CEP, fulfilling the entanglement transmission task at infinite scales. The authors of the 2007 paper~\cite{QEP_acl07} demonstrated that this is indeed achievable for specific topologies, such as a ``double''-honeycomb topology, where there are two links between every two adjacent nodes on a hexagonal network. The QEP scheme is equivalent to adding a preprocessing step, modifying the network into a triangular structure and thereby reducing the percolation threshold.

Note that despite being referred to as QEP, the mapping of the scheme is still aligned with classical percolation theory from a statistical physics point of view. The quantum aspect of this process is confined primarily to the preprocessing step, which is executed only at the local scale. Additionally, the QEP does not yield the optimal result either~\cite{QEP-detail_pcalw08}, which leaves open the question of whether a more effective entanglement transmission scheme might exist. It would be intriguing if this new scheme were guided by a fundamentally different statistical physical theory distinct from classical percolation. 
We will show that such a theory does exist.

\begin{figure}[t!]
	\centering
	\includegraphics[width=243pt]{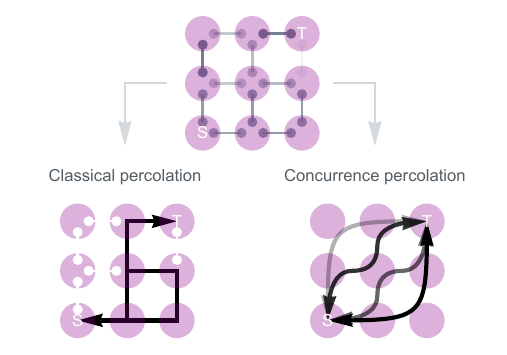}
	\caption{\label{fig_percolation}{Entanglement transmission on quantum networks can be understood from two different mappings.} In classical percolation {(bottom left)}, each link is present or absent with probability $p$ or $1-p$, respectively. Only the paths in which all links are fully present contribute to the connectivity. This forms the basis of the classical/quantum entanglement percolation (CEP/QEP) schemes~\cite{QEP_acl07},  with~the goal of securing a singlet between source $s$ and target $t$ through a ``gambling'' approach.
 In concurrence percolation {(bottom right)}, every path contributes to the connectivity. This mapping forms the basis of the deterministic entanglement transmission (DET) scheme~\cite{det_mcghr23}, where the aim is not to obtain a singlet probabilistically but to establish a partially entangled state deterministically.\hfill\hfill
}
\end{figure}

\section{Concurrence percolation in quantum networks}
\label{sec_quantum}

\subsection{No need to establish singlets}

The mapping of CEP/QEP to classical percolation is essentially established on the necessity that \emph{two nodes must be connected by at least one
path of singlets}. However, we have seen that in the QEP scheme, it is not mandatory for all links along the path to be converted to singlets. By bypassing some of the links, a more efficient scheme might be realized.

This observation leads us to a natural question: \emph{why not stop establishing singlets at all?} In other words, we would bypass not just some, but all links, resulting in a final state between the source $s$ and target $t$ that remains only partially, rather than maximally, entangled. This scenario is, of course, approachable, considering that one can always ``downgrade'' a singlet to a partially entangled state with no cost~\cite{nielsen_n99}.
What we truly seek, however, is a trade-off where we can achieve a much higher probability of obtaining such a partially entangled state instead of a singlet. By carefully weighing the compromise (having only partial entanglement) against the benefit (a significantly higher conversion probability), we might discover a more advantageous scheme for entanglement transmission overall. This revised approach challenges the conventional thinking in terms of classical percolation and could lead to new opportunities in developing new schemes on QN.

\subsection{Deterministic entanglement transmission (DET)}
Based on the above ideas, a new scheme named the deterministic entanglement transmission (DET) scheme is introduced~\cite{det_mcghr23}. The DET approaches the entanglement transmission task from a completely distinct perspective: the scheme demands that the conversion probability throughout the process always equals \emph{one}.  In other words, rather than ``gambling'' to increase the links' entanglement, we operate directly on partially entangled states in a deterministic fashion. The aim of DET is to maximize the final (partial) entanglement under the constraint of determinacy, contrasting with CEP/QEP's objective of always acquiring a singlet with high (but not unit) probability.

The DET involves two quantum communication protocols:  The first is a continuation of the swapping protocol~\cite{entangle-swap_zzhe93, QEP-series-rule_bvk99}. However, here the swapping protocol operates directly on partially entangled states. It can be shown that given entanglements $\theta_\text{AR}$ and $\theta_\text{RB}$ in the A--R--B configuration, one can tune the swapping protocol such that it deterministically yields a final state between A and B, having a new entanglement $\theta_\text{AB}$ that satisfies $\sin 2\theta_\text{AB} =\sin 2\theta_\text{AR}\sin 2\theta_\text{RB}$~\cite{det_mcghr23}. The second protocol is the entanglement concentration protocol~\cite{entangle-conc_bbps96, QEP-parallel-rule_v99}. This protocol takes two links between A and B (with entanglement $\theta_1$, $\theta_2$, respectively) as input. At the expense of the two links,  a new link that has a higher entanglement $\theta$ is produced between A and B, where $\cos \theta= \cos \theta_1 \cos \theta_2$ or $\sqrt{1/2}$, whichever is the largest.

The DET scheme is founded on the generalization of these two protocols to global scales. This is possible since the swapping and concentration protocols are fully equivalent to the series/parallel rules, respectively, as often employed in circuit network analysis~\cite{det_mcghr23}. Therefore,  the DET scheme becomes applicable if the network topology between the source $s$ and target $t$ is series-parallel~\cite{series-parallel-netw_d65}. In other words, the network can be fully reduced to a link between $s$ and $t$ using only series and parallel reductions. Note that there are already fast network algorithms designed specifically for detecting series-parallel networks and determining their feasible reductions, which will be explored further in the next section.

One of the most intriguing characteristics of the DET scheme is that, when applied to infinitely large series-parallel networks, a threshold similar to the CEP threshold is observed~\cite{conpt_mgh21}, below which the DET can never produce nonzero entanglement between $s$ and $t$. This threshold, however, is \emph{lower} than the CEP threshold, demonstrating a ``quantum advantage'' over the classical scheme.
The existence of such a threshold suggests that the DET may be globally governed by a statistical physics theory. In subsequent sections, we will explore this statistical theory in details.

\subsection{Concurrence percolation theory}
\label{sec_conpt}

To establish the statistical theory, recall that in classical percolation theory, given a regular lattice, one can define a ``sponge-crossing'' quantity, $P_\text{SC}$, 
as the probability that there is an open path connecting two far-apart boundaries~\cite{kesten1980critical}.   When the lattice becomes infinitely large (the number of nodes $N\to \infty$), it is known that a minimum value of $p$ exists, below which $P_\text{SC}$ becomes zero in the thermodynamic limit:
\begin{align}
\label{eq_pth}
p_{\text{th}} = \text{inf}\left\{p \subset [0, 1] | \text{lim}_{N \rightarrow \infty} P_\text{SC} > 0\right\}.
\end{align}
This minimum value coincides with the traditionally defined percolation threshold $p_{\text{th}}$ in Section~\ref{sec_percolation}, which is based on the size of the infinite cluster. 
As a result, Eq.~\eqref{eq_pth} offers an alternative definition of $p_{\text{th}}$.
In the special case of two-dimensional square lattices, Kesten proved that the ``sponge-crossing'' probability $P_\text{SC}$ of connecting the left and right boundaries is strictly zero until $p> 1/2$. Thus, $p_{\text{th}}=1/2$~\cite{kesten1980critical}.

Moreover, the existence of such a critical threshold is not limited to regular lattices but can also be observed for complex network topologies. All we need to do is to
generalize $P_\text{SC}$ from being defined between two apparent boundaries to two arbitrary sets of nodes, denoted $S$ and $T$. It is reasonable to believe that there still exists a nontrivial ConPT threshold $p_{\text{th}}$ for this generalized $P_\text{SC}$, as long as the minimum length of all paths connecting $S$ and $T$ increases with the network size $N$.
We contract the two sets $S$ and $T$ into two ``mega'' nodes, which amounts to erasing the internal network topologies of $S$ and $T$, and then calculate the ``sponge-crossing'' probability $P_\text{SC}$ between them. This provides us a definitive way of calculating $P_\text{SC}$ for arbitrary network topology and inferring $p_{\text{th}}$ from Eq.~\eqref{eq_pth}.

How to derive $p_{\text{th}}$? First, we need to know how $P_\text{SC}$ manifests as a function of $p$. The exact expression of $P_\text{SC}$ can be calculated by basic addition and multiplication rules of probability measures~\cite{stauffer92}. In general, $P_\text{SC}$ bears the form as a ratio of two large polynomials of $p$ (i.e.,~meromorphic in $p$), which quickly becomes complex when the number of links becomes large. Nevertheless, when the network topology between $S$ and $T$ is series-parallel, then $P_\text{SC}$ can be decomposed into the iteration of two connectivity rules, namely, the series and parallel rules (Table~\ref{table_rules}). these rules establish the probability of \emph{at least} one path connecting the two ends.

\begin{table}[t!]
	\centering
	\caption{\label{table_rules}{Connectivity rules that define the classical/concurrence percolation theories.}\hfill\hfill}
		\begin{tabular}{p{2.6cm}| p{4.0cm}| p{4.4cm}}
		\hline\hline
		& Classical & Concurrence \\
		\hline
		Series rule & $p=p_1 p_2\cdots$ & $c=c_1 c_2\cdots$ \\
		\hline
		Parallel rule & $1-p=$\newline
		$\left(1-p_1\right)\left(1-p_2\right)\cdots$ &  $\frac{1+\sqrt{1-c^2}}{2}=\max\{\frac{1}{2},$\newline
		$\frac{1+\sqrt{1-c_1^2}}{2} \frac{1+\sqrt{1-c_2^2}}{2}\cdots\}$ \\
		\hline

Higher-order~rules & 
\multicolumn{2}{p{8.7cm}}{
	 Can be approximated by the \emph{star-mesh transform}, by the following two-step argument:
}
\\
\cline{1-1}
\end{tabular}
\begin{tabular}{p{6.8cm}p{4.2cm}}
\vspace{-1.0mm}
1. The star-mesh transform can reduce an $N$-graph to an $(N-1)$-graph (right panel) and is solvable by applying the series and parallel rules recursively through a group of $N(N-1)/2$ coupled equations (see Section~\ref{sec_algorithm} for details).
\newline
2. Applying the transform consecutively on a network can reduce nodes one by one---and thus reduce any topology to two nodes, yielding the final (approximate) connectivity between them (bottom panel, i.~$\to$~viii.).
&
\begin{center}
	\vspace{-5mm}
	{\includegraphics[width=2.4cm]{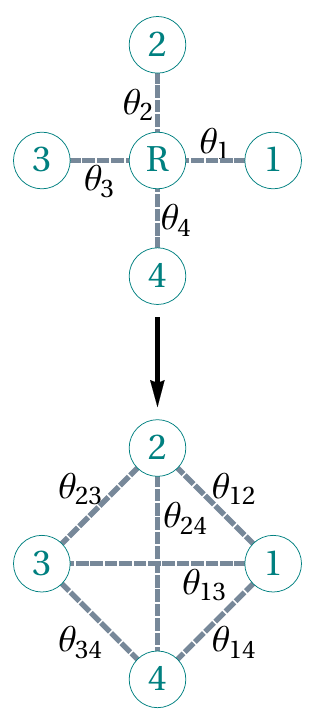}}
\end{center}
\\	
\multicolumn{2}{p{8.0cm}}{
	\begin{center}	
		\vspace{-17mm}	
		{\includegraphics[width=8.3cm]{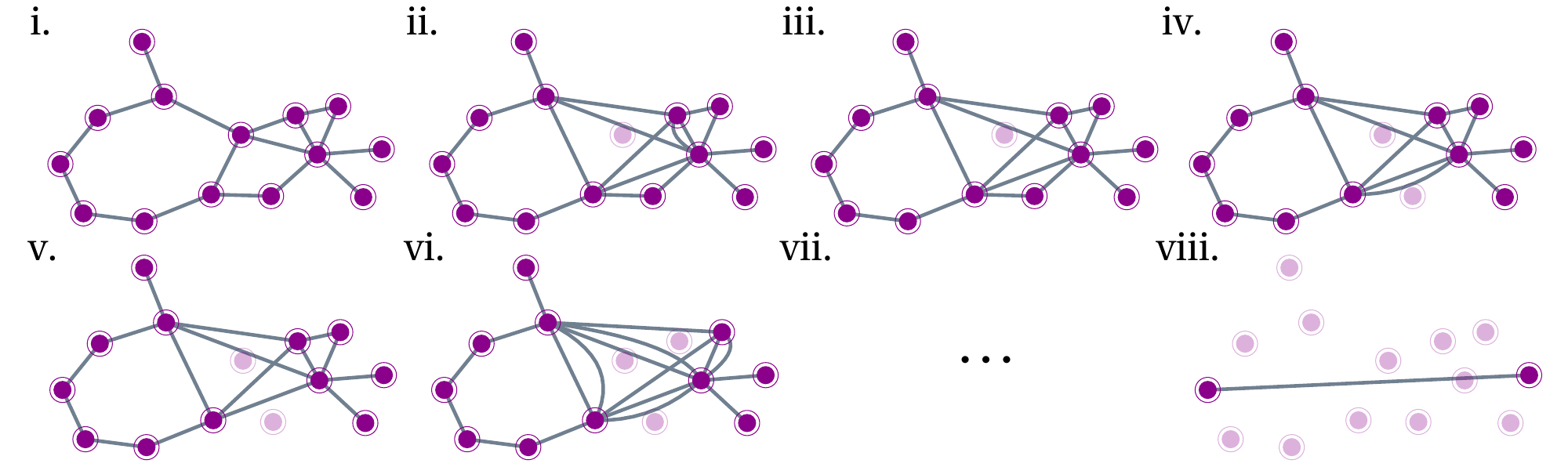}}
		\vspace{-7mm}
	\end{center}
}  \\		

		\hline\hline
	\end{tabular}
\end{table}

When the network topology between $S$ and $T$ is not series-parallel (such as a bridge circuit~\cite{series-parallel-netw_d65}), higher-order connectivity rules are needed.
Unlike the series/parallel rules, these higher-order rules do not follow a general form. That being said, there is a way to approximate (see Section~\ref{sec_algorithm} for a detailed discussion about the nature of this approximation)  these higher-order rules using \emph{only} the series/parallel rules. This technique is know as the star-mesh transform, originating from the study of circuit analysis. We will include more details on this technique in the Algorithms section (Section~\ref{sec_algorithm}).

It becomes clear that the series/parallel rules play a very special rule in defining percolation connectivity. Given that the DET scheme is also founded on series/parallel rules, a natural and intriguing question arises: can we define a new statistical theory \emph{reversely}, starting directly from the exact forms of series/parallel rules? Such a definition may not be complete, since we do \emph{not} know the exact forms for higher-order rules (which may not even have a closed mathematical form). Yet, using the star-mesh transform technique, it may be possible that we can approximate these higher-order rules using the series/parallel rules that we have known.

This is the motivation of the concurrence percolation theory. Recall that the DET series/parallel rules are given by $\sin 2\theta =\sin 2 \theta_1 \sin 2 \theta_2$ and $\cos \theta=\max\{\sqrt{1/2},\cos \theta_1\cos \theta_2\}$, respectively. Under a change of variable $c\equiv \sin 2\theta$, the rules can be rewritten in the form presented in Table~\ref{table_rules}. 
Note that the new series rule in terms of $c$ bears the same nominal form as the classical series rule in terms of $p$. This variable $c$, indeed, has a physical meaning, known as the \emph{concurrence}, a specific measure of bipartite entanglement~\cite{concurrence_hw97}. This
inspires the new theory to be termed ``concurrence percolation.'' In comparison to classical percolation where the variable of interest is the probability $p$, in concurrence percolation, choosing $c$ as the variable of interest is appropriate.

After fixing all connectivity rules (series + parallel + star-mesh), an analogous quantity, $C_\text{SC}$, referred to as the sponge-crossing concurrence can be defined as the weighted sum of all paths in terms of this new weight $c$~\cite{conpt_mgh21}. It is believed that a nontrivial threshold on $c$ also exists:
\begin{align}
\label{eq_cth}
c_{\text{th}} = \text{inf}\left\{c \subset [0, 1] | \text{lim}_{N \rightarrow \infty} C_\text{SC} > 0\right\},
\end{align}
such that $c_{\text{th}}$ is the minimum value of the concurrence $c$ per link, below which $C_\text{SC}$ becomes zero when $S$ and $T$ become infinitely distant.

\subsection{Results}

\begin{figure}[t!]	
	\centering
	\includegraphics[width=243pt]{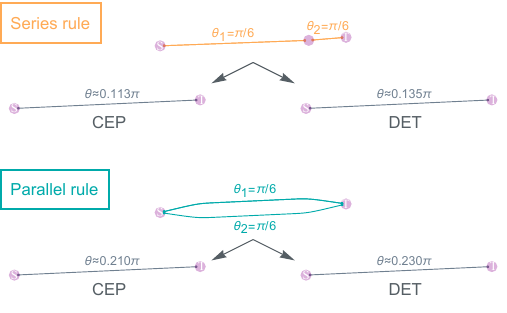}
	\caption{\label{fig_cep_det_example}The DET series/parallel rules outperform the CEP series/parallel rules.
    \hfill\hfill}
\end{figure}

\emph{DET outperforms CEP}.---Utilizing the framework of concurrence percolation, we can derive an essential and powerful result: the DET scheme always \emph{outperforms} the CEP scheme on any series-parallel QN. 
To rigorously demonstrate this comparative superiority, we rewrite both the classical and concurrence series/parallel rules in terms of the entanglement variable $\theta$ (Table~\ref{table_rules_scheme}).
These rules correspond to the entanglement transmission rules for CEP and DET, respectively (as illustrated in Fig.~\ref{fig_cep_det_example}).

\begin{table}[h!]
	\centering
	\caption{\label{table_rules_scheme}{Entanglement transmission rules for the CEP and DET schemes, derived from the classical and concurrence percolation rules (Table~\ref{table_rules}), respectively.}\hfill\hfill}
		\begin{tabular}{l| l| l}
		\hline\hline
		& CEP & DET \\
		\hline
		Series & $2 \sin^2 \theta= \left(2 \sin^2 \theta_1\right) \left(2 \sin^2 \theta_2\right) \cdots$ & $\sin 2\theta =\left(\sin 2\theta_1\right)\left(\sin 2\theta_2\right)\cdots$ \\
		\hline
		Parallel & $\cos 2\theta =\left(\cos 2\theta_1\right)\left(\cos 2\theta_2\right)\cdots$ &  $\cos \theta=\max\{\sqrt{1/2},\left(\cos \theta_1\right)\left(\cos \theta_2\right)\cdots\}$ \\
		\hline\hline
	\end{tabular}
\end{table}	

Now, for the series rule, we have 
\begin{eqnarray}
c^2&=&\prod_i \left(\sin 2\theta_i\right)^2\nonumber\\
&=&\prod_i \left(2 \sin^2 \theta_i\right)\left(2-2 \sin^2 \theta_i\right)\nonumber\\
&\ge&\left[\prod_i \left(2 \sin^2 \theta_i\right)\right] \left[2-\prod_i \left(2 \sin^2 \theta_i\right)\right]\nonumber\\
&=&p\left(2-p\right),
\end{eqnarray}
where the inequality is supported by the subadditivity of $f(x)=\ln(2-e^{-x})$, namely, 
\begin{equation}
    f(x_1+x_2+\cdots)\le f(x_1)+f(x_2)+\cdots
\end{equation}
for $x=-\ln \left(2\sin^2\theta\right) \ge0$. This leads to
\begin{equation}
    1-c^2\le 1-p\left(2-p\right)=\left(1-p\right)^2.
\end{equation}
This final inequality underscores that the $\theta$ obtained from the CEP series rule (under a change of variable $p=2\sin^2 \theta$) is never greater than the $\theta$ obtained  from the DET series rule (under a change of variable $c=\sin 2 \theta$).

For the parallel rule, similarly we have
\begin{eqnarray}
\frac{1}{2}+\frac{1}{2}\sqrt{1-c^2}&=&\prod_i \cos^2\theta_i\nonumber\\
&=&\prod_i\left(\frac{1}{2}+\frac{1-2\sin^2\theta_i}{2}\right)\nonumber\\
&\le&\frac{1}{2}+\frac{1}{2}\prod_i\left(1-2\sin^2\theta_i\right)\nonumber\\
&=&1-\frac{p}{2},
\end{eqnarray}
where the inequality is supported by the subadditivity of $f(x)=-\ln(1/2+e^{-x}/2)$ for $x=-\ln \left(1-2\sin^2\theta\right) \ge0$. This further leads to
\begin{equation}
    \frac{\sqrt{1-c^2}}{2}\le 1-\frac{p}{2}-\frac{1}{2}=\frac{1-p}{2},
\end{equation}
which, again, underscores that the $\theta$ obtained from the CEP parallel rule is never greater than the $\theta$ obtained from the DET parallel rule. Together, it can be established that the DET rules consistently yield superior results to those of the CEP rules, both in series and parallel configurations. This underlines the potential of DET as a valuable tool in the ongoing development of large-scale QN and adds a new dimension to our understanding of quantum connectivity.

\begin{figure}[t!]	
	\centering
	\begin{minipage}[b]{7cm}
		\centering\subcaption{\label{fig_bethe_lattice}$\qquad\qquad\qquad\qquad\qquad\qquad\qquad$}
        \vspace{1mm}
		\includegraphics[width=5.5cm]{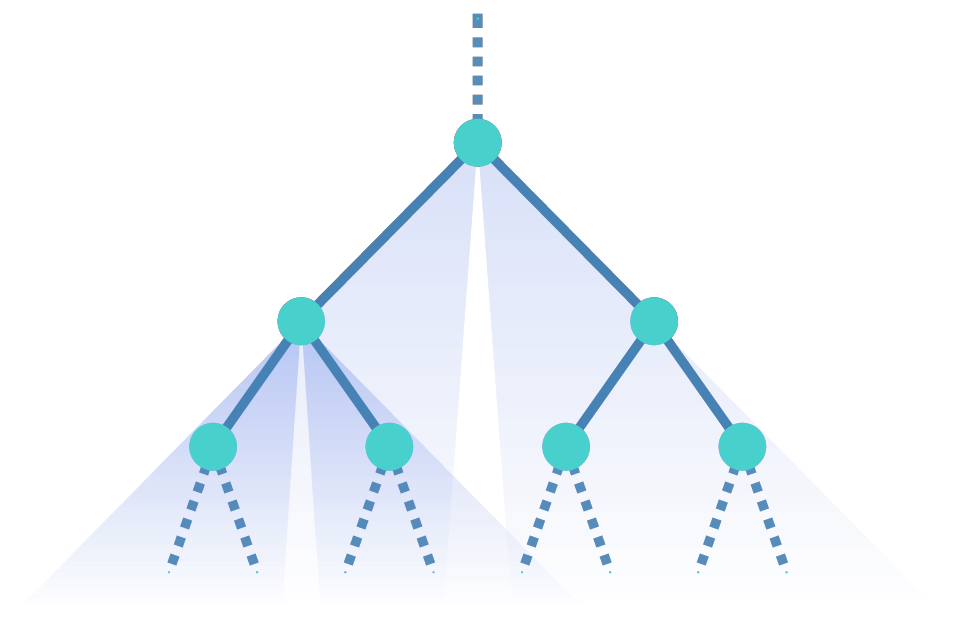}
	\end{minipage}
	\begin{minipage}[b]{6cm}
		\centering
        \subcaption{\label{fig_bethe_threshold}$\qquad\qquad\qquad\qquad\qquad\qquad\qquad$}
        \vspace{1mm}
        \includegraphics[width=6cm]{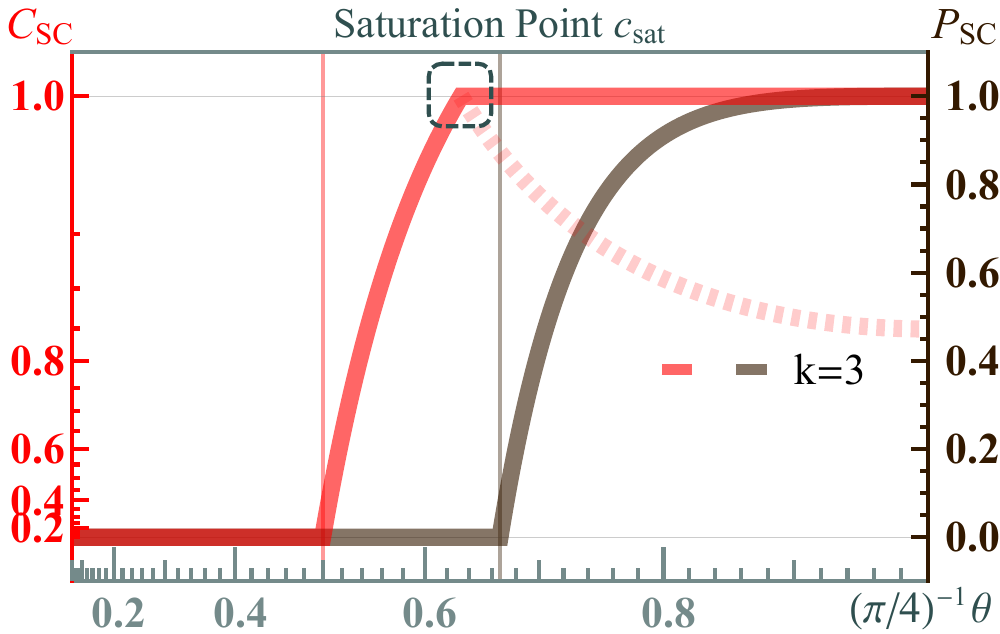}	
	\end{minipage}
	\caption{\label{fig_bethe}Classical percolation and concurrence percolation on the Bethe lattice. \subref{fig_bethe_lattice} The Bethe lattice ($k=3$). 
    \subref{fig_bethe_threshold} The sponge-crossing probability $P_\text{SC}$ (brown) between sets $S$ (the root) and $T$ (the collection of all leaf nodes) as a function of $\theta$. Driven by classical percolation, a transition threshold is found at $\theta=\pi/6$, or $p_\text{th}=2\sin^2\theta=1/2$. As a comparison, the sponge-crossing concurrence $C_\text{SC}$ (red), driven by concurrence percolation, shows a similar but lower threshold at $\theta=\pi/8$, or $c_\text{th}=\sin 2\theta=1/\sqrt{2}$. Moreover, a saturation point at $\theta=0.633\left(\pi/4\right)$, or $c_\text{sat}\approx 0.838$ also exists, beyond which we already have $C_\text{SC}=1$. This saturating feature has no counterpart in classical percolation. (The pink dashed line represents another nonphysical solution.)
    \hfill\hfill}
\end{figure}

\begin{figure}[t!]	
	\centering
	\begin{minipage}[b]{7cm}
		\centering
        \subcaption{\label{fig_square_lattice}$\qquad\qquad\qquad\qquad\qquad\qquad\qquad$}
        \includegraphics[width=3.7cm]{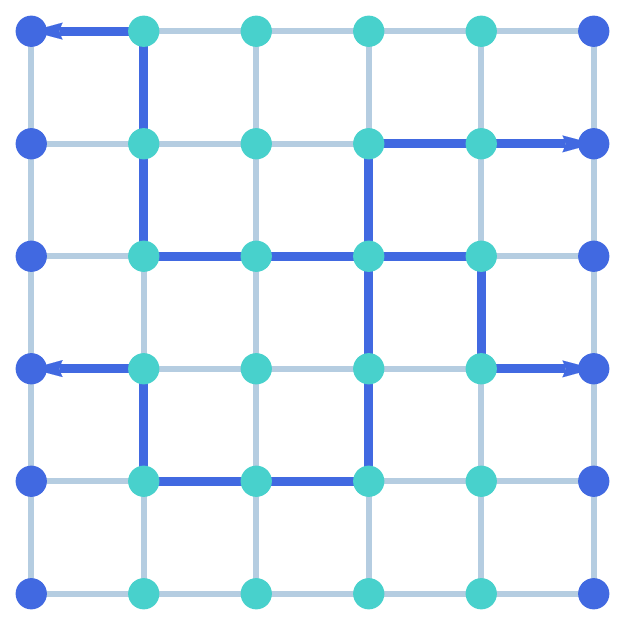}
	\end{minipage}
	\begin{minipage}[b]{6cm}
		\centering
        \subcaption{\label{fig_square_threshold}$\qquad\qquad\qquad\qquad\qquad\qquad\qquad$}
        \vspace{1mm}
        \includegraphics[trim={0 4cm 0 0},clip,width=6cm]{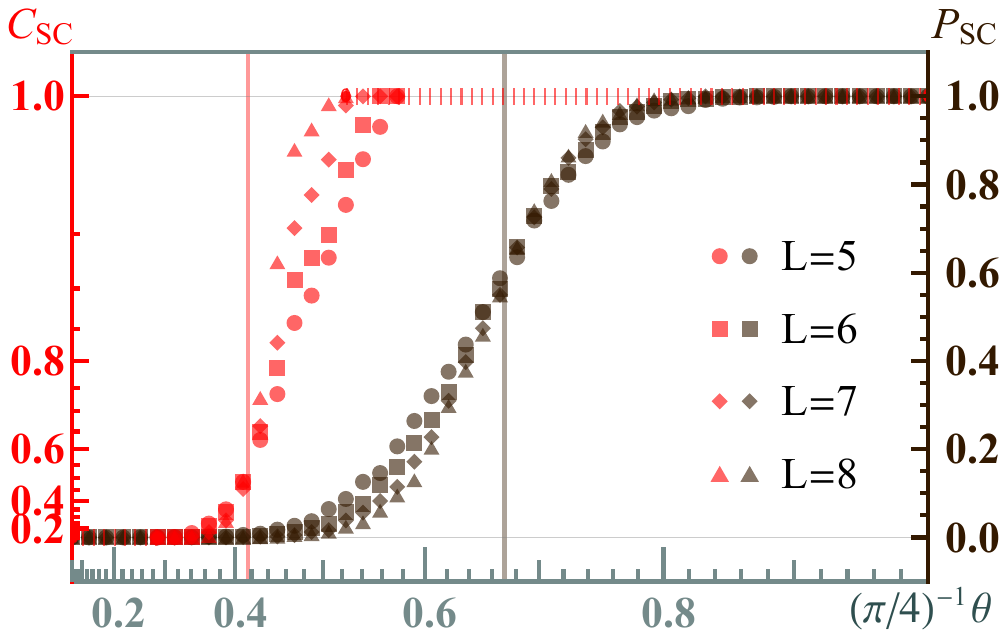}	
	\end{minipage}
    \begin{minipage}[b]{7cm}
		\centering
        \subcaption{\label{fig_hex_lattice}$\qquad\qquad\qquad\qquad\qquad\qquad\qquad$}
        \includegraphics[width=4.5cm]{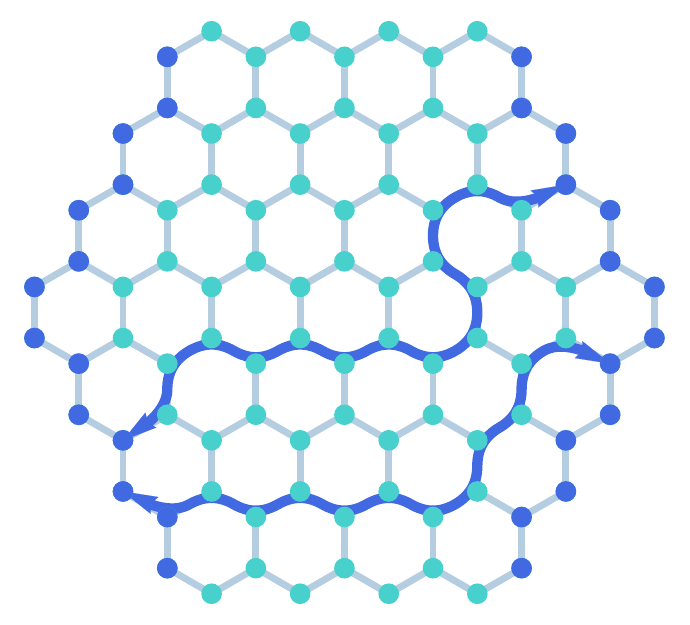}
	\end{minipage}
	\begin{minipage}[b]{6cm}
		\centering
        \subcaption{\label{fig_hex_threshold}$\qquad\qquad\qquad\qquad\qquad\qquad\qquad$}
        \vspace{1mm}
        \includegraphics[width=6cm]{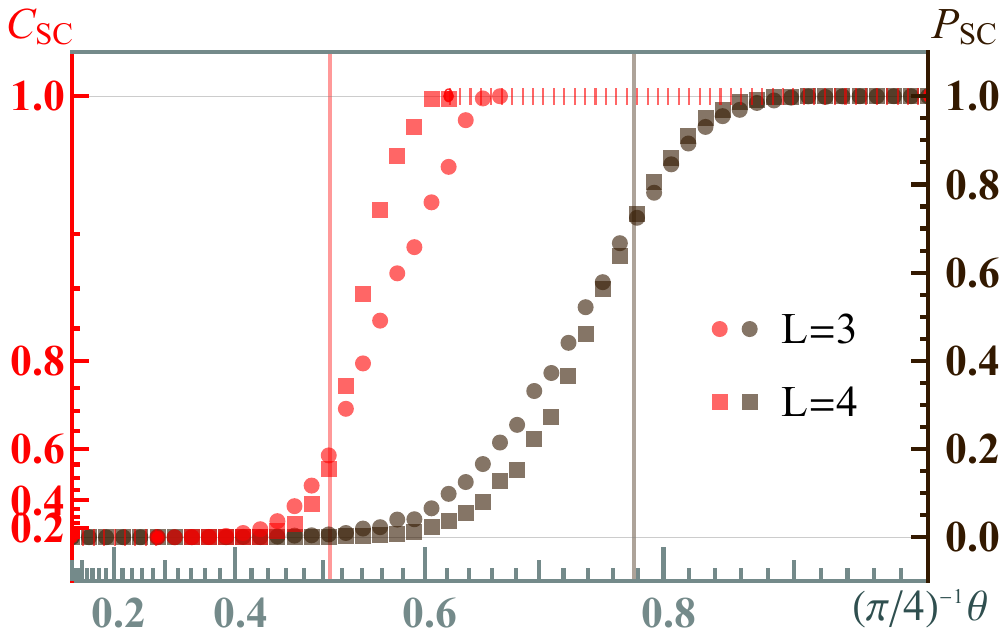}	
	\end{minipage}
    \begin{minipage}[b]{7cm}
		\centering
        \subcaption{\label{fig_tri_lattice}$\qquad\qquad\qquad\qquad\qquad\qquad\qquad$}
        \includegraphics[width=4.5cm]{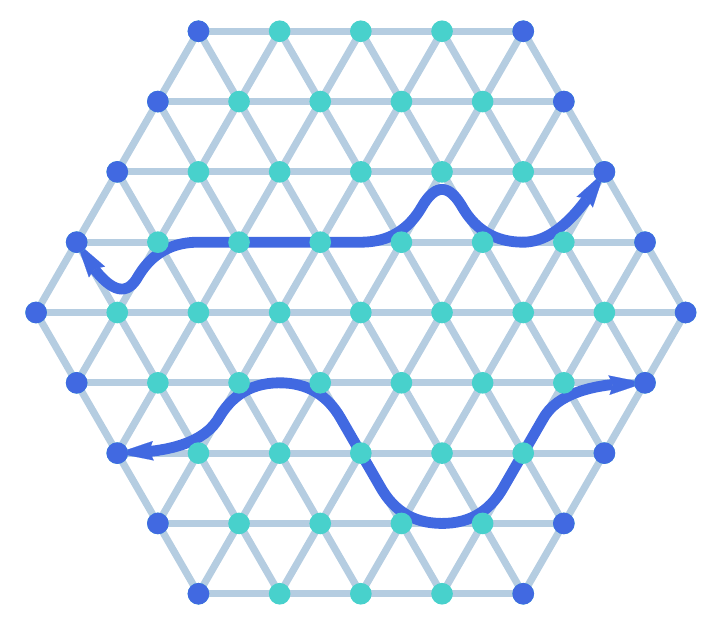}
	\end{minipage}
	\begin{minipage}[b]{6cm}
		\centering
        \subcaption{\label{fig_tri_threshold}$\qquad\qquad\qquad\qquad\qquad\qquad\qquad$}
        \vspace{1mm}
        \includegraphics[trim={0 4cm 0 0},clip,width=6cm]{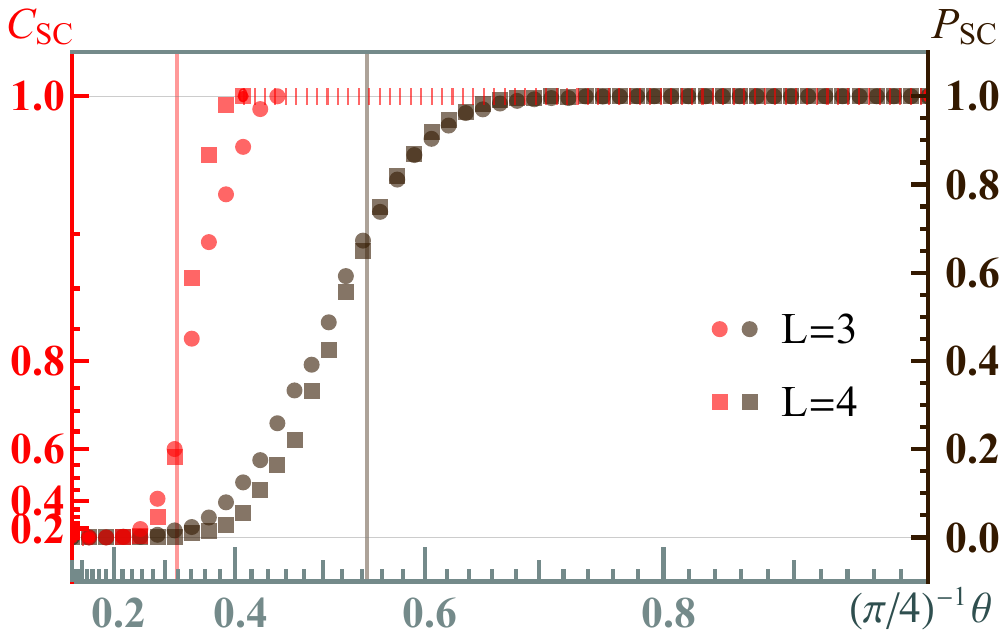}	
	\end{minipage}
	\caption{\label{fig_regular}Classical percolation and concurrence percolation on \subref{fig_square_lattice}-\subref{fig_square_threshold} the square lattice, 
    \subref{fig_hex_lattice}-\subref{fig_hex_threshold} the honeycomb lattice, and 
    \subref{fig_tri_lattice}-\subref{fig_tri_threshold} the triangular lattice, 
    The sponge-crossing probability $P_\text{SC}$ (brown) and sponge-crossing concurrence $C_\text{SC}$ (red) are defined between sets $S$ (the collection of nodes on the left boundary) and $T$ (the collection of nodes on the right boundary), as a function of $\theta$. The brown and red vertical lines denote the finite-size thresholds $p_\text{th}$ and $c_\text{th}$, respectively.
    \hfill\hfill}
\end{figure}

\emph{Concurrence percolation threshold}.---In infinite-size QNs, both a classical percolation threshold $p_\text{th}$ and a concurrence percolation threshold $c_\text{th}$ exist. 
This leads us to the second insightful finding within the realm of concurrence percolation: the prediction of a lower threshold compared to what was known from earlier classical-percolation-theory-based schemes, including CEP and its variants (such as QEP). 
What makes this result particularly interesting is that the improvement exists across various network topologies. Table~\ref{table_thresholds} shows these findings, detailing tests conducted on different network topologies, including the Bethe lattice (Fig.~\ref{fig_bethe}) as well as other regular lattices such as the square, honeycomb, and triangular lattices (Fig.~\ref{fig_regular}). 
This consistency across multiple configurations underscores the robustness of the concurrence percolation method, demonstrating its potential to redefine our understanding of entanglement transmission within large-scale QNs.

On series-parallel QN, this predicted concurrence percolation threshold is readily achievable using the DET scheme. On general network topologies, however, it is unknown if the higher-order connectivity rules produced by the star-mesh transform is realizable by LOCC. They are only approximations of the true LOCC-allowing rules. The study of the higher-order rules of concurrence percolation remains a difficult task that could be handled by multipartite strategies~\cite{QEP-GHZ_pclla10}, QN routing~\cite{q-netw-route_p19,q-netw-route_pkttjbeg19}, or QN coding~\cite{q-netw-code_klgnr10}.

\begin{table*}[h!]
	\centering
	\caption{\label{table_thresholds}
		The concurrence percolation threshold is the lowest threshold compared to earlier known classical-percolation-theory-based schemes, including CEP and its variants. Particularly, for the Bethe lattice, one has $P_\text{swap}(k)=2x-x^2$, where $x(k)$ is the solution of $2x+x^k(kx-x-k-1)-(1-x)/(k-1)=0$ by the $q$-swapping strategy~\cite{QEP-q-swap_cc09}, and $P_\text{GHZ}(k)=y$, where $y(k)$ is the solution of $1-(1-y)\sum_{i=0}^{\lfloor k/2-1\rfloor}{\binom{2i}{i} 4^{-i}(2y-y^2)^i}-1/(k-1)=0$ where $\lfloor \cdot\rfloor$ is the floor function~\cite{QEP-GHZ_pclla10}. For the square and triangular lattices, QEP yields the same thresholds as of CEP.
		\hfill\hfill}
	\medskip
	\vspace{-3mm}
	\begin{tabular}{p{2.8cm}|p{4.cm} p{1.3cm} p{1.9cm} p{1.7cm}}
		\hline\hline
		[unit: $\left(\pi/4\right)^{-1}\theta$] & Bethe lattice (degree $k$)& Square  & Honeycomb  & Triangular \\
		\hline\
		CEP~\cite{QEP_acl07}
		&$(4/\pi)\sin^{-1}[1/{\sqrt{2\left(k-1\right)}}]$ & $0.670$ &$0.777$ & $0.545$ \\
		QEP~\cite{QEP_acl07,QEP-q-swap_cc09,QEP-detail_pcalw08} 
		&$(4/\pi)\sin^{-1}\sqrt{P_\text{swap}(k)/2}$\footnotemark[1] & $0.670$ & $0.761$ & $0.545$\\
		QEP-GHZ~\cite{QEP-GHZ_pclla10} 
		&$(4/\pi)\sin^{-1}\sqrt{P_\text{GHZ}(k)/2}$\footnotemark[2] & $0.584$ & $0.745$ & $0.481$ \\
		Concurrence~\cite{conpt_mgh21}
		&$(2/\pi)\sin^{-1}(1/{\sqrt{k-1}})$ 
		& $0.42(8)$ & $0.51(8)$ & $0.32(8)$ \\
		\hline\hline
	\end{tabular}
\end{table*}

\emph{Saturation}.---Concurrence percolation also differs from classical percolation with the existence of a saturation point $c_{\text{sat}}$. Whenever $c\ge c_{\text{sat}}$, the sponge-crossing $C_\text{SC}$ consistently equals one [Fig.~\ref{fig_bethe_threshold}].
For example, basic calculations show that the exact value of the saturation point for a Bethe lattice of degree $k$ is given by~\cite{conpt_mgh21}
\begin{equation}
    c_{\text{sat}}=\sqrt{(1/2)^{1/k}-(1/4)^{1/k}}/\sqrt{(1/2)^{(k-1)/k}-(1/4)^{(k-1)/k}}.
\end{equation}
In contrast, in classical percolation, $P_\text{SC}$ equals one if and only if $p=1$.
This phenomenon originates from the anomaly of the parallel rule (Table~\ref{table_rules}) being not a smooth function. The presence of this saturation point unveils a new ``quantum advantage'' originating only from concurrence percolation: \emph{one can deterministically establish a singlet as long as the entanglement in each link surpasses the saturation point.} This advantage stands in contrast to schemes based on classical percolation, where a singlet can only be established with certainty if each link is perfectly entangled.

\begin{figure}[t!]
	\centering
	\hspace{-4mm}
	\begin{minipage}[t]{32.3mm}
		\subcaption{\label{fig_bethe_scaling_p10}$\qquad\qquad\qquad\qquad$}
        \includegraphics[height=26.35mm]{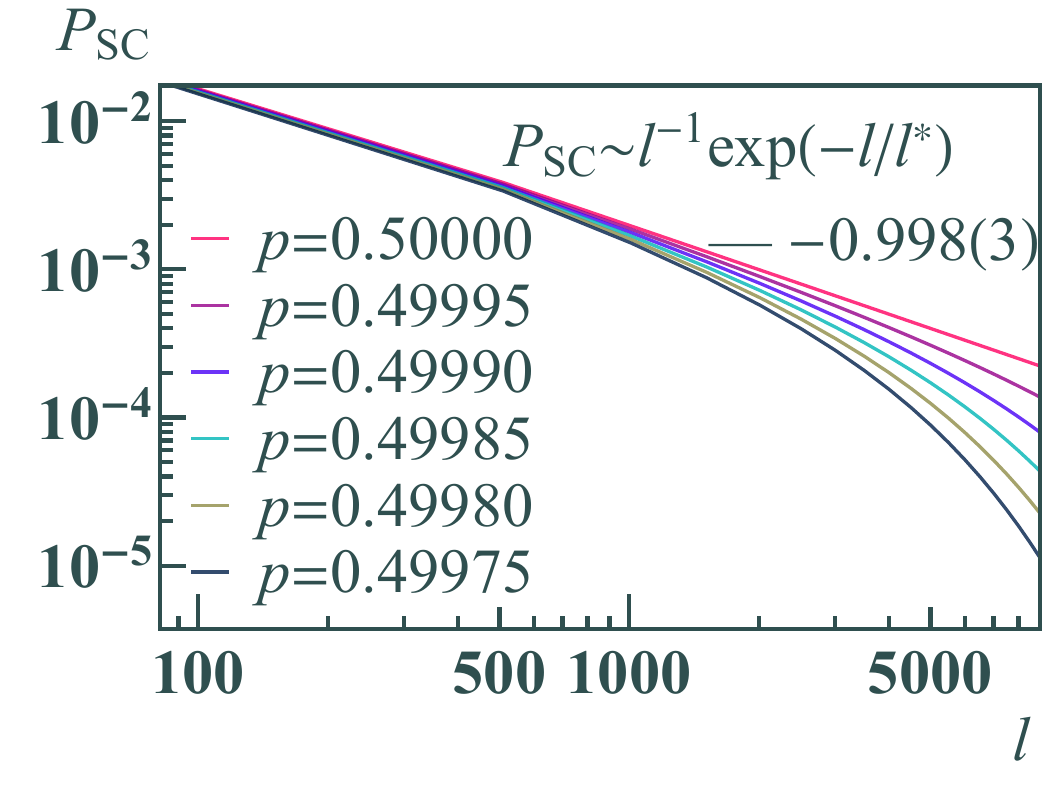}
		\vspace{-6mm}
	\end{minipage}
	\begin{minipage}[t]{32.3mm}
		\subcaption{\label{fig_bethe_scaling_p11}$\qquad\qquad\qquad\qquad$}
        \includegraphics[height=26.35mm]{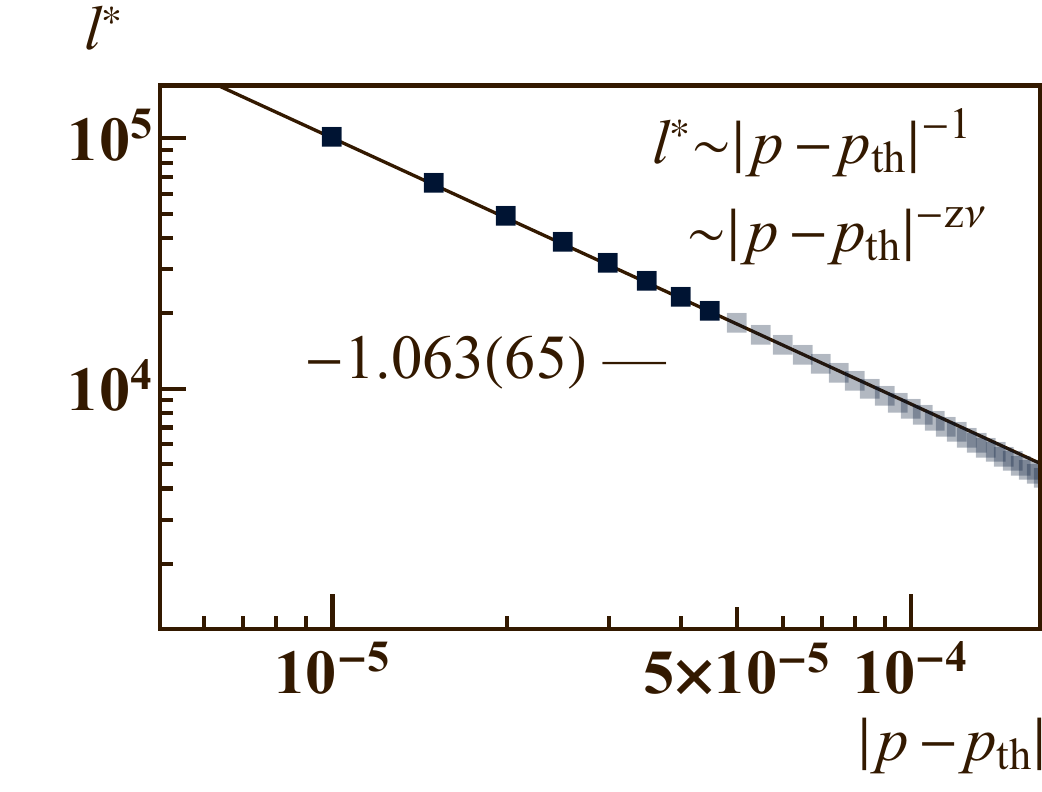}
		\vspace{-6mm}
	\end{minipage}
    \hspace{4mm}
	\addtocounter{subfigure}{3}
	\begin{minipage}[t]{32.3mm}
		\subcaption{\label{fig_bethe_scaling_c10}$\qquad\qquad\qquad\qquad$}
        \includegraphics[height=26.35mm]{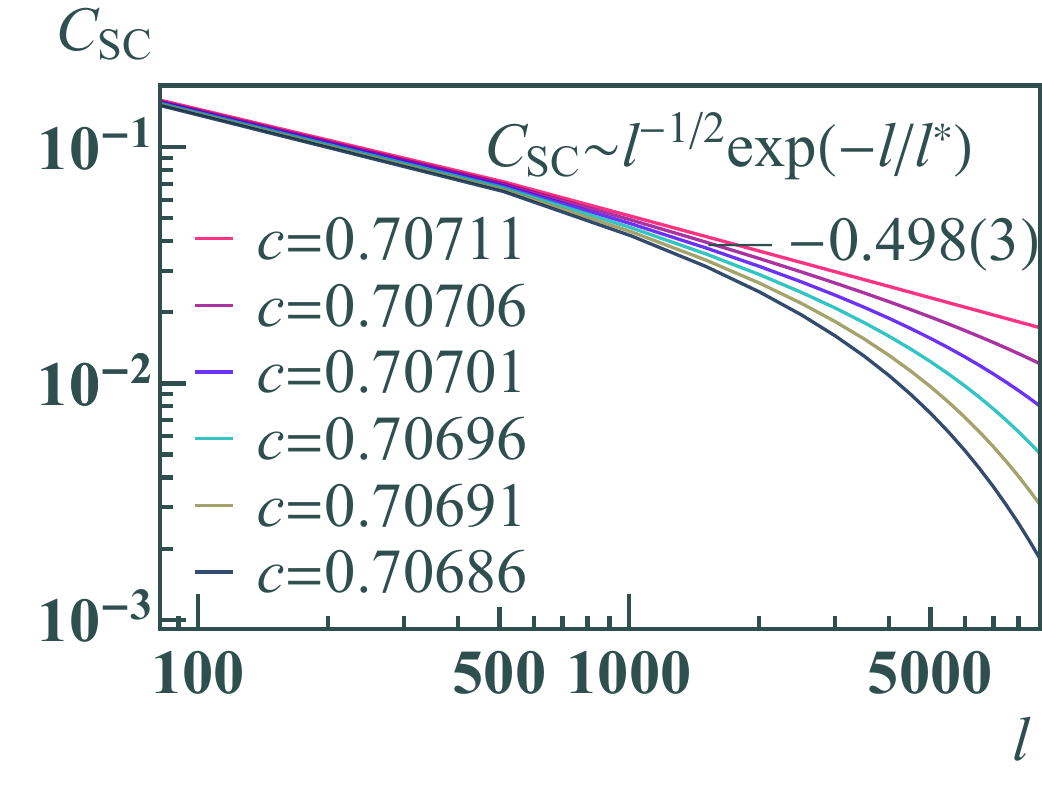}
		\vspace{-6mm}
	\end{minipage}
	\begin{minipage}[t]{32.3mm}
		\subcaption{\label{fig_bethe_scaling_c11}$\qquad\qquad\qquad\qquad$}
        \includegraphics[height=26.35mm]{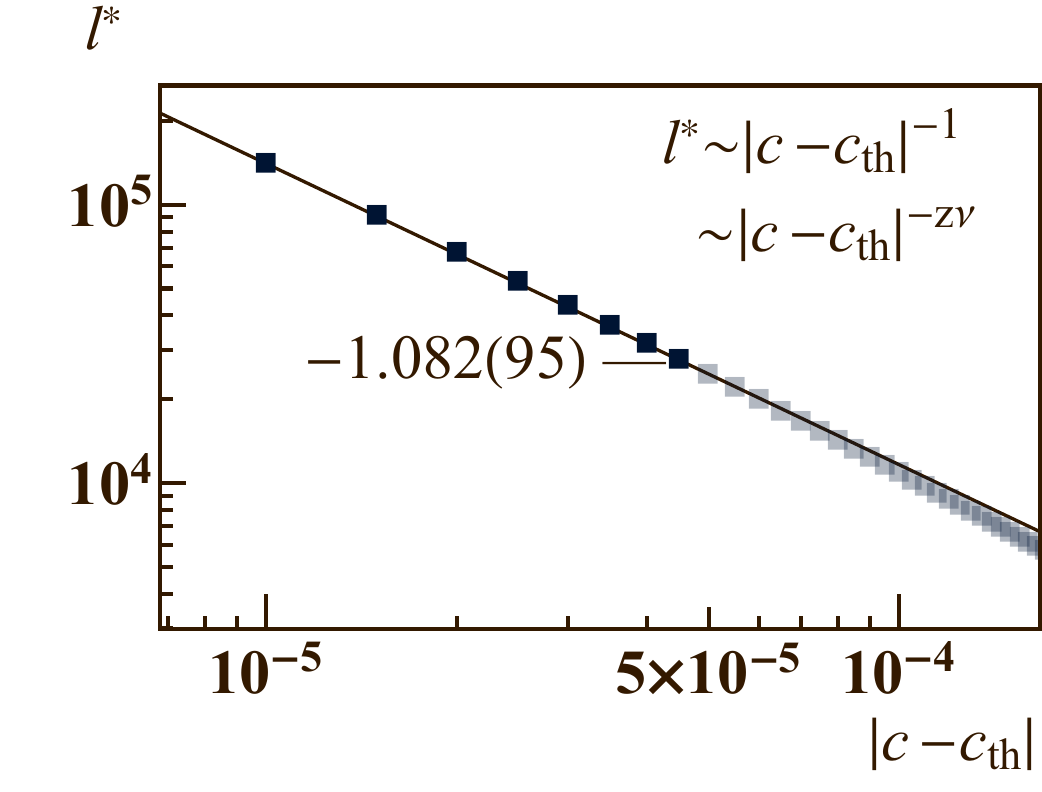}
		\vspace{-6mm}
	\end{minipage}
	
	\hspace{-4mm}
	\addtocounter{subfigure}{-5}
	\begin{minipage}[t]{32.3mm}
		\subcaption{\label{fig_bethe_scaling_p20}$\qquad\qquad\qquad\qquad$}
        \includegraphics[height=26.35mm]{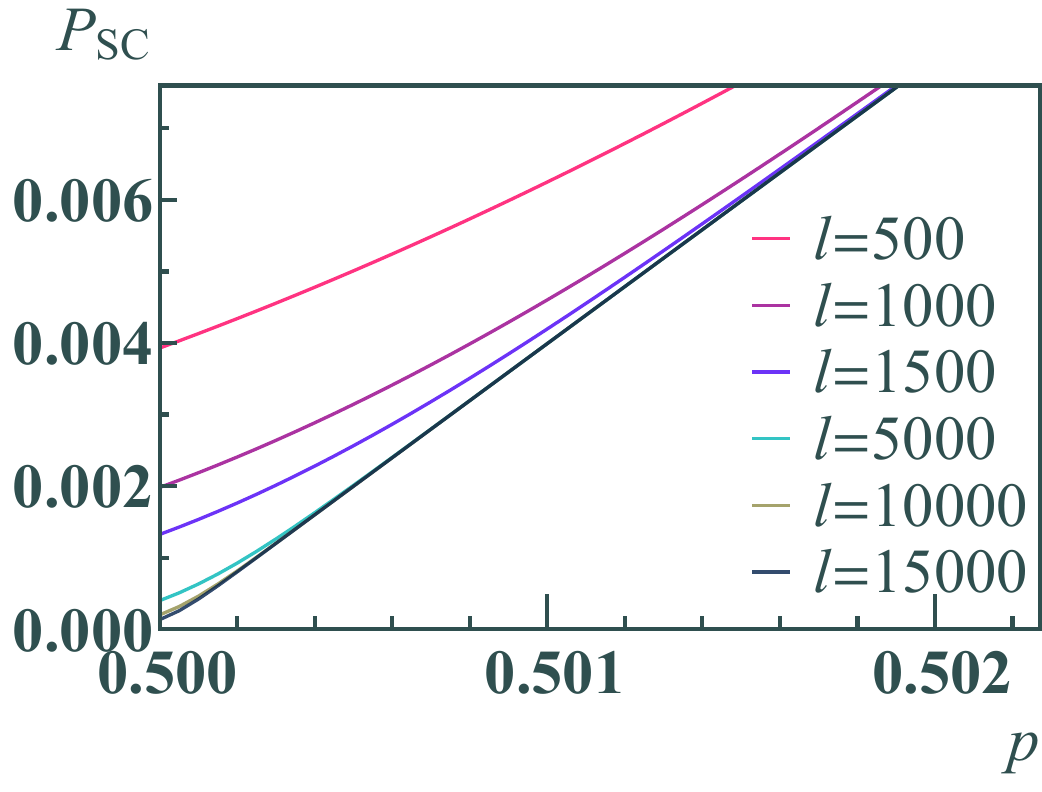}
		\vspace{-6mm}
	\end{minipage}
	\begin{minipage}[t]{32.3mm}
		\subcaption{\label{fig_bethe_scaling_p21}$\qquad\qquad\qquad\qquad$}
        \includegraphics[height=26.35mm]{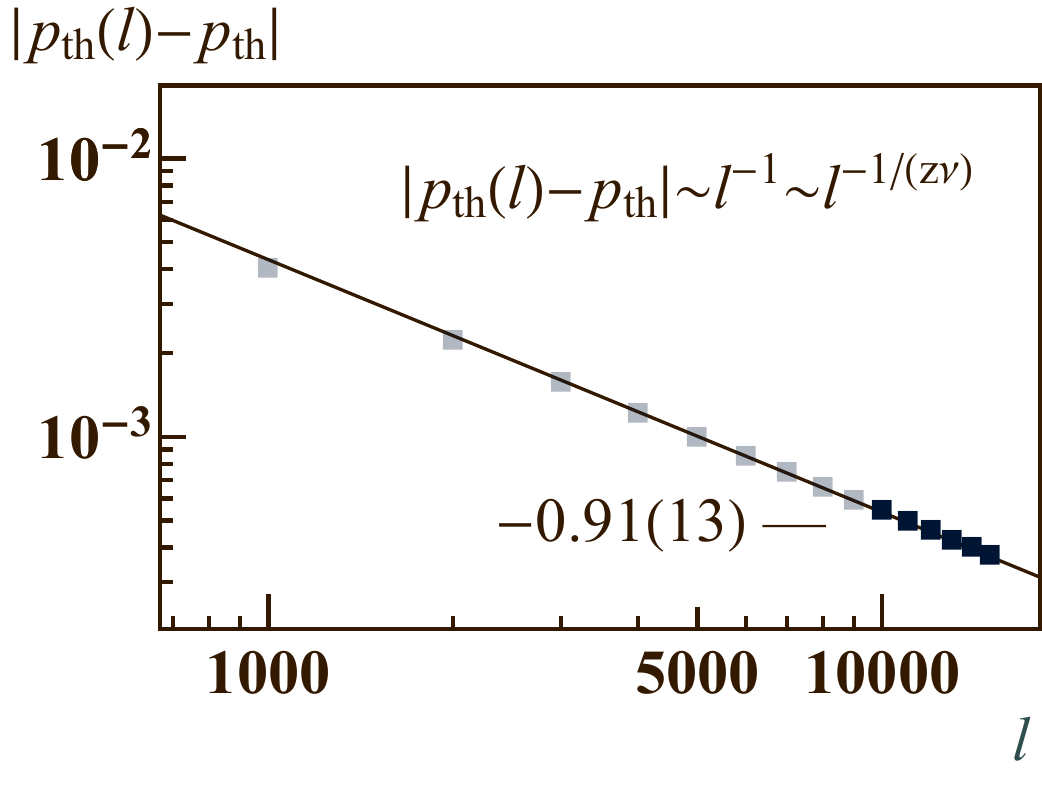}
		\vspace{-6mm}
	\end{minipage}
	\hspace{4mm}
	\addtocounter{subfigure}{3}
	\begin{minipage}[t]{32.3mm}
		\subcaption{\label{fig_bethe_scaling_c20}$\qquad\qquad\qquad\qquad$}
        \includegraphics[height=26.35mm]{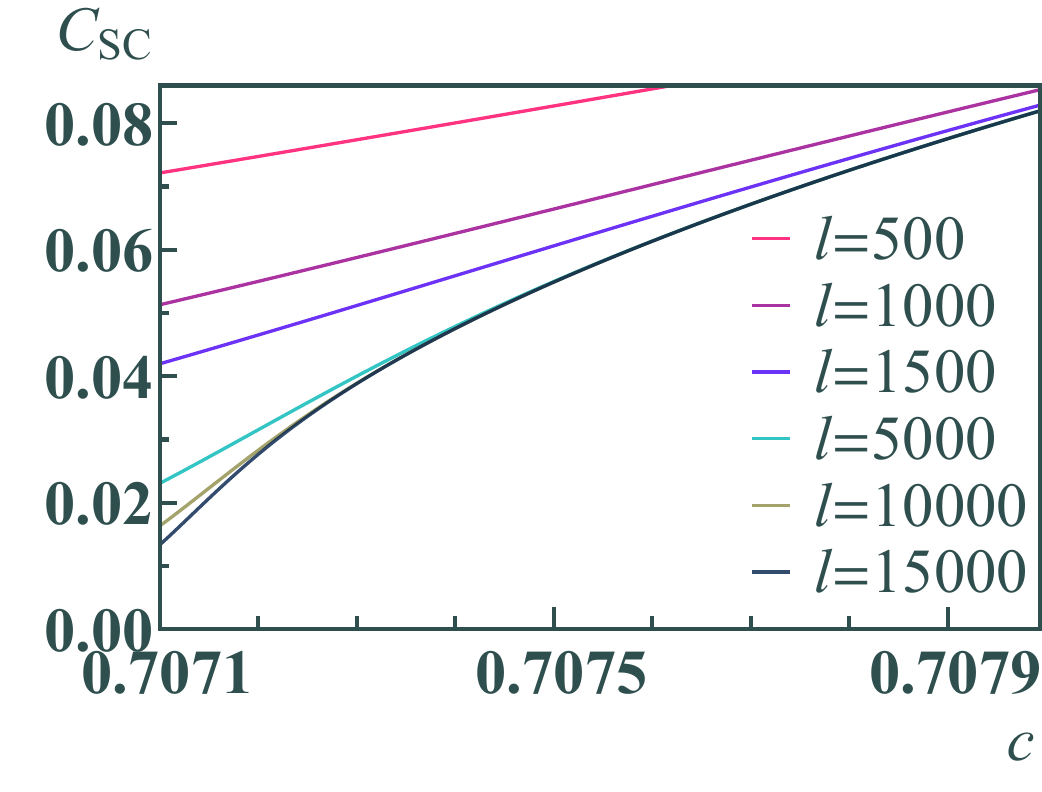}
		\vspace{-6mm}
	\end{minipage}
	\begin{minipage}[t]{32.3mm}
		\subcaption{\label{fig_bethe_scaling_c21}$\qquad\qquad\qquad\qquad$}
        \includegraphics[height=26.35mm]{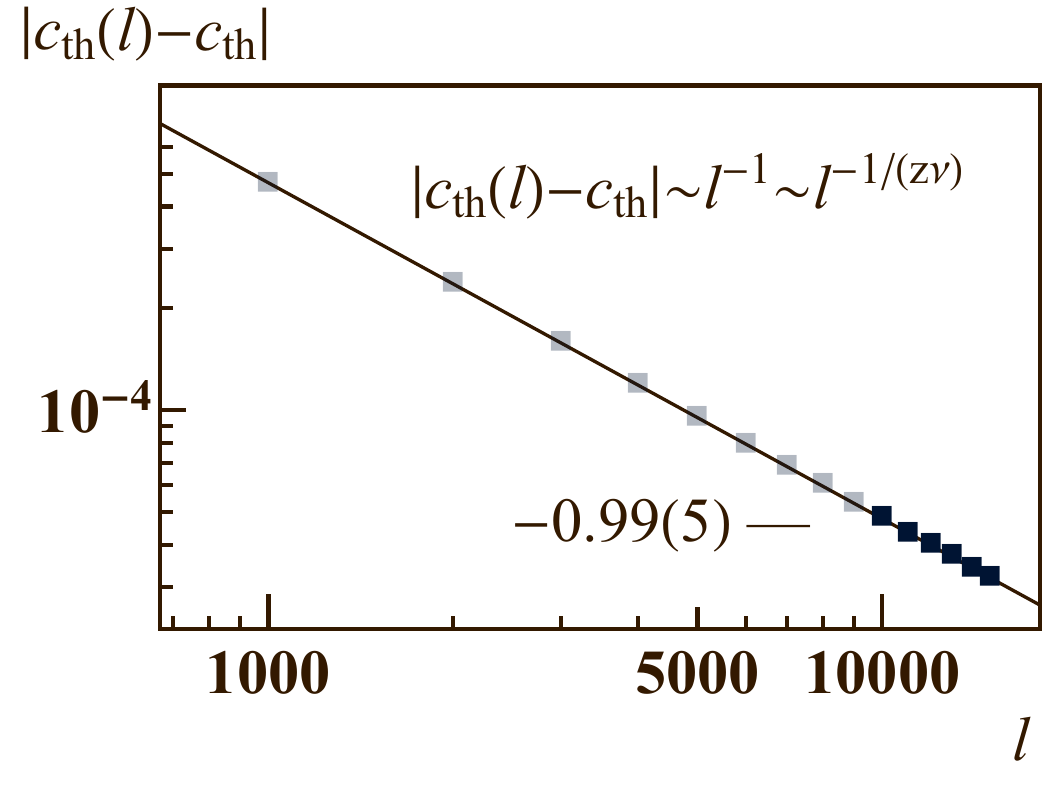}
		\vspace{-6mm}
	\end{minipage}
	
	\hspace{-4mm}
	\addtocounter{subfigure}{-5}
	\begin{minipage}[t]{61.6mm}
		\raggedright
		\subcaption{\label{fig_bethe_scaling_p3}$\qquad\qquad\qquad\qquad\qquad\qquad\qquad\qquad$}
        \includegraphics[height=26.35mm]{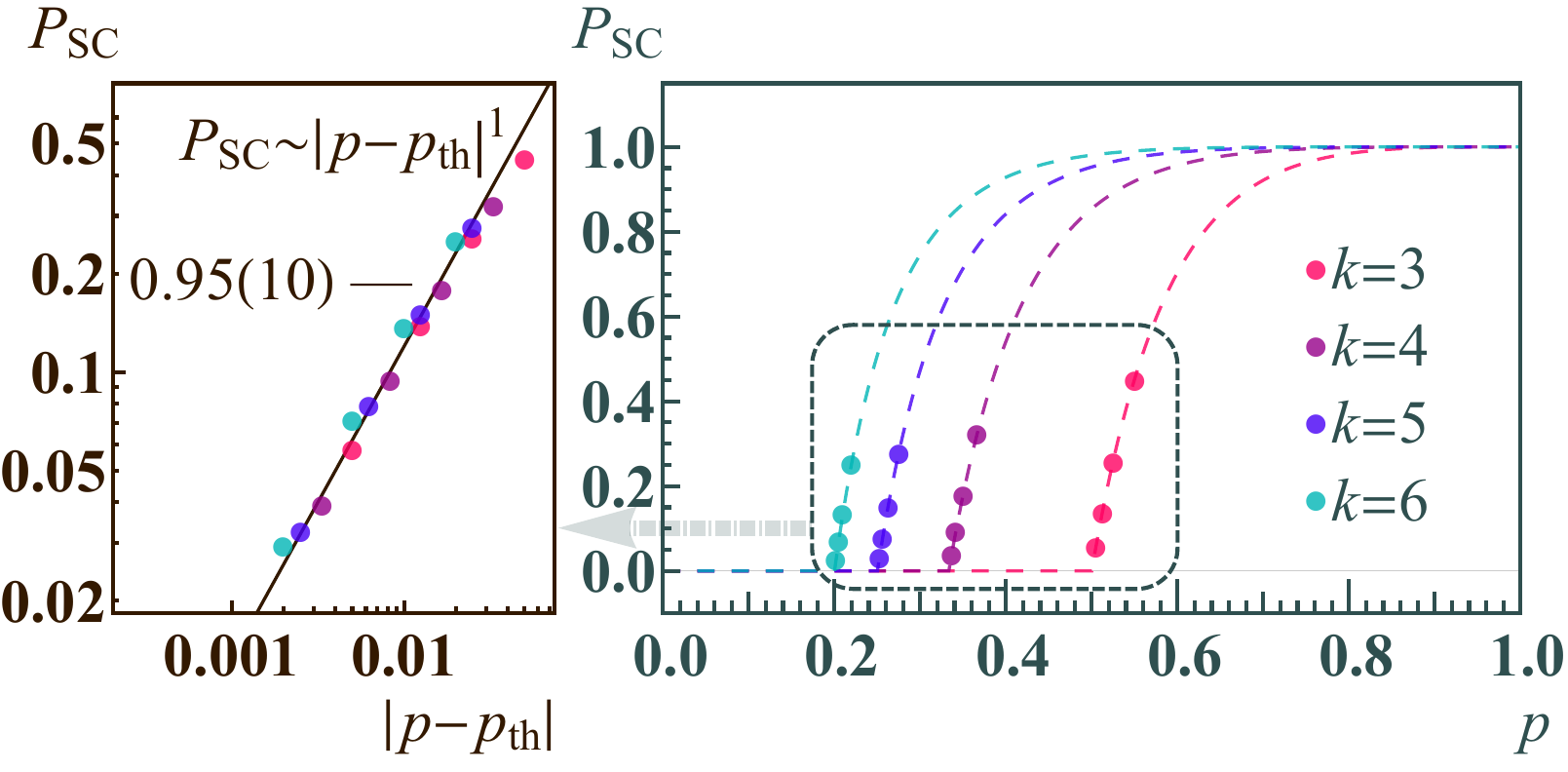}
		\vspace{-6mm}
	\end{minipage}
    \hspace{4mm}
	\addtocounter{subfigure}{+4}
	\begin{minipage}[t]{67.6mm}
		\subcaption{\label{fig_bethe_scaling_c3}$\qquad\qquad\qquad\qquad\qquad\qquad\qquad\qquad$}
        \includegraphics[height=26.35mm]{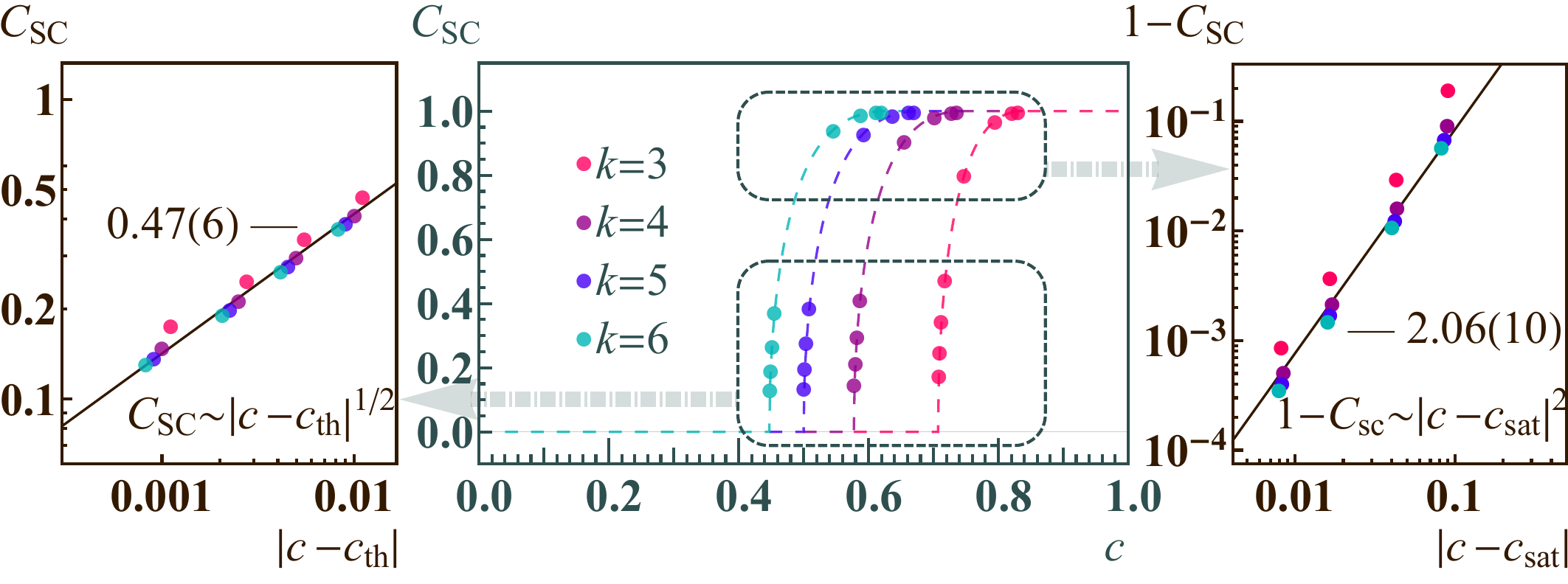}
		\vspace{-6mm}
	\end{minipage}	
	
	\caption{\label{fig_bethe_scaling}
		Critical phenomena of \subref{fig_bethe_scaling_p10}--\subref{fig_bethe_scaling_p3}~classical percolation and \subref{fig_bethe_scaling_c10}--\subref{fig_bethe_scaling_c3}~concurrence percolation theories in the Bethe lattice.
		\hfill\hfill}
\end{figure}

\emph{Critical exponents}.---Lastly, similar to classical percolation, concurrence percolation also shows critical phenomena, marked by a set of dependent or independent critical exponents~\cite{equilib-stat-phys}. However, it is important to note that concurrence percolation is defined based on \emph{connectivity rules} (Table~\ref{table_rules}), not \emph{clusters}. As a result, one cannot simply deduce a traditional cluster-based order parameter from the variable $c$ used in these rules. In fact, an effective cluster could be defined using $c$, $c^2$ or any other power of $c$. Altering the definition in this way essentially results in a variable change in the connectivity rules but does not change the underlying physics.

In the absence of a cluster definition, the sole critical exponent that can be determined is the dynamic thermal exponent, $\nu_\text{dyn}=z \nu$. This exponent is tied solely to length dimensions, reflecting how the system's correlation length diverges as $c$ approaches $c_\text{th}$. Note that the length in the context refers to chemical length, not the conventional Euclidean length. The two length scales are related by the dynamic critical exponent $z$~\cite{diffus-react-fractal-disorder-syst}. This is why $z\nu$, rather than $\nu$, is used in this case.

Importantly, the dynamic thermal exponent $z \nu$ can be retrieved from finite-size analysis~\cite{equilib-stat-phys}. Here, the idea is that the correlation length can be replaced by the system's finite length scale $l$ when $\left|c-c_\text{th}\right|\to 0$. Therefore, near the critical threshold, all dependence on $\xi$ can be deduced using $l$. The finite-size analysis results for both classical and concurrence percolation for the Bethe lattice are shown in Fig.~\ref{fig_bethe_scaling}. For concurrence percolation, it is found that $C_{\text{SC}}$ follows a power law with an exponential cutoff as a function of the number of layers $l$,
$C_{\text{SC}}\sim l^{-1/2}\exp(-l/l^*)$. Here, $l^*$ diverges as a power law as $c$ approaches $c_\text{th}$ [Fig.~\ref{fig_bethe_scaling_c10}]. The numerical value $z\nu=1.082(95)$ is obtained by fitting near $|c-c_\text{th}|\sim10^{-5}$ [Fig.~\ref{fig_bethe_scaling_c11}].

Alternatively, $z\nu$ can also be determined by looking at the \emph{finite-size} critical threshold $c_\text{th}(l)$, which is defined as the turning point of $C_{\text{SC}}$, 
\begin{equation}
    c_\text{th}(l)=c|_{\partial^2C_{\text{SC}}/\partial {c}^2=0},
\end{equation}
which deviates from $c_\text{th}$ as a power law with respect to $l$ [Fig.~\ref{fig_bethe_scaling_c20}].
Again, the numerical value $1/(z\nu)=0.99(5)$ is obtained near $l\sim10^4$ [Fig.~\ref{fig_bethe_scaling_c21}]. 

For general $k$, different $c_\text{th}$ and $c_\text{sat}$ are also presented [Fig.~\ref{fig_bethe_scaling_c3}], revealing two universal (i.e.,~independent of $k$) power laws of $C_{\text{SC}}$ near $c\to c_{\text{th}}$ and $c\to c_{\text{sat}}$, respectively, supported by numerical results (dots) on a finite Bethe lattice of $l=500$. In particular, the power-law relation $C_{\text{SC}}\sim|c-c_\text{th}|^{1/2}$ is reminiscent of the critical exponent $\beta$ in classical percolation, which follows $P_{\inf}\sim|p-p_\text{th}|^{\beta}$ [Eq.~\eqref{EqBeta}]. Yet, as previously discussed, $C_{\text{SC}}$ cannot be uniquely equated to a ``cluster-based'' order parameter. Thus, it would be premature to assert that $\beta=1/2$ without accounting for certain nuances.

Note that the above results also have their counterparts in classical percolation [Figs.~\ref{fig_bethe_scaling_p10}-\ref{fig_bethe_scaling_p3}] except for the saturation point $c_\text{sat}$. It is found that the critical exponent $z\nu$ is the same for both classical and concurrence percolation theories on the Bethe lattice. It thus remains unknown whether the classical and concurrence percolation theories belong to the same universality class or not.

In conclusion, the concept of concurrence percolation establishes a new theory that governs the behavior of entanglement transmission across large-scale QN. This novel theory brings in several unique characteristics that distinguish it from classical percolation, thereby providing a refreshing and rich perspective on QN.
We believe that the theoretical framework set by concurrence percolation may also open doors to practical applications, such as more efficient entanglement transmission schemes or novel protocols for quantum communication and computation.
In essence, concurrence percolation not only enriches our comprehension of the inherent complexity of QN but also signifies a leap towards a more refined and versatile understanding of the statistical physics that governs QN.

\section{Algorithms}
\label{sec_algorithm}

This section is dedicated to a comprehensive exploration of the fundamental algorithms that have played a pivotal role in our investigation of concurrence percolation theory. Not only are these algorithms instrumental in the analysis and understanding of QN, but they also serve as essential tools for modeling and simulating the complex behaviors within the network.

\begin{figure}[t]
	\centering
	\begin{minipage}[b]{1.65in}	
		\centering
		{\includegraphics[width=1.65in]{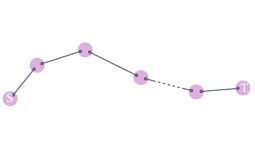}
			\vspace{-0mm}\subcaption{Simple series QN.\label{fig_seri}}}
	\end{minipage}
	\begin{minipage}[b]{1.65in}
		{\includegraphics[width=1.65in]{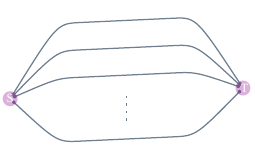}
			\vspace{-0mm}\subcaption{Simple parallel QN.\label{fig_para}}}
	\end{minipage}

	\begin{minipage}[b]{1.65in}	
		\centering
		{\includegraphics[width=1.65in]{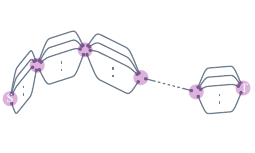}
			\vspace{-0mm}\subcaption{Parallel-then-series QN.\label{fig_para-then-seri}}}
	\end{minipage}
	\begin{minipage}[b]{1.65in}
		{\includegraphics[width=1.65in]{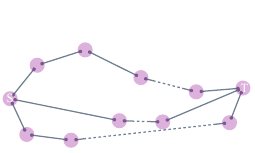}
			\vspace{-0mm}\subcaption{Series-then-parallel QN.\label{fig_seri-then-para}}}
	\end{minipage}

	\begin{minipage}[b]{1.65in}	
		\centering
		{\includegraphics[width=1.65in]{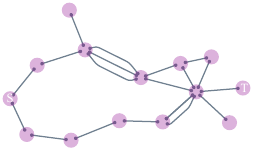}
			\vspace{-0mm}\subcaption{Series-parallel QN.\label{fig_seri-para}}}
	\end{minipage}
	\begin{minipage}[b]{1.65in}
		{\includegraphics[width=1.65in]{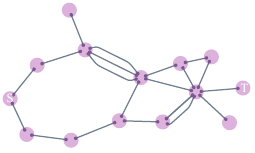}
			\vspace{-0mm}\subcaption{General QN.\label{fig_general}}}
	\end{minipage}	
	\caption{\label{fig_seri-para-qn}Different QN topologies between $S$ and $T$. 
		\subref{fig_seri}~Series. \subref{fig_para}~Parallel. \subref{fig_para-then-seri}~Parallel-then-series. \subref{fig_seri-then-para}~Series-then-parallel. \subref{fig_seri-para}~Series-parallel. \subref{fig_general}~Non-series-parallel.  
		\hfill\hfill}
\end{figure}

\subsection{Identification of series-parallel networks}

Series-parallel networks were introduced by Duffin~\cite{series-parallel-netw_d65} as a mathematical model of electrical networks, and a general version was introduced later by Lawler~\cite{lawler1976matroids} and Monma and Sidney~\cite{monma1979Algorithm} as a model for scheduling problems.
The classification of a network as series-parallel depends on the choice of two specific nodes of interest~\cite{series-parallel-netw_d65}. Given two source and target nodes $S$ and $T$, 
the network topology can be grouped into different categories (Fig.~\ref{fig_seri-para-qn}). All topologies between $S$ and $T$ given in Figs.~\ref{fig_seri}--\ref{fig_seri-para} are considered series-parallel, except Fig.~\ref{fig_general}, due to an existing ``bridge'' in addition to Fig.~\ref{fig_seri-para}. 
Importantly, it is worth highlighting that many realistic complex networks can be approximated as series-parallel. This is because, in infinite-dimensional systems, cycles can typically be ignored through the Bethe approximation~\cite{netw-percolation_ceah00}.

\begin{figure}[t!]	
	\centering
	\includegraphics[width=397pt]{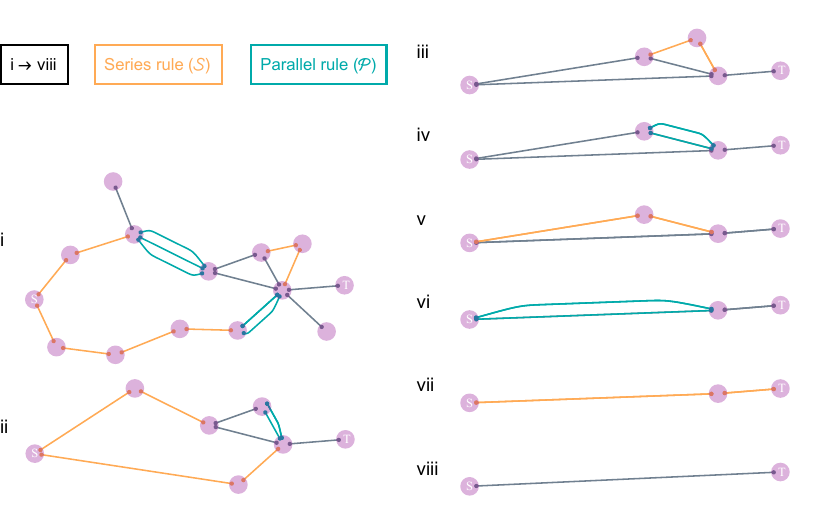}
	\caption{\label{fig_series_parallel_example} Decomposition of a series-parallel network to the final base graph (from i to viii). At each step, the links that the series rule and the parallel rule are applied to are highlighted in orange and cyan, respectively.
    \hfill\hfill}
\end{figure}

It is known that when a ``decomposition tree'' (Fig.~\ref{fig_series_parallel_example}) for a series-parallel graph is given, many problems, including those that are NP-hard for arbitrary graphs~\cite{bern1987subgraph,borie1987subgraph,kikuno1983number,takamizawa1982linear}, can be solved in linear time.
While series-parallel networks continue to play an important role in various applications, they have been extensively studied in their own right as well as in relation to other optimization problems (cf.~\cite{baffi1995parallel,bertolazzi1992draw,rendl1986assignment,steiner1985sp}). We also refer to~\cite{bg1976digraphs} for more results in series-parallel graphs.

Series-parallel networks enjoy nice algorithmic properties. There is a fast algorithm that determines whether any given network is series-parallel, and if it is, also returns the decomposition tree that is suitable to be used in the following applications~\cite{valdes1982sp}. Following this work, researchers have further developed parallelized algorithms to determine the important class of series-parallel networks~\cite{bodlaender2001parallel,eppstein1992parallel,he1991parallel,he1987parallel}.

\subsection{Star-mesh transform}
The star-mesh transform (also known as the Kron reduction~\cite{kron1939kron}) was originally developed as a circuit analysis technique for calculating the effective resistance in resistor networks. The star-mesh transform replaces a local star network topology by a mesh topology (a complete graph).
Importantly, the equivalent resistance between each pair of nodes remains consistent before and after the transformation.
Here, we generalize this idea to offer an approximation method for percolation on networks. This approach bears similarity to the real-space renormalization group (RG) methods used in percolation theory. However, the star-mesh transform is more versatile, applicable to various types of networks beyond regular lattices.

A star-mesh transform~\cite{star-mesh_v70} can be built upon only series and parallel rules but not higher-order rules to map an $N$-node star graph to an $(N-1)$-node complete graph, establishing a local equivalence (in terms of connectivity) between the two graphs. Mathematically, we denote $\mathcal{G}(N)$ to be a star graph with one root vertex and $N$ leaf vertices, where the weights of the $N$ edges are from $\theta_{1}$ to $\theta_{N}$. And the $N$-complete graph transformed from $\mathcal{G}(N)$ is denoted as $\mathcal{G}'(N)$, the weights of its $N(N-1)/2$ edges are $(\theta_{12},\theta_{13},\cdots,\theta_{1 N},\cdots,\theta_{N-1,N})$. The equivalence between $\mathcal{G}(N)$ and $\mathcal{G}'(N)$ are formatted $N(N-1)/2$ independent equations:
\begin{eqnarray}
\label{star-mesh-eqns}
\text{seri}\left(\theta_1,\theta_2\right) &=&c\left(1,2;\mathcal{G}'\left(N\right)\right),\nonumber\\
\text{seri}\left(\theta_1,\theta_3\right) &=&c\left(1,3;\mathcal{G}'\left(N\right)\right),\nonumber\\
&\cdots,&\nonumber\\
\text{seri}\left(\theta_1,\theta_N\right) &=&c\left(1,N;\mathcal{G}'\left(N\right)\right),\nonumber\\
&\cdots,&\nonumber\\
\text{seri}\left(\theta_{N-1},\theta_N\right) &=&c\left(N-1,N;\mathcal{G}'\left(N\right)\right),
\end{eqnarray}
in which, $\text{seri}(\theta_i,\theta_j)$ is the series-sum of $\theta_i$ and $\theta_j$ based on the series rule, and $c(i,j;\mathcal{G}'(N))$ is the net weight between vertices $i$ and $j$ of the complete graph $\mathcal{G}'(N)$.

\begin{figure}[t]
    \centering
    \includegraphics[width=\textwidth]{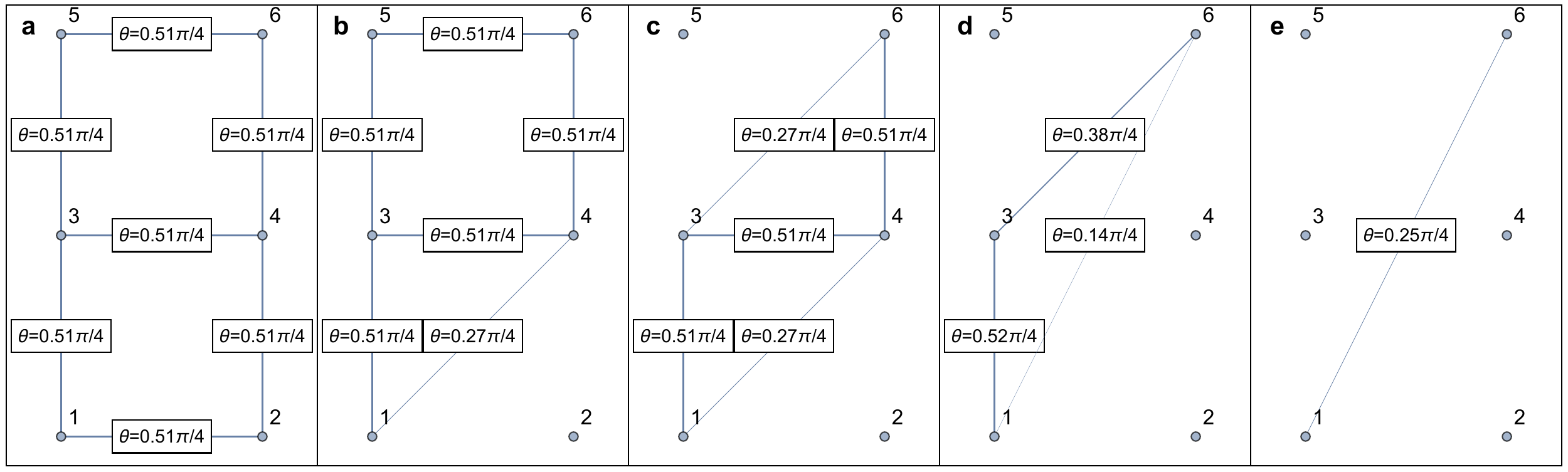}
    \caption{\label{fig_classical_example} Demonstration of calculating the classical percolation between nodes $1$ and $6$, in the following steps: (a) Original lattice. (b) and (c) Series rules. (d) Star-mesh transform on the star graph (with edges $4\leftrightarrow1$, $4\leftrightarrow3$, $4\leftrightarrow6$), converting it to a complete graph (with edges $1\leftrightarrow3$, $3\leftrightarrow6$, $6\leftrightarrow1$), then parallel rule for the double edges $1\leftrightarrow3$ and $3\leftrightarrow6$. (e) Series rule for edges $1\leftrightarrow3$ and $3\leftrightarrow6$, then parallel rule for edge $1\leftrightarrow6$.\hfill\hfill}
\end{figure}
\begin{figure}[t]
    \centering
    \includegraphics[width=\textwidth]{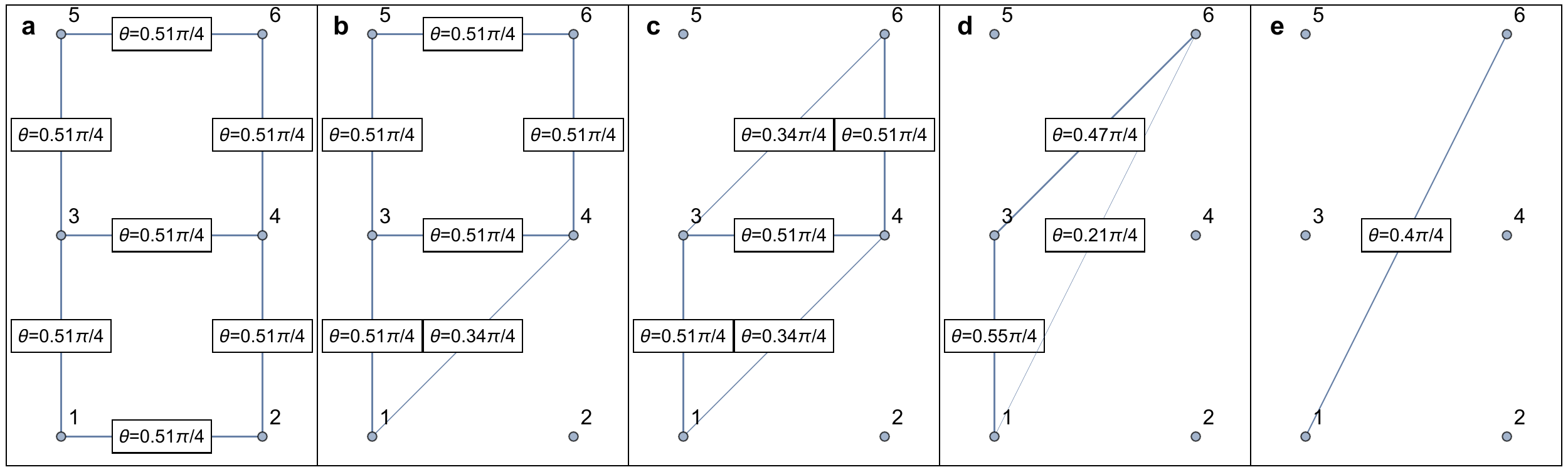}
    \caption{\label{fig_concurrence_example} Demonstration of calculating the concurrence percolation between nodes $1$ and $6$ (cf.~Fig.~\ref{fig_classical_example}).\hfill\hfill}
\end{figure}

To calculate the sponge-crossing percolation between the source $S$ and target $T$ in a certain network, we approximately obtain the equivalent simplified network by consecutively applying the star-mesh transform technique, where one node is degraded for each application. Specifically, we arbitrarily choose a vertex from $\mathcal{G}'(N)$ (w.l.o.g., the last one, $N$) to be the new root of a sub-star-graph of $\mathcal{G}'(N)$ constructed from the $N-1$ edges that connect the root to the other $N-1$ vertices. Next, we transform this sub-star-graph $(\text{sub}\mathcal{G}')(N-1)$ into a $(N-1)$-complete graph, denoted by $(\text{sub}\mathcal{G}')'(N-1)$, and combine it with what is left untransformed, {$\mathcal{G}'(N)\setminus\nobreak(\text{sub}\mathcal{G}')(N-\nobreak1)$. The new graph denoted as $\text{Comb}\left(\mathcal{G}_{\alpha},\mathcal{G}_{\beta}\right)$ is derived by setting each edge weight to be $\theta_{ij}=\text{para}(\alpha_{ij},\beta_{ij})$, which is the parallel-sum of $\alpha_{ij}\in \mathcal{G}_{\alpha} $ and $\beta_{ij}\in\mathcal{G}_{\beta}$ based on the parallel rule. Note that, because of the lack of closed-form solution for concurrence percolation, we use Broyden’s root-finding algorithm to numerically solve the $N(N-1)/2$ weights $\theta_{ij}$ that satisfy Eq.~(\ref{star-mesh-eqns}). In a word, we can calculate $c(i,j;\mathcal{G}'(N))$ by first solving a $(N-1)$-complete graph, 
\begin{equation}
\label{c-recur}
c(i,j;\mathcal{G}'(N))=c(i,j;\text{Comb}\left((\text{sub}\mathcal{G}')'(N-1),\mathcal{G}'(N)\setminus(\text{sub}\mathcal{G}')(N-1)\right)).
\end{equation}
By applying Eq.~(\ref{c-recur}) one after the other on all but boundary nodes, the network can be eventually reduced to two nodes and one link between them (Table~\ref{table_rules}), the final weight $\theta$ of which shall be approximately equivalent to the percolation of initial network. For demonstrative purposes, here we show how the star-mesh transform works for both the classical percolation (Fig.~\ref{fig_classical_example}) and concurrence percolation (Fig.~\ref{fig_concurrence_example}) on a small square lattice.

Since $c(i,j;\mathcal{G}'(N))$ is calculable through recursions and Eq.~(\ref{c-recur}) involves a $(N-1)$-level star-mesh transform, the entire procedure is a double recursion, the cost growing faster than exponential. But by practically carrying out the recursive computation using symbolic expressions in Mathematica and other numerical techniques, the solutions can be found within a sufficiently small error range~\cite{conpt_mgh21}. 

Note that the star-mesh transform functions as an approximation rather than an exact representation of higher-order connectivity rules. To see this, consider the example of classical percolation given in Fig.~\ref{fig_classical_example}. Under the change of variable $p\equiv 2\sin^2\theta$, the actual higher-order connectivity, i.e.,~the probability of at least one path connecting nodes $1$ and $6$, can be expressed as follows:
\begin{eqnarray}
    &&p_{34} \left[1-\left(1-p_{35}p_{56}\right)\left(1-p_{46}\right)\right]
    \left[1-\left(1-p_{12}p_{24}\right)\left(1-p_{13}\right)\right]\nonumber\\
    &+&\left(1-p_{34}\right) \left[1-\left(1-p_{13}p_{35}p_{56}\right)\left(1-p_{12}p_{24}p_{46}\right)\right]\nonumber\\
    &\approx& 0.0799,
\end{eqnarray}
where $p_{ij} \equiv 2\sin^2\theta_{ij}\approx0.304$ represents the singlet conversion probability for each link $i\leftrightarrow j$. The final probability ($\approx0.0799$) translates to a final entanglement $\theta \approx 0.256 \pi/4$, which is very close to the star-mesh approximation result of $\theta \approx 0.25 \pi/4$ (Fig.~\ref{fig_classical_example}). The closeness of these values supports our confidence that the star-mesh transform can offer a reasonably accurate approximation. Also, note that the process of reducing a network using star-mesh transforms is not unique in sequence. For example, in the fourth step of Fig.~\ref{fig_classical_example}, one can transform the star graph ($3 \leftrightarrow 1$, $3 \leftrightarrow 4$, $3 \leftrightarrow 6$) instead. Different sequences of reduction might lead to varying approximate results, but they tend to stay close to each other and to the exact value~\cite{conpt_mgh21}.

\subsection{Fast numerical approximation for concurrence percolation}
The heuristic approximation (star-mesh transform) used for higher-order connectivity rules
can be quite demanding in terms of computational resources. To address this challenge, in this section, we discuss a more efficient approach to calculate the sponge-crossing concurrence $C_\text{SC}$~\cite{conpt_mmhksg22}. This acceleration in computation is achieved through the incorporation of two key simplifying approximations: the parallel approximation and the $S_m$ approximation (Fig.~\ref{fig_concurrence_approximation}).

\begin{figure}[t!]
    \centering
    \includegraphics[width=397pt]{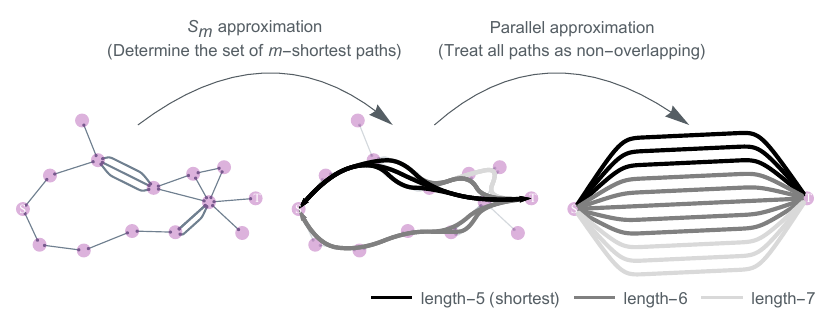}
    \caption{\label{fig_concurrence_approximation} {Approximations for calculating the sponge-crossing concurrence $C_\text{SC}$.} In the $S_m$ approximation, one defines $S_m$ as the set which contains up to the $m$-th shortest paths (i.e.,~the shortest paths, the $2$nd shortest paths, and so on up to the $m$-th shortest paths) between $s$ and $t$ for all $s\in S$ and $t\in T$. In the parallel approximation, one treats all paths in $S_m$ as parallel and non-overlapping. Thus, the network topology reduces to series-then-parallel, and the sponge-crossing concurrence can be calculated using the series/parallel rules (Table~\ref{table_rules}).\hfill\hfill}
\end{figure}

\subsubsection{Parallel approximation}
First, we introduce the parallel approximation: we treat all paths connecting nodes of interest to be parallel, i.e.,~treating them as if they have no overlap. For an arbitrary network with $N$ nodes and uniform concurrence $c$ per link, the parallel approximation $C_\text{SC}'$ of the true sponge-crossing concurrence $C_\text{SC}$ between two sets of nodes, $S$ and $T$, is given by
\begin{align}
\frac{1 + \sqrt{1 - C_\text{SC}'^{2}}}{2} &= \max\left\{ \prod_{l}\left(\frac{1 + \sqrt{1 - c^{2l}}}{2}\right)^{N_{l}}, \frac{1}{2}\right\},
\label{eq:conPar}
\end{align}
where $N_{l}$ is the total number of self-avoiding paths of length $l$ that connect the source/target nodes $s$ and $t$ for all $s\in S$ and $t\in T$, respectively. Equation~\eqref{eq:conPar} is the mathematical statement of the parallel approximation, indicating that we are taking each of the $N_l$ paths to be parallel and non-overlapping (Fig.~\ref{fig_concurrence_approximation}). 

Now, we will show that on series-parallel networks~\cite{series-parallel-netw_d65} the concurrence calculated under the parallel approximation forms an \emph{upper} bound to the true concurrence. First, we consider the case where our network is essentially parallel, i.e.,~it can be expressed as the parallel combination of $k$ subnetworks, each with concurrence $c_i$. In this case, the parallel approximation is exact:
\begin{align*}
   C_{\text{SC}}' = C_{\text{SC}} = \text{para}(c_1, c_2,\dots c_k).
\end{align*}

The more interesting case is that of an essentially series network, i.e.,~a network that can be decomposed as a combination of subnetworks in series. We consider an exemplary network that splits into $k$ branches, each with concurrence $c_{p_i}$. The concurrence of the segment before branching is $c_s$. Following the series and parallel rules (Table~\ref{table_rules}), the sponge-crossing concurrence from the left of this network segment to the right is
\begin{align*}
   C_\text{SC} &= 
\begin{cases} 
     c_s\left(2\sqrt{f(c_{p_0},\dots c_{p_k}) - f(c_{p_0},\dots c_{p_k})^2}\right) & f(c_{p_0},\dots c_{p_k})>1/2, \\
      c_s & f(c_{p_0},\dots c_{p_k})\le 1/2,
\end{cases}
\end{align*}
where $f(c_{p_0},\dots c_{p_k}) = \prod_{i=1}^k g(c_{p_i}) =  \prod_{i=1}^k\left(\dfrac{1 + \sqrt{1 - c_{p_i}^2}}{2} \right)$. Under the parallel approximation, the network is transformed so that the concurrence of the segment is given by
\begin{align*}
   C'_\text{SC} &= 
\begin{cases} 
      2\sqrt{f(c_sc_{p_0},\dots c_sc_{p_k}) - f(c_sc_{p_0},\dots c_sc_{p_k})^2} & f(c_sc_{p_0},\dots c_sc_{p_k})>1/2, \\
      1 & f(c_sc_{p_0},\dots c_sc_{p_k})\le 1/2.
\end{cases}
\end{align*}
Since $c_s c_{p_i} \le c_{p_i}$, it follows that $g(c_s c_{p_i}) \ge g(c_{p_i})$ and thus $f(c_sc_{p_0},\dots c_sc_{p_k}) \ge f(c_{p_0},\dots c_{p_k})$. After some calculations~\cite{conpt_mmhksg22} one can show that $C_\text{SC}'\ge C_\text{SC}$.

Taken together, since every series-parallel network can be decomposed into essentially series or parallel configurations, we have shown that $C_\text{SC}'$ is an upper bound for $C_\text{SC}$ on series-parallel networks. Interestingly, as we will see, this upper bound seemingly becomes tighter as the network becomes larger. We hence expect that a new concurrence threshold on $C_\text{SC}'$ can emerge, which should numerically approach the true $c_\text{th}$ from below and match $c_\text{th}$ in the thermodynamic limit $N\to\infty$.

\subsubsection{$S_m$ approximation}
For most regular lattices and complex networks, however, it is a nontrivial task to determine the number of paths $N_{l}$ of length $l$. When we look at arbitrary networks, the calculation for the sponge-crossing concurrence is essentially a path-counting problem which may require approximation as well. 

Although the literature of path counting on graphs is rich and well studied, we are not aware of any closed-form solutions for enumeration of self-avoiding walks of arbitrary length for even the simplest network (like 2D lattices)~\cite{krattenthaler2015}. While approximate path enumerations exist for both 2D lattices~\cite{Jensen_2004} and random networks~\cite{Roberts2007}, we find them impractical, since the concurrence calculation is very sensitive to $N_l$ for small $l$. 
Based on this, if we define $S_m$ as the set which contains up to the $m$-th shortest paths (i.e.,~the shortest paths, the $2$nd shortest paths, and so on up to the $m$-th shortest paths) between $s$ and $t$ for all $s\in S$ and $t\in T$, then it is possible to approximate the sponge-crossing concurrence between $S$ and $T$ using only these paths.

\subsubsection{Results}

\begin{figure}[t]
    \centering
	\includegraphics[width=397pt]{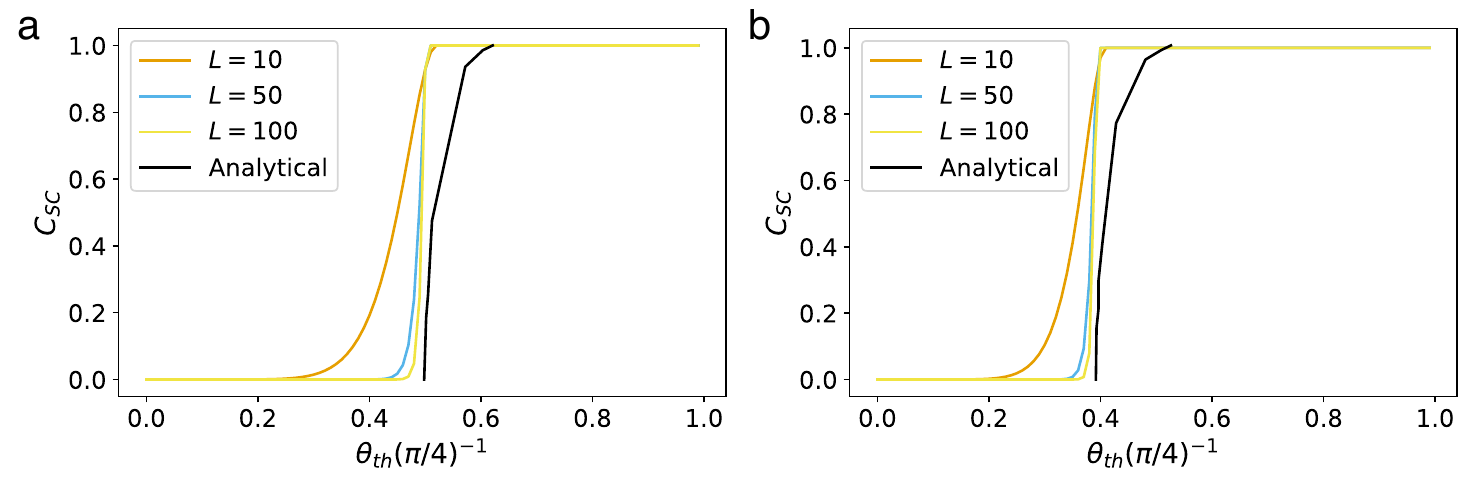}
	\caption{\label{figure4}{The sponge-crossing concurrence $C_\text{SC}$ for the Bethe lattice under the parallel approximation.} Results are shown for coordination numbers (a) $k = 3$ and (b) $k = 4$. As the number of layers, $L$, in the network become larger the numerical concurrence percolation threshold approaches the analytical value, $\theta_{\text{th}}=(2/\pi)\sin^{-1}(1/{\sqrt{k-1}})$ . The solid black lines represent the exact  $C_\text{SC}$ for the Bethe lattice (cf.~Fig.~\ref{fig_bethe}).\hfill\hfill}
\end{figure}

In this section, using the $S_m$ and parallel approximations, we present numerical results for calculating $C_\text{SC}$ in different networks of large size $N$. From that, we can numerically estimate the finite-size concurrence percolation threshold $c_\text{th}\equiv \sin 2\theta_\text{th}$, determining its position on the critical curve by matching the corresponding sponge-crossing concurrence at the half point, namely,
\begin{equation}
    c_\text{th}\equiv \sin 2\theta_\text{th}\approx c|_{C_\text{SC}=1/2}.
\end{equation}
Next, we show how to apply our approach to Bethe Lattice and 2D Square Lattices.

\emph{Bethe lattice}.---Given a finite Bethe lattice (i.e.,~a Cayley tree) [Fig.~\ref{fig_bethe_lattice}] with $L$ layers and coordination number $k$~\cite{diffus-react-fractal-disorder-syst, bunde2012}, all paths from the root node to any one of the boundary nodes have the same length, $L$. Since only one path exists from the root node to any node on the boundary, the number of paths of length $L$ is 
\begin{align}
N_L = k(k-1)^{L-1}.
\end{align}
There is no need to employ the $S_m$ approximation since all paths are exactly known. Only the parallel approximation $C_\text{SC}'$ of the sponge-crossing concurrence $C_{\text{SC}}$ is to be calculated, which is given by [following Eq.~\eqref{eq:conPar}]
\begin{align}
\frac{1 + \sqrt{1 - C_\text{SC}'^{2}}}{2} &= \max\left\{ \left(\frac{1 + \sqrt{1 - c^{2L}}}{2}\right)^{N_{L}}, \frac{1}{2}\right\}.
\label{eq:conParBethe}
\end{align}
To solve for $c_{\text{th}}$, near $C'_{\text{SC}}=0$ we let 
\begin{align}
\left(\frac{1 + \sqrt{1 - c_{\text{th}}^{2L}}}{2}\right)^{N_{L}} &= 1-\epsilon
\end{align}
given an arbitrarily small positive $\epsilon$. This gives rise to
\begin{align}
c_{\text{th}}^{2L} = 1-\left[2\left(1-\epsilon\right)^{1/N_L}-1\right]^2 \simeq -4N_L^{-1}\ln\left(1-\epsilon\right)+ O(N_L^{-2}),
\end{align}
and thus
\begin{align}
c_{\text{th}} \simeq  \left(\frac{4\epsilon}{k}\right)^{\frac{1}{2L}} \left(\frac{1}{k-1}\right)^{\frac{L-1}{2L}} \simeq \frac{1}{\sqrt{k-1}}
\end{align}
in the limit of large $L$. The result is identical to the exact concurrence percolation threshold for Bethe lattices (Table~\ref{table_thresholds}).

For validation purposes, numerical results of the parallel approximation $C'_\text{SC}$ versus the exact values $C_\text{SC}$ are shown in Fig.~\ref{figure4}. We see that as $L$ increases, the threshold $c_{\text{th}}$ approaches ${1}/{\sqrt{k-1}}$ from below, consistent with our theoretical result. Hence, it is highly suggested that the parallel approximation can correctly estimate the true concurrence percolation threshold $c_{\text{th}}$ in the limit $N \to \infty$.

It is known that a saturation point $c_\text{sat}<1$ also exists in the Bethe lattice~\cite{conpt_mgh21}, namely, before $c$ reaches unity, $C_{\text{SC}}$ will already reach unity at $c=c_\text{sat}$. It is obvious that $c_\text{sat}\ge c_\text{th}$, given the monotonicity of the series and parallel rules (Table~\ref{table_rules}).
To see if we can recover $c_\text{sat}$ using the parallel approximation too, let
\begin{align}
\left(\frac{1 + \sqrt{1 - c_{\text{sat}}^{2L}}}{2}\right)^{N_{L}} &= \frac{1}{2},
\end{align}
set by $C'_{\text{SC}} = 1$.
This yields
\begin{align}
c_{\text{sat}}^{2L} = 1-\left[2\left(1/2\right)^{1/N_L}-1\right]^2 \simeq 4N_L^{-1}\ln 2+ O(N_L^{-2}),
\end{align}
and thus
\begin{align}
c_{\text{sat}} \simeq  \left(\frac{4\ln 2}{k}\right)^{\frac{1}{2L}} \left(\frac{1}{k-1}\right)^{\frac{L-1}{2L}} \simeq \frac{1}{\sqrt{k-1}}.
\end{align}
We see that the saturation point calculated using the parallel approximation is equal to $c_{\text{th}}$ but different from the exact value [Fig.~\ref{fig_bethe_threshold}].

\begin{figure}[t]
	\centering
	\includegraphics[width=397pt]{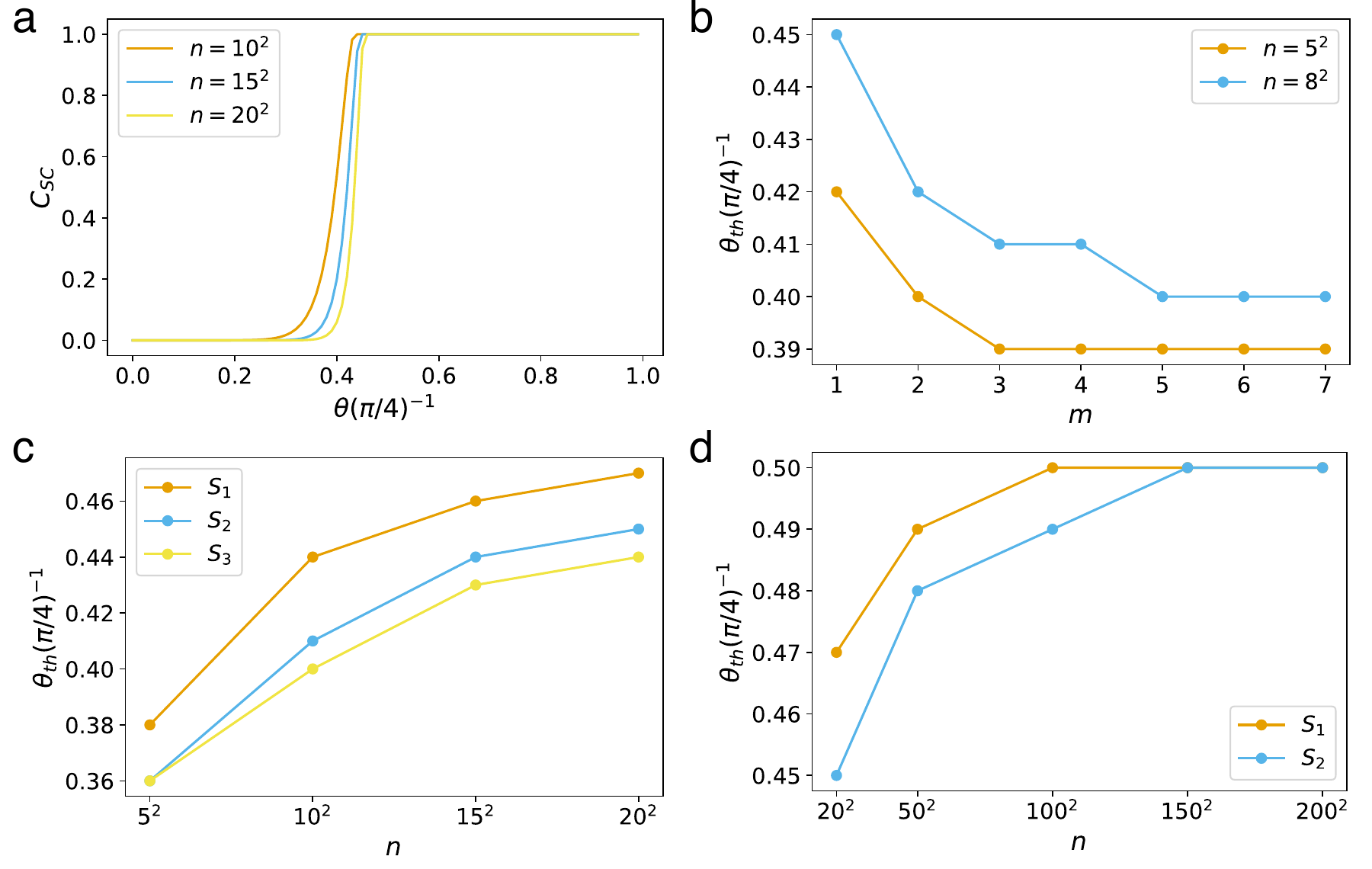}
	\caption{\label{figure6}{The sponge-crossing concurrence $C_\text{SC}$ for 2D square lattices under the $S_m$ and parallel approximations.} (a) Sponge-crossing concurrence $C_\text{SC}$ as a function of link's entanglement $\theta$ under the $S_3$ approximation. The results of $S_1$ and $S_2$ are nearly identical to $S_3$ and not plotted. (b) Numerical concurrence percolation threshold $\theta_{\text{th}}$ under the $S_m$ approximation. As the approximation order $m$ increases, $\theta_{\text{th}}$ approaches a constant value. (c) $\theta_{\text{th}}$ for different size $N$. (d) Same as (c) but for larger $N$. The results of $S_3$ are not shown because it becomes too computationally intensive to calculate for $N > 20^2$. As $N$ increases, $\theta_{\text{th}}$ also approaches a constant value.\hfill\hfill}
\end{figure}

\begin{figure}[t]
	\centering
	\includegraphics[width=198pt]{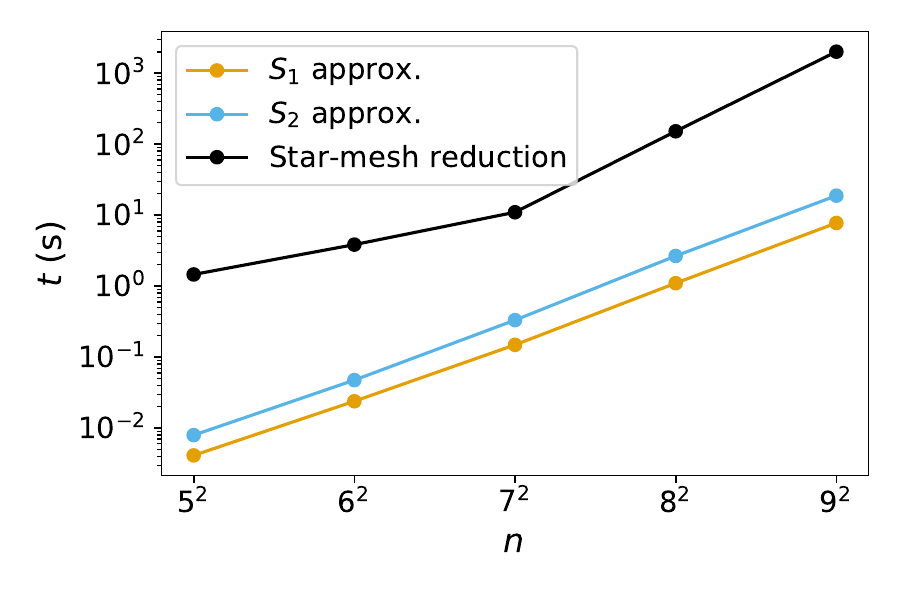}
	\caption{\label{figure7}{The speed-up obtained by the approximations over the star-mesh transform.} The figure shows the computing time (in seconds) to calculate the sponge-crossing concurrence between two nodes $s$ and $t$ on 2D square lattices with $N$ nodes, using the $S_1$ and $S_2$ approximation. In contrast to the star-mesh transform,  We can see that the approximations speed up the calculation over the star-mesh transform approach by two orders of magnitude.\hfill\hfill}	
\end{figure}

\emph{Two-dimensional square lattice}.---In a 2D square lattice of $N$ nodes ($\sqrt{N}\in \mathbb{Z}$),   the length of the $m$th shortest  self-avoiding path, between source and target nodes of coordinates $s=(x_s, y_s)$ and $t=(x_t, y_t)$ ($1\le x_s,x_t\le \sqrt{N}$ and $1\le y_s,y_t\le \sqrt{N}$), is simply  \[l_m = |x_s - x_t| + |y_s - y_t|+ 2\left(m-1\right).\]
Now, let $S$ and $T$ denote the left ($x_s=1$) and right ($x_t=\sqrt{N}$) boundaries. Let $s=(1,y_s)\in S$ and $t=(\sqrt{N},y_t)\in T$.
Under the $S_m$ approximation, the total number of self-avoiding paths of length $l$ between $S$ and $T$ is given by
\begin{align}
N_l\approx \sum_{y_s=1}^{\sqrt{N}} \sum_{y_t=1}^{\sqrt{N}} \delta_{l_1 l}  N_{l_1}(s\rightarrow t) +\delta_{l_2 l} N_{l_2}(s\rightarrow t) + \cdots + \delta_{l_m l} N_{l_m}(s\rightarrow t),
\end{align}
where $\delta_{ij}$ is the Kronecker delta. This approximation of $N_l$ is then substituted into the parallel approximation [Eq.~\eqref{eq:conPar}] to calculate $C_\text{SC}$ between $S$ and $T$. 

For $m\le2$, it is possible to directly enumerate the $1$st and $2$nd shortest self-avoiding paths between every pair of $s$ and $t$.
For $ m > 2$, however, it becomes difficult to write down a closed-form combinatorial expression for $N_{l_m}(s\rightarrow t)$. A path enumeration algorithm is thus needed. We treat paths of length $ l_m$ with $m > 2$ as deviations from the $1$st and $2$nd shortest paths. For a given $m$, these deviations can only take a finite number of shapes. Once we have identified these primitive deviations, we must next identify positions in the lattice where these deviations can be placed. Finally, we count the total number of paths by counting the number of shortest paths between deviations~\cite{conpt_mmhksg22}.

For example, given every pair of source and target nodes $s$ and $t$, all $3$rd-shortest paths ($m = 3$) have either two single-step deviations or one double-step deviation from the shortest path ($m = 1$). For the case where we have two single-step deviations, we first identify two sets of points, $D_1$ and $D_2$, where the first and second deviations can happen, respectively. Then we calculate $N_{s, D_1}$ (the number of shortest paths from $s$ to every point in $D_1$), $N_{D_1, D_2}$ (the number of shortest paths from every point in $D_1$ to every point in $D_2$), and $N_{D_2, t}$ (the number of shortest paths from every point in $D_2$ to $t$). The total number of $3$rd-shortest paths  is then given by $N_{l_2} (s \rightarrow t) = N_{s, D_1}N_{D_1, D_2}N_{D_2, t}$.

The final numerical results of the sponge-crossing concurrence  $C_\text{SC}$ are shown in Fig.~\ref{figure6}. We see that the transition in the value of $C_\text{SC}$ becomes sharper as the network size $N$ increases. Moreover, for higher-order approximation $S_m$ and/or larger $N$, the numerical threshold $\theta_{\text{th}}$ levels out at constant values that are very close to those calculated using the star-mesh transform. For example, for $N = 8^2$, the numerical approximation yields $\theta_{\text{th}} \approx 0.4$, closely mirroring the $\theta_{\text{th}} \approx 0.416$ result from the star-mesh transform technique. This evidence underscores the viability of the new approach for approximating the concurrence percolation threshold accurately.

One major benefit of this path-counting approach is its speed. As shown in Fig.~\ref{figure7}, the algorithm is over a hundred times faster than the heuristic star-mesh transform. This substantial increase in computational speed facilitates the extension of concurrence percolation threshold calculations to more complex network topologies, such as random networks.

\emph{Random networks}.---For random networks, the sponge-crossing concurrence is simply defined as the concurrence between $S=\{s\}$ and $T=\{t\}$, each set containing only one source (target) node $s$ ($t$), picked such that the shortest path between $s$ and $t$ is equal to the diameter of the network.
By randomly generating and averaging $10^2$ network realizations of certain sizes and degree distributions, the concurrence percolation threshold $\theta_\text{th}$ is obtained.

The outcomes of this numerical approach, applied across different topologies including the Erd\H{o}s--R\'{e}nyi (ER)~\cite{ER1960} and the Barab\'{a}si--Albert (BA)~\cite{Barabasi1999} random networks at large scales ($N\sim 10^4$), are summarized in Table~\ref{table_fast_concurrence}. These findings shed light on the inherent capacities of large-scale complex QNs, opening new avenues for exploration.

\begin{table}[t]
	\centering
	\begin{tabular}{l|l|l}
		\hline\hline
		Network topology & $ (\pi/4)^{-1} \theta_{\text{th}}$  (fast approximation)  & $(\pi/4)^{-1} \theta_{\text{th}}$~\cite{conpt_mgh21}\\ \hline
		{Bethe Lattice}  ($L = 100, k = 3, S_{\infty}$) & 0.5 & 0.5 \\ 
		{Bethe Lattice} ($L = 100, k = 4, S_{\infty}$) & 0.39 & 0.3918 \\ 
		2D square ($N = 8^2, S_9$) &  0.40 & 0.416 \\ 
		2D square ($N = 20^2, S_3$) &  0.44 & n/a \\ 
		2D square ($N = 200^2, S_2$) &  0.5 & n/a \\ 
		ER ({$N = 10^3, k = 3, S_5$}) &  {$0.6 \pm 0.002$} & n/a \\ 
		ER ({$N = 10^3, k = 4, S_5$}) &  {$0.53 \pm 0.0019$} & n/a \\ 
		ER ({$N = 10^4, k = 2, S_1$}) &  {$0.85 \pm 0.0021$} & n/a \\ 
		BA ({$N = 10^3, z = 5, S_1$}) &  {$0.3 \pm 0.0018$} & n/a \\ 
		BA ({$N = 10^4, z = 1, S_5$}) &  {$0.86 \pm 0.0057$} & n/a \\ \hline\hline
	\end{tabular}
	\caption{\label{table_fast_concurrence}{Numerical concurrence percolation thresholds $\theta_{\text{th}}$ ($\equiv 2^{-1}\arcsin{c_\text{th}}$) of different network topologies, obtained by the $S_m$ and parallel approximations~\cite{conpt_mmhksg22}.} The results are compared to those provided in Meng~et~al.~\cite{conpt_mgh21} for the Bethe lattice and 2D square lattice. Results on Erd\H{o}s--R\'{e}nyi (ER) and Barab\'{a}si--Albert (BA) networks are also reported.\hfill\hfill}
\end{table}

\section{Discussion and Conclusions}
\label{sec_discussion}

Distributing entanglement throughout a quantum network (QN) is a critical and complex problem at the heart of quantum communications that has attracted significant attention and studies. This field has been further enriched from a statistical physics point of view, by the discovery of two percolation theories (classical versus concurrence) that, at first glance, appear to be remotely related but are, in fact, fundamentally distinct (Fig.~\ref{fig_percolation}). These theories have not only deepened our understanding but also raised many new questions for further exploration and potentially groundbreaking research. In the following, we will outline and discuss some of the open questions that have been brought to light by these developments:

\begin{itemize}
    \item \emph{Optimality}. Does there exist an optimal scheme for entanglement transmission? 
    In the context of classical percolation, both classical entanglement percolation (CEP) and quantum entanglement percolation (QEP) fall short of yielding the optimal singlet conversion probability, especially as network size scales up~\cite{QEP_acl07}. The deterministic entanglement transmission (DET)~\cite{det_mcghr23}, on the other hand,  focuses on improving not the singlet conversion probability but the entanglement that can be deterministically obtained. 
    It has been found that the DET optimizes the \emph{average} concurrence on either series [Fig.~\ref{fig_seri}] \emph{or} parallel topologies [Fig.~\ref{fig_para}], meaning that even when we \emph{relax the requirement of determinacy} and consider the average entanglement of general probabilistic outcomes, the DET results remain the optimal on series or parallel topologies~\cite{det_mcghr23}.
    However, this result does not generalize to general series-parallel topologies [Fig.~\ref{fig_seri-para}], where DET may not always offer the optimal average concurrence. 
    This prompts us to ask how effective the DET actually is across various QN topologies.
    Answering this question could substantially deepen our comprehension of the maximum entanglement capacity of QN.
    \item \emph{Universality.} As a statistical physics theory, the concurrence percolation also exhibits a second-order phase transition near the threshold $c_\text{th}$, similar to classical percolation near $p_\text{th}$.  So, what can be said about the universality of this phase transition?  It has been found that the thermal critical exponent $\nu$ remains the same on different 2D lattices (square versus honeycomb versus triangular), suggesting that universality is likely at play near the critical threshold. Yet, the current definition of percolation based on connectivity rules (Table~\ref{table_rules}) does not clearly define an order parameter~\cite{conpt_mgh21}. 
    This lack of an order parameter hints at a missing degree of freedom in the connectivity-based model. This omission makes it challenging to determine other critical exponents besides $\nu$ (or its dynamic counterpart, $z\nu$). 
    Additionally, the existing data on $\nu$ (or $z\nu$) do not allow us to distinguish between the universality classes of concurrence percolation and classical percolation~\cite{conpt_mgh21}.
    Thus, an open question remains: are concurrence percolation and classical percolation simply two facets of a single underlying statistical theory if we overlook short-range details, or are they genuinely distinct theories at a macroscopic level?
    \item \emph{Experimental implementation.} One of the greatest challenges of quantum information experiments is to achieve high-efficiency multi-body operations. 
    For instance, two-body quantum gates like CNOT are considerably more challenging and less efficient to implement compared to single-body gates such as the rotation gates RX, RY, and RZ. In fact, the number of two-body gates often serves as a benchmark for gauging the computational complexity of quantum algorithms.
    Interestingly, a practical QN offers an easier path to scalability compared to universal quantum computers. This is since in a QN, only local operations are allowed on qubits across different nodes. This eliminates the need for complex gates like CNOT between qubits from different nodes. This design constraint substantially simplifies QN implementation and boosts its scalability.
    Recently, experimental feasibility of the series/parallel rules of the DET scheme (Table~\ref{table_rules_scheme}) has also been demonstrated on IBM's quantum computation platform \emph{Qiskit}~\cite{qiskit_a21}. The series and parallel rules typically perform with fidelity rates of $92.4\%$ and $78.2\%$, respectively~\cite{det_mcghr23}. 
    These rates are expected to improve, given ongoing advancements in two-qubit gate fidelity~\cite{q-cnot-fidel_cbdsbe19,q-cnot-fidel_kwsmckkdm21}.
    Compared to the CEP/QEP schemes, the DET scheme has its advantages and drawbacks. On the upside, the DET inputs/outputs are only partially entangled, which generally makes them easier to produce and results in higher fidelity. On the flip side, circuit parameters are input-dependent, requiring precise initial state estimations through techniques like heralding~\cite{q-netw_hkmsvtmh18} or tomography~\cite{q-tomogr_bcb21} before deployment.
    This brings us to a crucial question: to what extent can we experimentally scale the DET scheme for larger QNs? More importantly, given that the current CEP/QEP/DET schemes focus solely on pure states, there is an urgent need to extend these results to mixed states that are affected by noise---a vital step for the practical implementation of QNs.
    \item \emph{Other network-based tasks enhanced by entanglement.} The feasibility of {establishing} entanglement over network structures also opens up further new possibilities of \emph{using} entanglement to enhance some more general, nontrivial network-based tasks. For example, in Refs.~\cite{q-game_lhj09,q-game_mps14}, researchers studied the application of entanglement to enhance quantum games on both regular lattices and complex network structures, demonstrating that entanglement is a crucial resource for achieving favorable outcomes in the realm of quantum game theory. Additionally, similar improvements have been noted, such as in a quantum adaptation of the card game, bridge, as highlighted in Ref.~\cite{q-game_mtkpb14}. 
    The discussion on networks of networks in Section~\ref{sec_non} also provides an alternative perspective regarding entanglement. Rather than regarding it solely as a resource, we can view entanglement as a control parameter that regulates the interdependency between multiple network layers, potentially giving rise to novel critical or multicritical behaviors.
    Indeed, recent theoretical advancements in quantum phase transitions~\cite{first-order-q-phase-transit_p05,first-order-q-phase-transit_lg08,first-order-q-phase-transit_ll08} have suggested that long-range entanglement among quantum spins at near absolute zero temperatures could trigger a shift from a second-order to a first-order phase transition. We hypothesize that this long-range entanglement may operate similarly to the introduction of interdependency among nodes across multiple layers, akin to classical networks-of-networks models. Yet, it is worth noting that the underlying physics governing this interdependency stems from entirely distinct principles within the quantum realm.
\end{itemize}

We have explored the far-reaching implications of entanglement transmission in large-scale quantum networks, all through the lens of the percolation frameworks. The presence of two distinct types of percolation---classical versus concurrence---clearly suggests that the statistical landscape of quantum networks is rather complex. As we look forward, we are enthusiastic that both theoretical and experimental progress in the field of quantum networks will enrich our understanding of this captivating area of study.

\bibliography{concurrence-bib}

\begin{thebibliography}{193}%
\makeatletter
\providecommand \@ifxundefined [1]{%
 \@ifx{#1\undefined}
}%
\providecommand \@ifnum [1]{%
 \ifnum #1\expandafter \@firstoftwo
 \else \expandafter \@secondoftwo
 \fi
}%
\providecommand \@ifx [1]{%
 \ifx #1\expandafter \@firstoftwo
 \else \expandafter \@secondoftwo
 \fi
}%
\providecommand \natexlab [1]{#1}%
\providecommand \enquote  [1]{``#1''}%
\providecommand \bibnamefont  [1]{#1}%
\providecommand \bibfnamefont [1]{#1}%
\providecommand \citenamefont [1]{#1}%
\providecommand \href@noop [0]{\@secondoftwo}%
\providecommand \href [0]{\begingroup \@sanitize@url \@href}%
\providecommand \@href[1]{\@@startlink{#1}\@@href}%
\providecommand \@@href[1]{\endgroup#1\@@endlink}%
\providecommand \@sanitize@url [0]{\catcode `\\12\catcode `\$12\catcode `\&12\catcode `\#12\catcode `\^12\catcode `\_12\catcode `\%12\relax}%
\providecommand \@@startlink[1]{}%
\providecommand \@@endlink[0]{}%
\providecommand \url  [0]{\begingroup\@sanitize@url \@url }%
\providecommand \@url [1]{\endgroup\@href {#1}{\urlprefix }}%
\providecommand \urlprefix  [0]{URL }%
\providecommand \Eprint [0]{\href }%
\providecommand \doibase [0]{https://doi.org/}%
\providecommand \selectlanguage [0]{\@gobble}%
\providecommand \bibinfo  [0]{\@secondoftwo}%
\providecommand \bibfield  [0]{\@secondoftwo}%
\providecommand \translation [1]{[#1]}%
\providecommand \BibitemOpen [0]{}%
\providecommand \bibitemStop [0]{}%
\providecommand \bibitemNoStop [0]{.\EOS\space}%
\providecommand \EOS [0]{\spacefactor3000\relax}%
\providecommand \BibitemShut  [1]{\csname bibitem#1\endcsname}%
\let\auto@bib@innerbib\@empty
\bibitem [{\citenamefont {Nielsen}\ and\ \citenamefont {Chuang}(2010)}]{q-comput-q-inf}%
  \BibitemOpen
  \bibfield  {author} {\bibinfo {author} {\bibfnamefont {M.~A.}\ \bibnamefont {Nielsen}}\ and\ \bibinfo {author} {\bibfnamefont {I.~L.}\ \bibnamefont {Chuang}},\ }\href@noop {} {\emph {\bibinfo {title} {Quantum {{Computation}} and {{Quantum Information}}}}},\ \bibinfo {edition} {10th}\ ed.\ (\bibinfo  {publisher} {{Cambridge University Press}},\ \bibinfo {address} {{New York}},\ \bibinfo {year} {2010})\BibitemShut {NoStop}%
\bibitem [{\citenamefont {Cohen}\ and\ \citenamefont {Louie}(2016)}]{fundam-condens-matter-phys}%
  \BibitemOpen
  \bibfield  {author} {\bibinfo {author} {\bibfnamefont {M.~L.}\ \bibnamefont {Cohen}}\ and\ \bibinfo {author} {\bibfnamefont {S.~G.}\ \bibnamefont {Louie}},\ }\href@noop {} {\emph {\bibinfo {title} {Fundamentals of {{Condensed Matter Physics}}}}},\ \bibinfo {edition} {1st}\ ed.\ (\bibinfo  {publisher} {{Cambridge University Press}},\ \bibinfo {address} {{Cambridge, UK}},\ \bibinfo {year} {2016})\BibitemShut {NoStop}%
\bibitem [{\citenamefont {Plischke}\ and\ \citenamefont {Bergersen}(1994)}]{equilib-stat-phys}%
  \BibitemOpen
  \bibfield  {author} {\bibinfo {author} {\bibfnamefont {M.}~\bibnamefont {Plischke}}\ and\ \bibinfo {author} {\bibfnamefont {B.}~\bibnamefont {Bergersen}},\ }\href@noop {} {\emph {\bibinfo {title} {Equilibrium {{Statistical Physics}}}}},\ \bibinfo {edition} {2nd}\ ed.\ (\bibinfo  {publisher} {{World Scientific}},\ \bibinfo {address} {{Singapore}},\ \bibinfo {year} {1994})\BibitemShut {NoStop}%
\bibitem [{\citenamefont {Itzykson}\ and\ \citenamefont {Drouffe}(1989{\natexlab{a}})}]{stat-field-theor-1}%
  \BibitemOpen
  \bibfield  {author} {\bibinfo {author} {\bibfnamefont {C.}~\bibnamefont {Itzykson}}\ and\ \bibinfo {author} {\bibfnamefont {J.-M.}\ \bibnamefont {Drouffe}},\ }\href@noop {} {\emph {\bibinfo {title} {Statistical {{Field Theory}}: {{Volume}} 1, {{From Brownian Motion}} to {{Renormalization}} and {{Lattice Gauge Theory}}}}},\ \bibinfo {edition} {1st}\ ed.\ (\bibinfo  {publisher} {{Cambridge University Press}},\ \bibinfo {address} {{New York}},\ \bibinfo {year} {1989})\BibitemShut {NoStop}%
\bibitem [{\citenamefont {Itzykson}\ and\ \citenamefont {Drouffe}(1989{\natexlab{b}})}]{stat-field-theor-2}%
  \BibitemOpen
  \bibfield  {author} {\bibinfo {author} {\bibfnamefont {C.}~\bibnamefont {Itzykson}}\ and\ \bibinfo {author} {\bibfnamefont {J.-M.}\ \bibnamefont {Drouffe}},\ }\href@noop {} {\emph {\bibinfo {title} {Statistical {{Field Theory}}: {{Volume}} 2, {{Strong Coupling}}, {{Monte Carlo Methods}}, {{Conformal Field Theory}} and {{Random Systems}}}}},\ \bibinfo {edition} {1st}\ ed.\ (\bibinfo  {publisher} {{Cambridge University Press}},\ \bibinfo {address} {{New York}},\ \bibinfo {year} {1989})\BibitemShut {NoStop}%
\bibitem [{\citenamefont {Albert}\ and\ \citenamefont {Barab{\'a}si}(2002)}]{stat-of-netw_ab02}%
  \BibitemOpen
  \bibfield  {author} {\bibinfo {author} {\bibfnamefont {R.}~\bibnamefont {Albert}}\ and\ \bibinfo {author} {\bibfnamefont {A.-L.}\ \bibnamefont {Barab{\'a}si}},\ }\bibfield  {title} {\bibinfo {title} {Statistical mechanics of complex networks},\ }\href {https://doi.org/10.1103/RevModPhys.74.47} {\bibfield  {journal} {\bibinfo  {journal} {Rev. Mod. Phys.}\ }\textbf {\bibinfo {volume} {74}},\ \bibinfo {pages} {47} (\bibinfo {year} {2002})}\BibitemShut {NoStop}%
\bibitem [{\citenamefont {Newman}\ \emph {et~al.}(2006)\citenamefont {Newman}, \citenamefont {Barab{\'a}si},\ and\ \citenamefont {Watts}}]{struct-dyn-netw}%
  \BibitemOpen
  \bibfield  {author} {\bibinfo {author} {\bibfnamefont {M.}~\bibnamefont {Newman}}, \bibinfo {author} {\bibfnamefont {A.-L.}\ \bibnamefont {Barab{\'a}si}},\ and\ \bibinfo {author} {\bibfnamefont {D.~J.}\ \bibnamefont {Watts}},\ }\href@noop {} {\emph {\bibinfo {title} {The {{Structure}} and {{Dynamics}} of {{Networks}}}}},\ \bibinfo {edition} {1st}\ ed.\ (\bibinfo  {publisher} {{Princeton University Press}},\ \bibinfo {address} {{Princeton}},\ \bibinfo {year} {2006})\BibitemShut {NoStop}%
\bibitem [{\citenamefont {Chitambar}\ and\ \citenamefont {Gour}(2019)}]{q-resour_cg19}%
  \BibitemOpen
  \bibfield  {author} {\bibinfo {author} {\bibfnamefont {E.}~\bibnamefont {Chitambar}}\ and\ \bibinfo {author} {\bibfnamefont {G.}~\bibnamefont {Gour}},\ }\bibfield  {title} {\bibinfo {title} {Quantum resource theories},\ }\href {https://doi.org/10.1103/RevModPhys.91.025001} {\bibfield  {journal} {\bibinfo  {journal} {Rev. Mod. Phys.}\ }\textbf {\bibinfo {volume} {91}},\ \bibinfo {pages} {025001} (\bibinfo {year} {2019})}\BibitemShut {NoStop}%
\bibitem [{\citenamefont {Cirac}\ \emph {et~al.}(1997)\citenamefont {Cirac}, \citenamefont {Zoller}, \citenamefont {Kimble},\ and\ \citenamefont {Mabuchi}}]{entangle-distrib_czkm97}%
  \BibitemOpen
  \bibfield  {author} {\bibinfo {author} {\bibfnamefont {J.~I.}\ \bibnamefont {Cirac}}, \bibinfo {author} {\bibfnamefont {P.}~\bibnamefont {Zoller}}, \bibinfo {author} {\bibfnamefont {H.~J.}\ \bibnamefont {Kimble}},\ and\ \bibinfo {author} {\bibfnamefont {H.}~\bibnamefont {Mabuchi}},\ }\bibfield  {title} {\bibinfo {title} {Quantum {{State Transfer}} and {{Entanglement Distribution Among Distant Nodes}} in a {{Quantum Network}}},\ }\href {https://doi.org/10.1103/PhysRevLett.78.3221} {\bibfield  {journal} {\bibinfo  {journal} {Phys. Rev. Lett.}\ }\textbf {\bibinfo {volume} {78}},\ \bibinfo {pages} {3221} (\bibinfo {year} {1997})}\BibitemShut {NoStop}%
\bibitem [{\citenamefont {Ac{\'i}n}\ \emph {et~al.}(2007)\citenamefont {Ac{\'i}n}, \citenamefont {Cirac},\ and\ \citenamefont {Lewenstein}}]{QEP_acl07}%
  \BibitemOpen
  \bibfield  {author} {\bibinfo {author} {\bibfnamefont {A.}~\bibnamefont {Ac{\'i}n}}, \bibinfo {author} {\bibfnamefont {J.~I.}\ \bibnamefont {Cirac}},\ and\ \bibinfo {author} {\bibfnamefont {M.}~\bibnamefont {Lewenstein}},\ }\bibfield  {title} {\bibinfo {title} {Entanglement percolation in quantum networks},\ }\href {https://doi.org/10.1038/nphys549} {\bibfield  {journal} {\bibinfo  {journal} {Nat. Phys.}\ }\textbf {\bibinfo {volume} {3}},\ \bibinfo {pages} {256} (\bibinfo {year} {2007})}\BibitemShut {NoStop}%
\bibitem [{\citenamefont {Kimble}(2008)}]{q-internet_k08}%
  \BibitemOpen
  \bibfield  {author} {\bibinfo {author} {\bibfnamefont {H.~J.}\ \bibnamefont {Kimble}},\ }\bibfield  {title} {\bibinfo {title} {The quantum {{I}}nternet},\ }\href {https://doi.org/10.1038/nature07127} {\bibfield  {journal} {\bibinfo  {journal} {Nature}\ }\textbf {\bibinfo {volume} {453}},\ \bibinfo {pages} {1023} (\bibinfo {year} {2008})}\BibitemShut {NoStop}%
\bibitem [{\citenamefont {Broadfoot}\ \emph {et~al.}(2009)\citenamefont {Broadfoot}, \citenamefont {Dorner},\ and\ \citenamefont {Jaksch}}]{QEP-mix-state_bdj09}%
  \BibitemOpen
  \bibfield  {author} {\bibinfo {author} {\bibfnamefont {S.}~\bibnamefont {Broadfoot}}, \bibinfo {author} {\bibfnamefont {U.}~\bibnamefont {Dorner}},\ and\ \bibinfo {author} {\bibfnamefont {D.}~\bibnamefont {Jaksch}},\ }\bibfield  {title} {\bibinfo {title} {Entanglement percolation with bipartite mixed states},\ }\href {https://doi.org/10.1209/0295-5075/88/50002} {\bibfield  {journal} {\bibinfo  {journal} {EPL}\ }\textbf {\bibinfo {volume} {88}},\ \bibinfo {pages} {50002} (\bibinfo {year} {2009})}\BibitemShut {NoStop}%
\bibitem [{\citenamefont {Lapeyre}\ \emph {et~al.}(2009)\citenamefont {Lapeyre}, \citenamefont {Wehr},\ and\ \citenamefont {Lewenstein}}]{QEP-lattice_lwl09}%
  \BibitemOpen
  \bibfield  {author} {\bibinfo {author} {\bibfnamefont {G.~J.}\ \bibnamefont {Lapeyre}}, \bibinfo {author} {\bibfnamefont {J.}~\bibnamefont {Wehr}},\ and\ \bibinfo {author} {\bibfnamefont {M.}~\bibnamefont {Lewenstein}},\ }\bibfield  {title} {\bibinfo {title} {Enhancement of entanglement percolation in quantum networks via lattice transformations},\ }\href {https://doi.org/10.1103/PhysRevA.79.042324} {\bibfield  {journal} {\bibinfo  {journal} {Phys. Rev. A}\ }\textbf {\bibinfo {volume} {79}},\ \bibinfo {pages} {042324} (\bibinfo {year} {2009})}\BibitemShut {NoStop}%
\bibitem [{\citenamefont {Broadfoot}\ \emph {et~al.}(2010)\citenamefont {Broadfoot}, \citenamefont {Dorner},\ and\ \citenamefont {Jaksch}}]{QEP-mix-state_bdj10}%
  \BibitemOpen
  \bibfield  {author} {\bibinfo {author} {\bibfnamefont {S.}~\bibnamefont {Broadfoot}}, \bibinfo {author} {\bibfnamefont {U.}~\bibnamefont {Dorner}},\ and\ \bibinfo {author} {\bibfnamefont {D.}~\bibnamefont {Jaksch}},\ }\bibfield  {title} {\bibinfo {title} {Singlet generation in mixed-state quantum networks},\ }\href {https://doi.org/10.1103/PhysRevA.81.042316} {\bibfield  {journal} {\bibinfo  {journal} {Phys. Rev. A}\ }\textbf {\bibinfo {volume} {81}},\ \bibinfo {pages} {042316} (\bibinfo {year} {2010})}\BibitemShut {NoStop}%
\bibitem [{\citenamefont {Albert}\ \emph {et~al.}(2000)\citenamefont {Albert}, \citenamefont {Jeong},\ and\ \citenamefont {Barab{\'a}si}}]{albert2000error}%
  \BibitemOpen
  \bibfield  {author} {\bibinfo {author} {\bibfnamefont {R.}~\bibnamefont {Albert}}, \bibinfo {author} {\bibfnamefont {H.}~\bibnamefont {Jeong}},\ and\ \bibinfo {author} {\bibfnamefont {A.-L.}\ \bibnamefont {Barab{\'a}si}},\ }\bibfield  {title} {\bibinfo {title} {Error and attack tolerance of complex networks},\ }\href {https://doi.org/10.1038/35019019} {\bibfield  {journal} {\bibinfo  {journal} {Nature}\ }\textbf {\bibinfo {volume} {406}},\ \bibinfo {pages} {378} (\bibinfo {year} {2000})}\BibitemShut {NoStop}%
\bibitem [{\citenamefont {Wu}\ and\ \citenamefont {Zhu}(2011)}]{QEP-complex-netw_wz11}%
  \BibitemOpen
  \bibfield  {author} {\bibinfo {author} {\bibfnamefont {L.}~\bibnamefont {Wu}}\ and\ \bibinfo {author} {\bibfnamefont {S.}~\bibnamefont {Zhu}},\ }\bibfield  {title} {\bibinfo {title} {Entanglement percolation on a quantum internet with scale-free and clustering characters},\ }\href {https://doi.org/10.1103/PhysRevA.84.052304} {\bibfield  {journal} {\bibinfo  {journal} {Phys. Rev. A}\ }\textbf {\bibinfo {volume} {84}},\ \bibinfo {pages} {052304} (\bibinfo {year} {2011})}\BibitemShut {NoStop}%
\bibitem [{\citenamefont {Li}\ \emph {et~al.}(2012)\citenamefont {Li}, \citenamefont {Cavalcanti},\ and\ \citenamefont {Kwek}}]{q-netw-simul_lck12}%
  \BibitemOpen
  \bibfield  {author} {\bibinfo {author} {\bibfnamefont {Y.}~\bibnamefont {Li}}, \bibinfo {author} {\bibfnamefont {D.}~\bibnamefont {Cavalcanti}},\ and\ \bibinfo {author} {\bibfnamefont {L.~C.}\ \bibnamefont {Kwek}},\ }\bibfield  {title} {\bibinfo {title} {Long-distance entanglement generation with scalable and robust two-dimensional quantum network},\ }\href {https://doi.org/10.1103/PhysRevA.85.062330} {\bibfield  {journal} {\bibinfo  {journal} {Phys. Rev. A}\ }\textbf {\bibinfo {volume} {85}},\ \bibinfo {pages} {062330} (\bibinfo {year} {2012})}\BibitemShut {NoStop}%
\bibitem [{\citenamefont {Perseguers}\ \emph {et~al.}(2013)\citenamefont {Perseguers}, \citenamefont {Lapeyre~Jr.}, \citenamefont {Cavalcanti}, \citenamefont {Lewenstein},\ and\ \citenamefont {Ac{\'i}n}}]{q-netw-summ_pjcla13}%
  \BibitemOpen
  \bibfield  {author} {\bibinfo {author} {\bibfnamefont {S.}~\bibnamefont {Perseguers}}, \bibinfo {author} {\bibfnamefont {G.~J.}\ \bibnamefont {Lapeyre~Jr.}}, \bibinfo {author} {\bibfnamefont {D.}~\bibnamefont {Cavalcanti}}, \bibinfo {author} {\bibfnamefont {M.}~\bibnamefont {Lewenstein}},\ and\ \bibinfo {author} {\bibfnamefont {A.}~\bibnamefont {Ac{\'i}n}},\ }\bibfield  {title} {\bibinfo {title} {Distribution of entanglement in large-scale quantum networks},\ }\href {https://doi.org/10.1088/0034-4885/76/9/096001} {\bibfield  {journal} {\bibinfo  {journal} {Rep. Prog. Phys.}\ }\textbf {\bibinfo {volume} {76}},\ \bibinfo {pages} {096001} (\bibinfo {year} {2013})}\BibitemShut {NoStop}%
\bibitem [{\citenamefont {Siomau}(2017)}]{QEP-gossip_s17}%
  \BibitemOpen
  \bibfield  {author} {\bibinfo {author} {\bibfnamefont {M.}~\bibnamefont {Siomau}},\ }\bibfield  {title} {\bibinfo {title} {Gossip algorithms in quantum networks},\ }\href {https://doi.org/10.1016/j.physleta.2016.10.057} {\bibfield  {journal} {\bibinfo  {journal} {Phys. Lett. A}\ }\textbf {\bibinfo {volume} {381}},\ \bibinfo {pages} {136} (\bibinfo {year} {2017})}\BibitemShut {NoStop}%
\bibitem [{\citenamefont {Carvacho}\ \emph {et~al.}(2017)\citenamefont {Carvacho}, \citenamefont {Andreoli}, \citenamefont {Santodonato}, \citenamefont {Bentivegna}, \citenamefont {Chaves},\ and\ \citenamefont {Sciarrino}}]{q-netw_casbcs17}%
  \BibitemOpen
  \bibfield  {author} {\bibinfo {author} {\bibfnamefont {G.}~\bibnamefont {Carvacho}}, \bibinfo {author} {\bibfnamefont {F.}~\bibnamefont {Andreoli}}, \bibinfo {author} {\bibfnamefont {L.}~\bibnamefont {Santodonato}}, \bibinfo {author} {\bibfnamefont {M.}~\bibnamefont {Bentivegna}}, \bibinfo {author} {\bibfnamefont {R.}~\bibnamefont {Chaves}},\ and\ \bibinfo {author} {\bibfnamefont {F.}~\bibnamefont {Sciarrino}},\ }\bibfield  {title} {\bibinfo {title} {Experimental violation of local causality in a quantum network},\ }\href {https://doi.org/10.1038/ncomms14775} {\bibfield  {journal} {\bibinfo  {journal} {Nat. Commun.}\ }\textbf {\bibinfo {volume} {8}},\ \bibinfo {pages} {14775} (\bibinfo {year} {2017})}\BibitemShut {NoStop}%
\bibitem [{\citenamefont {Das}\ \emph {et~al.}(2018)\citenamefont {Das}, \citenamefont {Khatri},\ and\ \citenamefont {Dowling}}]{q-netw_dkd18}%
  \BibitemOpen
  \bibfield  {author} {\bibinfo {author} {\bibfnamefont {S.}~\bibnamefont {Das}}, \bibinfo {author} {\bibfnamefont {S.}~\bibnamefont {Khatri}},\ and\ \bibinfo {author} {\bibfnamefont {J.~P.}\ \bibnamefont {Dowling}},\ }\bibfield  {title} {\bibinfo {title} {Robust quantum network architectures and topologies for entanglement distribution},\ }\href {https://doi.org/10.1103/PhysRevA.97.012335} {\bibfield  {journal} {\bibinfo  {journal} {Phys. Rev. A}\ }\textbf {\bibinfo {volume} {97}},\ \bibinfo {pages} {012335} (\bibinfo {year} {2018})}\BibitemShut {NoStop}%
\bibitem [{\citenamefont {Pirandola}(2019)}]{q-netw-route_p19}%
  \BibitemOpen
  \bibfield  {author} {\bibinfo {author} {\bibfnamefont {S.}~\bibnamefont {Pirandola}},\ }\bibfield  {title} {\bibinfo {title} {End-to-end capacities of a quantum communication network},\ }\href {https://doi.org/10.1038/s42005-019-0147-3} {\bibfield  {journal} {\bibinfo  {journal} {Commun. Phys.}\ }\textbf {\bibinfo {volume} {2}},\ \bibinfo {pages} {1} (\bibinfo {year} {2019})}\BibitemShut {NoStop}%
\bibitem [{\citenamefont {Biamonte}\ \emph {et~al.}(2019)\citenamefont {Biamonte}, \citenamefont {Faccin},\ and\ \citenamefont {Domenico}}]{q-netw-summ_bfd19}%
  \BibitemOpen
  \bibfield  {author} {\bibinfo {author} {\bibfnamefont {J.}~\bibnamefont {Biamonte}}, \bibinfo {author} {\bibfnamefont {M.}~\bibnamefont {Faccin}},\ and\ \bibinfo {author} {\bibfnamefont {M.~D.}\ \bibnamefont {Domenico}},\ }\bibfield  {title} {\bibinfo {title} {Complex networks from classical to quantum},\ }\href {https://doi.org/10.1038/s42005-019-0152-6} {\bibfield  {journal} {\bibinfo  {journal} {Commun. Phys.}\ }\textbf {\bibinfo {volume} {2}},\ \bibinfo {pages} {53} (\bibinfo {year} {2019})}\BibitemShut {NoStop}%
\bibitem [{\citenamefont {Unnikrishnan}\ \emph {et~al.}(2019)\citenamefont {Unnikrishnan}, \citenamefont {MacFarlane}, \citenamefont {Yi}, \citenamefont {Diamanti}, \citenamefont {Markham},\ and\ \citenamefont {Kerenidis}}]{q-netw_umfydmk19}%
  \BibitemOpen
  \bibfield  {author} {\bibinfo {author} {\bibfnamefont {A.}~\bibnamefont {Unnikrishnan}}, \bibinfo {author} {\bibfnamefont {I.~J.}\ \bibnamefont {MacFarlane}}, \bibinfo {author} {\bibfnamefont {R.}~\bibnamefont {Yi}}, \bibinfo {author} {\bibfnamefont {E.}~\bibnamefont {Diamanti}}, \bibinfo {author} {\bibfnamefont {D.}~\bibnamefont {Markham}},\ and\ \bibinfo {author} {\bibfnamefont {I.}~\bibnamefont {Kerenidis}},\ }\bibfield  {title} {\bibinfo {title} {Anonymity for {{Practical Quantum Networks}}},\ }\href {https://doi.org/10.1103/PhysRevLett.122.240501} {\bibfield  {journal} {\bibinfo  {journal} {Phys. Rev. Lett.}\ }\textbf {\bibinfo {volume} {122}},\ \bibinfo {pages} {240501} (\bibinfo {year} {2019})}\BibitemShut {NoStop}%
\bibitem [{\citenamefont {Castellini}\ \emph {et~al.}(2019)\citenamefont {Castellini}, \citenamefont {Bellomo}, \citenamefont {Compagno},\ and\ \citenamefont {Lo~Franco}}]{q-netw_cbclf19}%
  \BibitemOpen
  \bibfield  {author} {\bibinfo {author} {\bibfnamefont {A.}~\bibnamefont {Castellini}}, \bibinfo {author} {\bibfnamefont {B.}~\bibnamefont {Bellomo}}, \bibinfo {author} {\bibfnamefont {G.}~\bibnamefont {Compagno}},\ and\ \bibinfo {author} {\bibfnamefont {R.}~\bibnamefont {Lo~Franco}},\ }\bibfield  {title} {\bibinfo {title} {Activating remote entanglement in a quantum network by local counting of identical particles},\ }\href {https://doi.org/10.1103/PhysRevA.99.062322} {\bibfield  {journal} {\bibinfo  {journal} {Phys. Rev. A}\ }\textbf {\bibinfo {volume} {99}},\ \bibinfo {pages} {062322} (\bibinfo {year} {2019})}\BibitemShut {NoStop}%
\bibitem [{\citenamefont {Khabiboulline}\ \emph {et~al.}(2019)\citenamefont {Khabiboulline}, \citenamefont {Borregaard}, \citenamefont {De~Greve},\ and\ \citenamefont {Lukin}}]{q-netw_kbdgl19}%
  \BibitemOpen
  \bibfield  {author} {\bibinfo {author} {\bibfnamefont {E.~T.}\ \bibnamefont {Khabiboulline}}, \bibinfo {author} {\bibfnamefont {J.}~\bibnamefont {Borregaard}}, \bibinfo {author} {\bibfnamefont {K.}~\bibnamefont {De~Greve}},\ and\ \bibinfo {author} {\bibfnamefont {M.~D.}\ \bibnamefont {Lukin}},\ }\bibfield  {title} {\bibinfo {title} {Optical {{Interferometry}} with {{Quantum Networks}}},\ }\href {https://doi.org/10.1103/PhysRevLett.123.070504} {\bibfield  {journal} {\bibinfo  {journal} {Phys. Rev. Lett.}\ }\textbf {\bibinfo {volume} {123}},\ \bibinfo {pages} {070504} (\bibinfo {year} {2019})}\BibitemShut {NoStop}%
\bibitem [{\citenamefont {Renou}\ \emph {et~al.}(2019)\citenamefont {Renou}, \citenamefont {Wang}, \citenamefont {Boreiri}, \citenamefont {Beigi}, \citenamefont {Gisin},\ and\ \citenamefont {Brunner}}]{q-netw_rwbbgb19}%
  \BibitemOpen
  \bibfield  {author} {\bibinfo {author} {\bibfnamefont {M.-O.}\ \bibnamefont {Renou}}, \bibinfo {author} {\bibfnamefont {Y.}~\bibnamefont {Wang}}, \bibinfo {author} {\bibfnamefont {S.}~\bibnamefont {Boreiri}}, \bibinfo {author} {\bibfnamefont {S.}~\bibnamefont {Beigi}}, \bibinfo {author} {\bibfnamefont {N.}~\bibnamefont {Gisin}},\ and\ \bibinfo {author} {\bibfnamefont {N.}~\bibnamefont {Brunner}},\ }\bibfield  {title} {\bibinfo {title} {Limits on {{Correlations}} in {{Networks}} for {{Quantum}} and {{No-Signaling Resources}}},\ }\href {https://doi.org/10.1103/PhysRevLett.123.070403} {\bibfield  {journal} {\bibinfo  {journal} {Phys. Rev. Lett.}\ }\textbf {\bibinfo {volume} {123}},\ \bibinfo {pages} {070403} (\bibinfo {year} {2019})}\BibitemShut {NoStop}%
\bibitem [{\citenamefont {Hahn}\ \emph {et~al.}(2019)\citenamefont {Hahn}, \citenamefont {Pappa},\ and\ \citenamefont {Eisert}}]{multipartite-q-netw_hpe19}%
  \BibitemOpen
  \bibfield  {author} {\bibinfo {author} {\bibfnamefont {F.}~\bibnamefont {Hahn}}, \bibinfo {author} {\bibfnamefont {A.}~\bibnamefont {Pappa}},\ and\ \bibinfo {author} {\bibfnamefont {J.}~\bibnamefont {Eisert}},\ }\bibfield  {title} {\bibinfo {title} {Quantum network routing and local complementation},\ }\href {https://doi.org/10.1038/s41534-019-0191-6} {\bibfield  {journal} {\bibinfo  {journal} {npj Quantum Inf.}\ }\textbf {\bibinfo {volume} {5}},\ \bibinfo {pages} {1} (\bibinfo {year} {2019})}\BibitemShut {NoStop}%
\bibitem [{\citenamefont {Zhuang}\ and\ \citenamefont {Zhang}(2021)}]{Zhuang2021}%
  \BibitemOpen
  \bibfield  {author} {\bibinfo {author} {\bibfnamefont {Q.}~\bibnamefont {Zhuang}}\ and\ \bibinfo {author} {\bibfnamefont {B.}~\bibnamefont {Zhang}},\ }\bibfield  {title} {\bibinfo {title} {Quantum communication capacity transition of complex quantum networks},\ }\href {https://doi.org/10.1103/physreva.104.022608} {\bibfield  {journal} {\bibinfo  {journal} {Phys. Rev. A}\ }\textbf {\bibinfo {volume} {104}},\ \bibinfo {pages} {022608} (\bibinfo {year} {2021})}\BibitemShut {NoStop}%
\bibitem [{\citenamefont {Sadhu}\ \emph {et~al.}(2023)\citenamefont {Sadhu}, \citenamefont {Somayajula}, \citenamefont {Horodecki},\ and\ \citenamefont {Das}}]{q-netw_sshd23}%
  \BibitemOpen
  \bibfield  {author} {\bibinfo {author} {\bibfnamefont {A.}~\bibnamefont {Sadhu}}, \bibinfo {author} {\bibfnamefont {M.~A.}\ \bibnamefont {Somayajula}}, \bibinfo {author} {\bibfnamefont {K.}~\bibnamefont {Horodecki}},\ and\ \bibinfo {author} {\bibfnamefont {S.}~\bibnamefont {Das}},\ }\bibfield  {title} {\bibinfo {title} {Practical limitations on robustness and scalability of quantum {{Internet}}},\ }\href@noop {} {\bibfield  {journal} {\bibinfo  {journal} {arXiv:2308.12739}\ } (\bibinfo {year} {2023})}\BibitemShut {NoStop}%
\bibitem [{\citenamefont {Kalb}\ \emph {et~al.}(2017)\citenamefont {Kalb}, \citenamefont {Reiserer}, \citenamefont {Humphreys}, \citenamefont {Bakermans}, \citenamefont {Kamerling}, \citenamefont {Nickerson}, \citenamefont {Benjamin}, \citenamefont {Twitchen}, \citenamefont {Markham},\ and\ \citenamefont {Hanson}}]{q-netw_krhbknbtmh17}%
  \BibitemOpen
  \bibfield  {author} {\bibinfo {author} {\bibfnamefont {N.}~\bibnamefont {Kalb}}, \bibinfo {author} {\bibfnamefont {A.~A.}\ \bibnamefont {Reiserer}}, \bibinfo {author} {\bibfnamefont {P.~C.}\ \bibnamefont {Humphreys}}, \bibinfo {author} {\bibfnamefont {J.~J.~W.}\ \bibnamefont {Bakermans}}, \bibinfo {author} {\bibfnamefont {S.~J.}\ \bibnamefont {Kamerling}}, \bibinfo {author} {\bibfnamefont {N.~H.}\ \bibnamefont {Nickerson}}, \bibinfo {author} {\bibfnamefont {S.~C.}\ \bibnamefont {Benjamin}}, \bibinfo {author} {\bibfnamefont {D.~J.}\ \bibnamefont {Twitchen}}, \bibinfo {author} {\bibfnamefont {M.}~\bibnamefont {Markham}},\ and\ \bibinfo {author} {\bibfnamefont {R.}~\bibnamefont {Hanson}},\ }\bibfield  {title} {\bibinfo {title} {Entanglement distillation between solid-state quantum network nodes},\ }\href {https://doi.org/10.1126/science.aan0070} {\bibfield  {journal} {\bibinfo  {journal} {Science}\ }\textbf {\bibinfo {volume} {356}},\ \bibinfo {pages} {928} (\bibinfo {year} {2017})}\BibitemShut {NoStop}%
\bibitem [{\citenamefont {Humphreys}\ \emph {et~al.}(2018)\citenamefont {Humphreys}, \citenamefont {Kalb}, \citenamefont {Morits}, \citenamefont {Schouten}, \citenamefont {Vermeulen}, \citenamefont {Twitchen}, \citenamefont {Markham},\ and\ \citenamefont {Hanson}}]{q-netw_hkmsvtmh18}%
  \BibitemOpen
  \bibfield  {author} {\bibinfo {author} {\bibfnamefont {P.~C.}\ \bibnamefont {Humphreys}}, \bibinfo {author} {\bibfnamefont {N.}~\bibnamefont {Kalb}}, \bibinfo {author} {\bibfnamefont {J.~P.~J.}\ \bibnamefont {Morits}}, \bibinfo {author} {\bibfnamefont {R.~N.}\ \bibnamefont {Schouten}}, \bibinfo {author} {\bibfnamefont {R.~F.~L.}\ \bibnamefont {Vermeulen}}, \bibinfo {author} {\bibfnamefont {D.~J.}\ \bibnamefont {Twitchen}}, \bibinfo {author} {\bibfnamefont {M.}~\bibnamefont {Markham}},\ and\ \bibinfo {author} {\bibfnamefont {R.}~\bibnamefont {Hanson}},\ }\bibfield  {title} {\bibinfo {title} {Deterministic delivery of remote entanglement on a quantum network},\ }\href {https://doi.org/10.1038/s41586-018-0200-5} {\bibfield  {journal} {\bibinfo  {journal} {Nature}\ }\textbf {\bibinfo {volume} {558}},\ \bibinfo {pages} {268} (\bibinfo {year} {2018})}\BibitemShut {NoStop}%
\bibitem [{\citenamefont {Pompili}\ \emph {et~al.}(2021)\citenamefont {Pompili}, \citenamefont {Hermans}, \citenamefont {Baier}, \citenamefont {Beukers}, \citenamefont {Humphreys}, \citenamefont {Schouten}, \citenamefont {Vermeulen}, \citenamefont {Tiggelman}, \citenamefont {{dos Santos Martins}}, \citenamefont {Dirkse}, \citenamefont {Wehner},\ and\ \citenamefont {Hanson}}]{q-netw_phbbhsvtmdwh21}%
  \BibitemOpen
  \bibfield  {author} {\bibinfo {author} {\bibfnamefont {M.}~\bibnamefont {Pompili}}, \bibinfo {author} {\bibfnamefont {S.~L.~N.}\ \bibnamefont {Hermans}}, \bibinfo {author} {\bibfnamefont {S.}~\bibnamefont {Baier}}, \bibinfo {author} {\bibfnamefont {H.~K.~C.}\ \bibnamefont {Beukers}}, \bibinfo {author} {\bibfnamefont {P.~C.}\ \bibnamefont {Humphreys}}, \bibinfo {author} {\bibfnamefont {R.~N.}\ \bibnamefont {Schouten}}, \bibinfo {author} {\bibfnamefont {R.~F.~L.}\ \bibnamefont {Vermeulen}}, \bibinfo {author} {\bibfnamefont {M.~J.}\ \bibnamefont {Tiggelman}}, \bibinfo {author} {\bibfnamefont {L.}~\bibnamefont {{dos Santos Martins}}}, \bibinfo {author} {\bibfnamefont {B.}~\bibnamefont {Dirkse}}, \bibinfo {author} {\bibfnamefont {S.}~\bibnamefont {Wehner}},\ and\ \bibinfo {author} {\bibfnamefont {R.}~\bibnamefont {Hanson}},\ }\bibfield  {title} {\bibinfo {title} {Realization of a multinode quantum network of remote solid-state qubits},\ }\href {https://doi.org/10.1126/science.abg1919} {\bibfield  {journal}
  {\bibinfo  {journal} {Science}\ }\textbf {\bibinfo {volume} {372}},\ \bibinfo {pages} {259} (\bibinfo {year} {2021})}\BibitemShut {NoStop}%
\bibitem [{\citenamefont {Riebe}\ \emph {et~al.}(2008)\citenamefont {Riebe}, \citenamefont {Monz}, \citenamefont {Kim}, \citenamefont {Villar}, \citenamefont {Schindler}, \citenamefont {Chwalla}, \citenamefont {Hennrich},\ and\ \citenamefont {Blatt}}]{entangle-swap_rmkvschb08}%
  \BibitemOpen
  \bibfield  {author} {\bibinfo {author} {\bibfnamefont {M.}~\bibnamefont {Riebe}}, \bibinfo {author} {\bibfnamefont {T.}~\bibnamefont {Monz}}, \bibinfo {author} {\bibfnamefont {K.}~\bibnamefont {Kim}}, \bibinfo {author} {\bibfnamefont {A.~S.}\ \bibnamefont {Villar}}, \bibinfo {author} {\bibfnamefont {P.}~\bibnamefont {Schindler}}, \bibinfo {author} {\bibfnamefont {M.}~\bibnamefont {Chwalla}}, \bibinfo {author} {\bibfnamefont {M.}~\bibnamefont {Hennrich}},\ and\ \bibinfo {author} {\bibfnamefont {R.}~\bibnamefont {Blatt}},\ }\bibfield  {title} {\bibinfo {title} {Deterministic entanglement swapping with an ion-trap quantum computer},\ }\href {https://doi.org/10.1038/nphys1107} {\bibfield  {journal} {\bibinfo  {journal} {Nat. Phys.}\ }\textbf {\bibinfo {volume} {4}},\ \bibinfo {pages} {839} (\bibinfo {year} {2008})}\BibitemShut {NoStop}%
\bibitem [{\citenamefont {Almendros}\ \emph {et~al.}(2009)\citenamefont {Almendros}, \citenamefont {Huwer}, \citenamefont {Piro}, \citenamefont {Rohde}, \citenamefont {Schuck}, \citenamefont {Hennrich}, \citenamefont {Dubin},\ and\ \citenamefont {Eschner}}]{q-netw-ion-trap_ahprshde09}%
  \BibitemOpen
  \bibfield  {author} {\bibinfo {author} {\bibfnamefont {M.}~\bibnamefont {Almendros}}, \bibinfo {author} {\bibfnamefont {J.}~\bibnamefont {Huwer}}, \bibinfo {author} {\bibfnamefont {N.}~\bibnamefont {Piro}}, \bibinfo {author} {\bibfnamefont {F.}~\bibnamefont {Rohde}}, \bibinfo {author} {\bibfnamefont {C.}~\bibnamefont {Schuck}}, \bibinfo {author} {\bibfnamefont {M.}~\bibnamefont {Hennrich}}, \bibinfo {author} {\bibfnamefont {F.}~\bibnamefont {Dubin}},\ and\ \bibinfo {author} {\bibfnamefont {J.}~\bibnamefont {Eschner}},\ }\bibfield  {title} {\bibinfo {title} {Bandwidth-{{Tunable Single-Photon Source}} in an {{Ion-Trap Quantum Network}}},\ }\href {https://doi.org/10.1103/PhysRevLett.103.213601} {\bibfield  {journal} {\bibinfo  {journal} {Phys. Rev. Lett.}\ }\textbf {\bibinfo {volume} {103}},\ \bibinfo {pages} {213601} (\bibinfo {year} {2009})}\BibitemShut {NoStop}%
\bibitem [{\citenamefont {Essam}(1980)}]{percolation-theor_e80}%
  \BibitemOpen
  \bibfield  {author} {\bibinfo {author} {\bibfnamefont {J.~W.}\ \bibnamefont {Essam}},\ }\bibfield  {title} {\bibinfo {title} {Percolation theory},\ }\href {https://doi.org/10.1088/0034-4885/43/7/001} {\bibfield  {journal} {\bibinfo  {journal} {Rep. Prog. Phys.}\ }\textbf {\bibinfo {volume} {43}},\ \bibinfo {pages} {833} (\bibinfo {year} {1980})}\BibitemShut {NoStop}%
\bibitem [{\citenamefont {Stauffer}\ and\ \citenamefont {Aharony}(1992)}]{stauffer92}%
  \BibitemOpen
  \bibfield  {author} {\bibinfo {author} {\bibfnamefont {D.}~\bibnamefont {Stauffer}}\ and\ \bibinfo {author} {\bibfnamefont {A.}~\bibnamefont {Aharony}},\ }\href@noop {} {\emph {\bibinfo {title} {Introduction to Percolation Theory}}},\ \bibinfo {edition} {2nd}\ ed.\ (\bibinfo  {publisher} {Taylor and Francis},\ \bibinfo {address} {London},\ \bibinfo {year} {1992})\BibitemShut {NoStop}%
\bibitem [{\citenamefont {{ben-Avraham}}\ and\ \citenamefont {Havlin}(2000)}]{diffus-react-fractal-disorder-syst}%
  \BibitemOpen
  \bibfield  {author} {\bibinfo {author} {\bibfnamefont {D.}~\bibnamefont {{ben-Avraham}}}\ and\ \bibinfo {author} {\bibfnamefont {S.}~\bibnamefont {Havlin}},\ }\href@noop {} {\emph {\bibinfo {title} {Diffusion and {{Reactions}} in {{Fractals}} and {{Disordered Systems}}}}},\ \bibinfo {edition} {1st}\ ed.\ (\bibinfo  {publisher} {{Cambridge University Press}},\ \bibinfo {address} {{Cambridge}},\ \bibinfo {year} {2000})\BibitemShut {NoStop}%
\bibitem [{\citenamefont {Bunde}\ and\ \citenamefont {Havlin}(2012)}]{bunde2012}%
  \BibitemOpen
  \bibfield  {author} {\bibinfo {author} {\bibfnamefont {A.}~\bibnamefont {Bunde}}\ and\ \bibinfo {author} {\bibfnamefont {S.}~\bibnamefont {Havlin}},\ }\href@noop {} {\emph {\bibinfo {title} {Fractals and Disordered Systems}}}\ (\bibinfo  {publisher} {Springer},\ \bibinfo {year} {2012})\BibitemShut {NoStop}%
\bibitem [{\citenamefont {Meng}\ \emph {et~al.}(2021)\citenamefont {Meng}, \citenamefont {Gao},\ and\ \citenamefont {Havlin}}]{conpt_mgh21}%
  \BibitemOpen
  \bibfield  {author} {\bibinfo {author} {\bibfnamefont {X.}~\bibnamefont {Meng}}, \bibinfo {author} {\bibfnamefont {J.}~\bibnamefont {Gao}},\ and\ \bibinfo {author} {\bibfnamefont {S.}~\bibnamefont {Havlin}},\ }\bibfield  {title} {\bibinfo {title} {Concurrence {{Percolation}} in {{Quantum Networks}}},\ }\href {https://doi.org/10.1103/PhysRevLett.126.170501} {\bibfield  {journal} {\bibinfo  {journal} {Phys. Rev. Lett.}\ }\textbf {\bibinfo {volume} {126}},\ \bibinfo {pages} {170501} (\bibinfo {year} {2021})}\BibitemShut {NoStop}%
\bibitem [{\citenamefont {Perseguers}(2010)}]{QEP_p10}%
  \BibitemOpen
  \bibfield  {author} {\bibinfo {author} {\bibfnamefont {S.}~\bibnamefont {Perseguers}},\ }\emph {\bibinfo {title} {Entanglement {{Distribution}} in {{Quantum Networks}}}},\ \href@noop {} {Ph.D. thesis},\ \bibinfo  {school} {Technische Universit\"at M\"unchen} (\bibinfo {year} {2010})\BibitemShut {NoStop}%
\bibitem [{\citenamefont {Chitambar}\ \emph {et~al.}(2014)\citenamefont {Chitambar}, \citenamefont {Leung}, \citenamefont {Man{\v c}inska}, \citenamefont {Ozols},\ and\ \citenamefont {Winter}}]{locc_clmow14}%
  \BibitemOpen
  \bibfield  {author} {\bibinfo {author} {\bibfnamefont {E.}~\bibnamefont {Chitambar}}, \bibinfo {author} {\bibfnamefont {D.}~\bibnamefont {Leung}}, \bibinfo {author} {\bibfnamefont {L.}~\bibnamefont {Man{\v c}inska}}, \bibinfo {author} {\bibfnamefont {M.}~\bibnamefont {Ozols}},\ and\ \bibinfo {author} {\bibfnamefont {A.}~\bibnamefont {Winter}},\ }\bibfield  {title} {\bibinfo {title} {Everything {{You Always Wanted}} to {{Know About LOCC}} ({{But Were Afraid}} to {{Ask}})},\ }\href {https://doi.org/10.1007/s00220-014-1953-9} {\bibfield  {journal} {\bibinfo  {journal} {Commun. Math. Phys.}\ }\textbf {\bibinfo {volume} {328}},\ \bibinfo {pages} {303} (\bibinfo {year} {2014})}\BibitemShut {NoStop}%
\bibitem [{\citenamefont {Blok}\ \emph {et~al.}(2021)\citenamefont {Blok}, \citenamefont {Ramasesh}, \citenamefont {Schuster}, \citenamefont {O'Brien}, \citenamefont {Kreikebaum}, \citenamefont {Dahlen}, \citenamefont {Morvan}, \citenamefont {Yoshida}, \citenamefont {Yao},\ and\ \citenamefont {Siddiqi}}]{qutrit-scramble_brsobkdmyys21}%
  \BibitemOpen
  \bibfield  {author} {\bibinfo {author} {\bibfnamefont {M.~S.}\ \bibnamefont {Blok}}, \bibinfo {author} {\bibfnamefont {V.~V.}\ \bibnamefont {Ramasesh}}, \bibinfo {author} {\bibfnamefont {T.}~\bibnamefont {Schuster}}, \bibinfo {author} {\bibfnamefont {K.}~\bibnamefont {O'Brien}}, \bibinfo {author} {\bibfnamefont {J.~M.}\ \bibnamefont {Kreikebaum}}, \bibinfo {author} {\bibfnamefont {D.}~\bibnamefont {Dahlen}}, \bibinfo {author} {\bibfnamefont {A.}~\bibnamefont {Morvan}}, \bibinfo {author} {\bibfnamefont {B.}~\bibnamefont {Yoshida}}, \bibinfo {author} {\bibfnamefont {N.~Y.}\ \bibnamefont {Yao}},\ and\ \bibinfo {author} {\bibfnamefont {I.}~\bibnamefont {Siddiqi}},\ }\bibfield  {title} {\bibinfo {title} {Quantum {{Information Scrambling}} on a {{Superconducting Qutrit Processor}}},\ }\href {https://doi.org/10.1103/PhysRevX.11.021010} {\bibfield  {journal} {\bibinfo  {journal} {Phys. Rev. X}\ }\textbf {\bibinfo {volume} {11}},\ \bibinfo {pages} {021010} (\bibinfo {year} {2021})}\BibitemShut {NoStop}%
\bibitem [{\citenamefont {Hu}\ \emph {et~al.}(2018)\citenamefont {Hu}, \citenamefont {Guo}, \citenamefont {Liu}, \citenamefont {Huang}, \citenamefont {Li},\ and\ \citenamefont {Guo}}]{ququart-superdense_hglhlg18}%
  \BibitemOpen
  \bibfield  {author} {\bibinfo {author} {\bibfnamefont {X.-M.}\ \bibnamefont {Hu}}, \bibinfo {author} {\bibfnamefont {Y.}~\bibnamefont {Guo}}, \bibinfo {author} {\bibfnamefont {B.-H.}\ \bibnamefont {Liu}}, \bibinfo {author} {\bibfnamefont {Y.-F.}\ \bibnamefont {Huang}}, \bibinfo {author} {\bibfnamefont {C.-F.}\ \bibnamefont {Li}},\ and\ \bibinfo {author} {\bibfnamefont {G.-C.}\ \bibnamefont {Guo}},\ }\bibfield  {title} {\bibinfo {title} {Beating the channel capacity limit for superdense coding with entangled ququarts},\ }\href {https://doi.org/10.1126/sciadv.aat9304} {\bibfield  {journal} {\bibinfo  {journal} {Sci. Adv.}\ }\textbf {\bibinfo {volume} {4}},\ \bibinfo {pages} {eaat9304} (\bibinfo {year} {2018})}\BibitemShut {NoStop}%
\bibitem [{\citenamefont {Or{\'u}s}(2014)}]{tensor-netw_o14}%
  \BibitemOpen
  \bibfield  {author} {\bibinfo {author} {\bibfnamefont {R.}~\bibnamefont {Or{\'u}s}},\ }\bibfield  {title} {\bibinfo {title} {A practical introduction to tensor networks: Matrix product states and projected entangled pair states},\ }\href {https://doi.org/10.1016/j.aop.2014.06.013} {\bibfield  {journal} {\bibinfo  {journal} {Ann. Phys.}\ }\textbf {\bibinfo {volume} {349}},\ \bibinfo {pages} {117} (\bibinfo {year} {2014})}\BibitemShut {NoStop}%
\bibitem [{\citenamefont {Meng}\ and\ \citenamefont {Yang}(2021)}]{lstm_my21}%
  \BibitemOpen
  \bibfield  {author} {\bibinfo {author} {\bibfnamefont {X.}~\bibnamefont {Meng}}\ and\ \bibinfo {author} {\bibfnamefont {T.}~\bibnamefont {Yang}},\ }\bibfield  {title} {\bibinfo {title} {Entanglement-{{Structured LSTM Boosts Chaotic Time Series Forecasting}}},\ }\href {https://doi.org/10.3390/e23111491} {\bibfield  {journal} {\bibinfo  {journal} {Entropy}\ }\textbf {\bibinfo {volume} {23}},\ \bibinfo {pages} {1491} (\bibinfo {year} {2021})}\BibitemShut {NoStop}%
\bibitem [{\citenamefont {Verstraete}\ and\ \citenamefont {Cirac}(2006)}]{mps_vc06}%
  \BibitemOpen
  \bibfield  {author} {\bibinfo {author} {\bibfnamefont {F.}~\bibnamefont {Verstraete}}\ and\ \bibinfo {author} {\bibfnamefont {J.~I.}\ \bibnamefont {Cirac}},\ }\bibfield  {title} {\bibinfo {title} {Matrix product states represent ground states faithfully},\ }\href {https://doi.org/10.1103/physrevb.73.094423} {\bibfield  {journal} {\bibinfo  {journal} {Phys. Rev. B}\ }\textbf {\bibinfo {volume} {73}},\ \bibinfo {pages} {094423} (\bibinfo {year} {2006})}\BibitemShut {NoStop}%
\bibitem [{\citenamefont {Bigoni}\ \emph {et~al.}(2016)\citenamefont {Bigoni}, \citenamefont {{Engsig-Karup}},\ and\ \citenamefont {Marzouk}}]{tensor-train-spectr_bekm16}%
  \BibitemOpen
  \bibfield  {author} {\bibinfo {author} {\bibfnamefont {D.}~\bibnamefont {Bigoni}}, \bibinfo {author} {\bibfnamefont {A.~P.}\ \bibnamefont {{Engsig-Karup}}},\ and\ \bibinfo {author} {\bibfnamefont {Y.~M.}\ \bibnamefont {Marzouk}},\ }\bibfield  {title} {\bibinfo {title} {Spectral {{Tensor-Train Decomposition}}},\ }\href {https://doi.org/10.1137/15m1036919} {\bibfield  {journal} {\bibinfo  {journal} {SIAM J. Sci. Comput.}\ }\textbf {\bibinfo {volume} {38}},\ \bibinfo {pages} {A2405} (\bibinfo {year} {2016})}\BibitemShut {NoStop}%
\bibitem [{\citenamefont {Cirac}\ \emph {et~al.}(2021)\citenamefont {Cirac}, \citenamefont {{P{\'e}rez-Garc{\'i}a}}, \citenamefont {Schuch},\ and\ \citenamefont {Verstraete}}]{mps_cpgsv21}%
  \BibitemOpen
  \bibfield  {author} {\bibinfo {author} {\bibfnamefont {J.~I.}\ \bibnamefont {Cirac}}, \bibinfo {author} {\bibfnamefont {D.}~\bibnamefont {{P{\'e}rez-Garc{\'i}a}}}, \bibinfo {author} {\bibfnamefont {N.}~\bibnamefont {Schuch}},\ and\ \bibinfo {author} {\bibfnamefont {F.}~\bibnamefont {Verstraete}},\ }\bibfield  {title} {\bibinfo {title} {Matrix product states and projected entangled pair states: {{Concepts}}, symmetries, theorems},\ }\href {https://doi.org/10.1103/RevModPhys.93.045003} {\bibfield  {journal} {\bibinfo  {journal} {Rev. Mod. Phys.}\ }\textbf {\bibinfo {volume} {93}},\ \bibinfo {pages} {045003} (\bibinfo {year} {2021})}\BibitemShut {NoStop}%
\bibitem [{\citenamefont {Vidal}(2008)}]{mera_v08}%
  \BibitemOpen
  \bibfield  {author} {\bibinfo {author} {\bibfnamefont {G.}~\bibnamefont {Vidal}},\ }\bibfield  {title} {\bibinfo {title} {Class of {{Quantum Many-Body States That Can Be Efficiently Simulated}}},\ }\href {https://doi.org/10.1103/physrevlett.101.110501} {\bibfield  {journal} {\bibinfo  {journal} {Phys. Rev. Lett.}\ }\textbf {\bibinfo {volume} {101}},\ \bibinfo {pages} {110501} (\bibinfo {year} {2008})}\BibitemShut {NoStop}%
\bibitem [{\citenamefont {Farr{\`a}s}\ \emph {et~al.}(2007)\citenamefont {Farr{\`a}s}, \citenamefont {{Mart{\'i}-Farr{\'e}}},\ and\ \citenamefont {Padr{\'o}}}]{multipartite-secret-share_fmfp07}%
  \BibitemOpen
  \bibfield  {author} {\bibinfo {author} {\bibfnamefont {O.}~\bibnamefont {Farr{\`a}s}}, \bibinfo {author} {\bibfnamefont {J.}~\bibnamefont {{Mart{\'i}-Farr{\'e}}}},\ and\ \bibinfo {author} {\bibfnamefont {C.}~\bibnamefont {Padr{\'o}}},\ }\bibfield  {title} {\bibinfo {title} {Ideal {{Multipartite Secret Sharing Schemes}}},\ }in\ \href {https://doi.org/10.1007/978-3-540-72540-4_26} {\emph {\bibinfo {booktitle} {Advances in {{Cryptology}} - {{EUROCRYPT}} 2007}}},\ \bibinfo {series and number} {Lecture {{Notes}} in {{Computer Science}}},\ \bibinfo {editor} {edited by\ \bibinfo {editor} {\bibfnamefont {M.}~\bibnamefont {Naor}}}\ (\bibinfo  {publisher} {{Springer}},\ \bibinfo {address} {{Berlin}},\ \bibinfo {year} {2007})\ pp.\ \bibinfo {pages} {448--465}\BibitemShut {NoStop}%
\bibitem [{\citenamefont {D{\"u}r}\ \emph {et~al.}(2000)\citenamefont {D{\"u}r}, \citenamefont {Vidal},\ and\ \citenamefont {Cirac}}]{tripartite_dvc00}%
  \BibitemOpen
  \bibfield  {author} {\bibinfo {author} {\bibfnamefont {W.}~\bibnamefont {D{\"u}r}}, \bibinfo {author} {\bibfnamefont {G.}~\bibnamefont {Vidal}},\ and\ \bibinfo {author} {\bibfnamefont {J.~I.}\ \bibnamefont {Cirac}},\ }\bibfield  {title} {\bibinfo {title} {Three qubits can be entangled in two inequivalent ways},\ }\href {https://doi.org/10.1103/PhysRevA.62.062314} {\bibfield  {journal} {\bibinfo  {journal} {Phys. Rev. A}\ }\textbf {\bibinfo {volume} {62}},\ \bibinfo {pages} {062314} (\bibinfo {year} {2000})}\BibitemShut {NoStop}%
\bibitem [{\citenamefont {Battiston}\ \emph {et~al.}(2021)\citenamefont {Battiston}, \citenamefont {Amico}, \citenamefont {Barrat}, \citenamefont {Bianconi}, \citenamefont {{Ferraz de Arruda}}, \citenamefont {Franceschiello}, \citenamefont {Iacopini}, \citenamefont {K{\'e}fi}, \citenamefont {Latora}, \citenamefont {Moreno}, \citenamefont {Murray}, \citenamefont {Peixoto}, \citenamefont {Vaccarino},\ and\ \citenamefont {Petri}}]{high-ord-netw_babbfafiklmmpvp21}%
  \BibitemOpen
  \bibfield  {author} {\bibinfo {author} {\bibfnamefont {F.}~\bibnamefont {Battiston}}, \bibinfo {author} {\bibfnamefont {E.}~\bibnamefont {Amico}}, \bibinfo {author} {\bibfnamefont {A.}~\bibnamefont {Barrat}}, \bibinfo {author} {\bibfnamefont {G.}~\bibnamefont {Bianconi}}, \bibinfo {author} {\bibfnamefont {G.}~\bibnamefont {{Ferraz de Arruda}}}, \bibinfo {author} {\bibfnamefont {B.}~\bibnamefont {Franceschiello}}, \bibinfo {author} {\bibfnamefont {I.}~\bibnamefont {Iacopini}}, \bibinfo {author} {\bibfnamefont {S.}~\bibnamefont {K{\'e}fi}}, \bibinfo {author} {\bibfnamefont {V.}~\bibnamefont {Latora}}, \bibinfo {author} {\bibfnamefont {Y.}~\bibnamefont {Moreno}}, \bibinfo {author} {\bibfnamefont {M.~M.}\ \bibnamefont {Murray}}, \bibinfo {author} {\bibfnamefont {T.~P.}\ \bibnamefont {Peixoto}}, \bibinfo {author} {\bibfnamefont {F.}~\bibnamefont {Vaccarino}},\ and\ \bibinfo {author} {\bibfnamefont {G.}~\bibnamefont {Petri}},\ }\bibfield  {title} {\bibinfo {title} {The physics of higher-order interactions in
  complex systems},\ }\href {https://doi.org/10.1038/s41567-021-01371-4} {\bibfield  {journal} {\bibinfo  {journal} {Nat. Phys.}\ }\textbf {\bibinfo {volume} {17}},\ \bibinfo {pages} {1093} (\bibinfo {year} {2021})}\BibitemShut {NoStop}%
\bibitem [{\citenamefont {Lambiotte}\ \emph {et~al.}(2019)\citenamefont {Lambiotte}, \citenamefont {Rosvall},\ and\ \citenamefont {Scholtes}}]{lambiotte2019networks}%
  \BibitemOpen
  \bibfield  {author} {\bibinfo {author} {\bibfnamefont {R.}~\bibnamefont {Lambiotte}}, \bibinfo {author} {\bibfnamefont {M.}~\bibnamefont {Rosvall}},\ and\ \bibinfo {author} {\bibfnamefont {I.}~\bibnamefont {Scholtes}},\ }\bibfield  {title} {\bibinfo {title} {From networks to optimal higher-order models of complex systems},\ }\href {https://doi.org/10.1038/s41567-019-0459-y} {\bibfield  {journal} {\bibinfo  {journal} {Nat. Phys.}\ }\textbf {\bibinfo {volume} {15}},\ \bibinfo {pages} {313} (\bibinfo {year} {2019})}\BibitemShut {NoStop}%
\bibitem [{\citenamefont {Bretto}(2013{\natexlab{a}})}]{hypergr-theor-introd}%
  \BibitemOpen
  \bibfield  {author} {\bibinfo {author} {\bibfnamefont {A.}~\bibnamefont {Bretto}},\ }\href@noop {} {\emph {\bibinfo {title} {Hypergraph {{Theory}}: {{An Introduction}}}}},\ \bibinfo {edition} {1st}\ ed.\ (\bibinfo  {publisher} {{Springer}},\ \bibinfo {address} {{Cham, Switzerland}},\ \bibinfo {year} {2013})\BibitemShut {NoStop}%
\bibitem [{\citenamefont {Kesten}(1982)}]{kesten1982percolation}%
  \BibitemOpen
  \bibfield  {author} {\bibinfo {author} {\bibfnamefont {H.}~\bibnamefont {Kesten}},\ }\href@noop {} {\emph {\bibinfo {title} {Percolation Theory for Mathematicians}}},\ Vol.~\bibinfo {volume} {2}\ (\bibinfo  {publisher} {{Springer}},\ \bibinfo {year} {1982})\BibitemShut {NoStop}%
\bibitem [{\citenamefont {Turcotte}(1997)}]{turcotte1997fractals}%
  \BibitemOpen
  \bibfield  {author} {\bibinfo {author} {\bibfnamefont {D.~L.}\ \bibnamefont {Turcotte}},\ }\href@noop {} {\emph {\bibinfo {title} {Fractals and Chaos in Geology and Geophysics}}}\ (\bibinfo  {publisher} {{Cambridge University Press}},\ \bibinfo {year} {1997})\BibitemShut {NoStop}%
\bibitem [{\citenamefont {Stanley}\ and\ \citenamefont {Ahlers}(1973)}]{stanley1973introduction}%
  \BibitemOpen
  \bibfield  {author} {\bibinfo {author} {\bibfnamefont {H.~E.}\ \bibnamefont {Stanley}}\ and\ \bibinfo {author} {\bibfnamefont {G.}~\bibnamefont {Ahlers}},\ }\href@noop {} {\emph {\bibinfo {title} {Introduction to Phase Transitions and Critical Phenomena}}}\ (\bibinfo  {publisher} {Oxford University Press},\ \bibinfo {year} {1973})\BibitemShut {NoStop}%
\bibitem [{\citenamefont {Grassberger}(1986)}]{grassberger1986spreading}%
  \BibitemOpen
  \bibfield  {author} {\bibinfo {author} {\bibfnamefont {P.}~\bibnamefont {Grassberger}},\ }\bibfield  {title} {\bibinfo {title} {Spreading of epidemic processes leading to fractal structures},\ }in\ \href@noop {} {\emph {\bibinfo {booktitle} {Fractals in Physics}}}\ (\bibinfo  {publisher} {{Elsevier}},\ \bibinfo {year} {1986})\ pp.\ \bibinfo {pages} {273--278}\BibitemShut {NoStop}%
\bibitem [{\citenamefont {Newman}(2002)}]{newman2002spread}%
  \BibitemOpen
  \bibfield  {author} {\bibinfo {author} {\bibfnamefont {M.~E.}\ \bibnamefont {Newman}},\ }\bibfield  {title} {\bibinfo {title} {Spread of epidemic disease on networks},\ }\href {https://doi.org/10.1103/PhysRevE.66.016128} {\bibfield  {journal} {\bibinfo  {journal} {Phys. Rev. E}\ }\textbf {\bibinfo {volume} {66}},\ \bibinfo {pages} {016128} (\bibinfo {year} {2002})}\BibitemShut {NoStop}%
\bibitem [{\citenamefont {{Pastor-Satorras}}\ and\ \citenamefont {Vespignani}(2002)}]{pastor2002immunization}%
  \BibitemOpen
  \bibfield  {author} {\bibinfo {author} {\bibfnamefont {R.}~\bibnamefont {{Pastor-Satorras}}}\ and\ \bibinfo {author} {\bibfnamefont {A.}~\bibnamefont {Vespignani}},\ }\bibfield  {title} {\bibinfo {title} {Immunization of complex networks},\ }\href {https://doi.org/10.1103/PhysRevE.65.036104} {\bibfield  {journal} {\bibinfo  {journal} {Phys. Rev. E}\ }\textbf {\bibinfo {volume} {65}},\ \bibinfo {pages} {036104} (\bibinfo {year} {2002})}\BibitemShut {NoStop}%
\bibitem [{\citenamefont {Parshani}\ \emph {et~al.}(2010{\natexlab{a}})\citenamefont {Parshani}, \citenamefont {Carmi},\ and\ \citenamefont {Havlin}}]{parshani2010epidemic}%
  \BibitemOpen
  \bibfield  {author} {\bibinfo {author} {\bibfnamefont {R.}~\bibnamefont {Parshani}}, \bibinfo {author} {\bibfnamefont {S.}~\bibnamefont {Carmi}},\ and\ \bibinfo {author} {\bibfnamefont {S.}~\bibnamefont {Havlin}},\ }\bibfield  {title} {\bibinfo {title} {Epidemic threshold for the susceptible-infectious-susceptible model on random networks},\ }\href {https://doi.org/10.1103/PhysRevLett.104.258701} {\bibfield  {journal} {\bibinfo  {journal} {Phys. Rev. Lett.}\ }\textbf {\bibinfo {volume} {104}},\ \bibinfo {pages} {258701} (\bibinfo {year} {2010}{\natexlab{a}})}\BibitemShut {NoStop}%
\bibitem [{\citenamefont {Lindquist}\ \emph {et~al.}(2011)\citenamefont {Lindquist}, \citenamefont {Ma}, \citenamefont {{Van den Driessche}},\ and\ \citenamefont {Willeboordse}}]{lindquist2011effective}%
  \BibitemOpen
  \bibfield  {author} {\bibinfo {author} {\bibfnamefont {J.}~\bibnamefont {Lindquist}}, \bibinfo {author} {\bibfnamefont {J.}~\bibnamefont {Ma}}, \bibinfo {author} {\bibfnamefont {P.}~\bibnamefont {{Van den Driessche}}},\ and\ \bibinfo {author} {\bibfnamefont {F.~H.}\ \bibnamefont {Willeboordse}},\ }\bibfield  {title} {\bibinfo {title} {Effective degree network disease models},\ }\href {https://doi.org/10.1007/s00285-010-0331-2} {\bibfield  {journal} {\bibinfo  {journal} {J. Math. Biol.}\ }\textbf {\bibinfo {volume} {62}},\ \bibinfo {pages} {143} (\bibinfo {year} {2011})}\BibitemShut {NoStop}%
\bibitem [{\citenamefont {Vespignani}(2012)}]{vespignani2012modelling}%
  \BibitemOpen
  \bibfield  {author} {\bibinfo {author} {\bibfnamefont {A.}~\bibnamefont {Vespignani}},\ }\bibfield  {title} {\bibinfo {title} {Modelling dynamical processes in complex socio-technical systems},\ }\href {https://doi.org/10.1038/nphys2160} {\bibfield  {journal} {\bibinfo  {journal} {Nat. Phys.}\ }\textbf {\bibinfo {volume} {8}},\ \bibinfo {pages} {32} (\bibinfo {year} {2012})}\BibitemShut {NoStop}%
\bibitem [{\citenamefont {{Pastor-Satorras}}\ \emph {et~al.}(2015)\citenamefont {{Pastor-Satorras}}, \citenamefont {Castellano}, \citenamefont {Van~Mieghem},\ and\ \citenamefont {Vespignani}}]{pastor2015epidemic}%
  \BibitemOpen
  \bibfield  {author} {\bibinfo {author} {\bibfnamefont {R.}~\bibnamefont {{Pastor-Satorras}}}, \bibinfo {author} {\bibfnamefont {C.}~\bibnamefont {Castellano}}, \bibinfo {author} {\bibfnamefont {P.}~\bibnamefont {Van~Mieghem}},\ and\ \bibinfo {author} {\bibfnamefont {A.}~\bibnamefont {Vespignani}},\ }\bibfield  {title} {\bibinfo {title} {Epidemic processes in complex networks},\ }\href {https://doi.org/10.1103/RevModPhys.87.925} {\bibfield  {journal} {\bibinfo  {journal} {Rev. Mod. Phys.}\ }\textbf {\bibinfo {volume} {87}},\ \bibinfo {pages} {925} (\bibinfo {year} {2015})}\BibitemShut {NoStop}%
\bibitem [{\citenamefont {Croccolo}\ and\ \citenamefont {Roman}(2020)}]{croccolo2020spreading}%
  \BibitemOpen
  \bibfield  {author} {\bibinfo {author} {\bibfnamefont {F.}~\bibnamefont {Croccolo}}\ and\ \bibinfo {author} {\bibfnamefont {H.~E.}\ \bibnamefont {Roman}},\ }\bibfield  {title} {\bibinfo {title} {Spreading of infections on random graphs: {{A}} percolation-type model for {{COVID-19}}},\ }\href {https://doi.org/10.1016/j.chaos.2020.110077} {\bibfield  {journal} {\bibinfo  {journal} {Chaos Solitons Fractals}\ }\textbf {\bibinfo {volume} {139}},\ \bibinfo {pages} {110077} (\bibinfo {year} {2020})}\BibitemShut {NoStop}%
\bibitem [{\citenamefont {MacKay}\ and\ \citenamefont {Jan}(1984)}]{mackay1984forest}%
  \BibitemOpen
  \bibfield  {author} {\bibinfo {author} {\bibfnamefont {G.}~\bibnamefont {MacKay}}\ and\ \bibinfo {author} {\bibfnamefont {N.}~\bibnamefont {Jan}},\ }\bibfield  {title} {\bibinfo {title} {Forest fires as critical phenomena},\ }\href {https://doi.org/10.1088/0305-4470/17/14/006} {\bibfield  {journal} {\bibinfo  {journal} {J. Phys. Math. Gen.}\ }\textbf {\bibinfo {volume} {17}},\ \bibinfo {pages} {L757} (\bibinfo {year} {1984})}\BibitemShut {NoStop}%
\bibitem [{\citenamefont {Ritzenberg}\ and\ \citenamefont {Cohen}(1984)}]{ritzenberg1984first}%
  \BibitemOpen
  \bibfield  {author} {\bibinfo {author} {\bibfnamefont {A.~L.}\ \bibnamefont {Ritzenberg}}\ and\ \bibinfo {author} {\bibfnamefont {R.~J.}\ \bibnamefont {Cohen}},\ }\bibfield  {title} {\bibinfo {title} {First passage percolation: {{Scaling}} and critical exponents},\ }\href {https://doi.org/10.1103/PhysRevB.30.4038} {\bibfield  {journal} {\bibinfo  {journal} {Phys. Rev. B}\ }\textbf {\bibinfo {volume} {30}},\ \bibinfo {pages} {4038} (\bibinfo {year} {1984})}\BibitemShut {NoStop}%
\bibitem [{\citenamefont {De~Gennes}(1979)}]{de1979scaling}%
  \BibitemOpen
  \bibfield  {author} {\bibinfo {author} {\bibfnamefont {P.-G.}\ \bibnamefont {De~Gennes}},\ }\href@noop {} {\emph {\bibinfo {title} {Scaling Concepts in Polymer Physics}}}\ (\bibinfo  {publisher} {{Cornell University Press}},\ \bibinfo {year} {1979})\BibitemShut {NoStop}%
\bibitem [{\citenamefont {Herrmann}(1986)}]{herrmann1986geometrical}%
  \BibitemOpen
  \bibfield  {author} {\bibinfo {author} {\bibfnamefont {H.~J.}\ \bibnamefont {Herrmann}},\ }\bibfield  {title} {\bibinfo {title} {Geometrical cluster growth models and kinetic gelation},\ }\href {https://doi.org/10.1016/0370-1573(86)90047-5} {\bibfield  {journal} {\bibinfo  {journal} {Phys. Rep.}\ }\textbf {\bibinfo {volume} {136}},\ \bibinfo {pages} {153} (\bibinfo {year} {1986})}\BibitemShut {NoStop}%
\bibitem [{\citenamefont {Shapir}\ \emph {et~al.}(1982)\citenamefont {Shapir}, \citenamefont {Aharony},\ and\ \citenamefont {Harris}}]{wave-localis_sah82}%
  \BibitemOpen
  \bibfield  {author} {\bibinfo {author} {\bibfnamefont {Y.}~\bibnamefont {Shapir}}, \bibinfo {author} {\bibfnamefont {A.}~\bibnamefont {Aharony}},\ and\ \bibinfo {author} {\bibfnamefont {A.~B.}\ \bibnamefont {Harris}},\ }\bibfield  {title} {\bibinfo {title} {Localization and {{Quantum Percolation}}},\ }\href {https://doi.org/10.1103/PhysRevLett.49.486} {\bibfield  {journal} {\bibinfo  {journal} {Phys. Rev. Lett.}\ }\textbf {\bibinfo {volume} {49}},\ \bibinfo {pages} {486} (\bibinfo {year} {1982})}\BibitemShut {NoStop}%
\bibitem [{\citenamefont {Soukoulis}\ \emph {et~al.}(1992)\citenamefont {Soukoulis}, \citenamefont {Li},\ and\ \citenamefont {Grest}}]{wave-localis_slg92}%
  \BibitemOpen
  \bibfield  {author} {\bibinfo {author} {\bibfnamefont {C.~M.}\ \bibnamefont {Soukoulis}}, \bibinfo {author} {\bibfnamefont {Q.}~\bibnamefont {Li}},\ and\ \bibinfo {author} {\bibfnamefont {G.~S.}\ \bibnamefont {Grest}},\ }\bibfield  {title} {\bibinfo {title} {Quantum percolation in three-dimensional systems},\ }\href {https://doi.org/10.1103/PhysRevB.45.7724} {\bibfield  {journal} {\bibinfo  {journal} {Phys. Rev. B}\ }\textbf {\bibinfo {volume} {45}},\ \bibinfo {pages} {7724} (\bibinfo {year} {1992})}\BibitemShut {NoStop}%
\bibitem [{\citenamefont {Motrunich}\ \emph {et~al.}(2000)\citenamefont {Motrunich}, \citenamefont {Mau}, \citenamefont {Huse},\ and\ \citenamefont {Fisher}}]{strong-disord_mmhf00}%
  \BibitemOpen
  \bibfield  {author} {\bibinfo {author} {\bibfnamefont {O.}~\bibnamefont {Motrunich}}, \bibinfo {author} {\bibfnamefont {S.-C.}\ \bibnamefont {Mau}}, \bibinfo {author} {\bibfnamefont {D.~A.}\ \bibnamefont {Huse}},\ and\ \bibinfo {author} {\bibfnamefont {D.~S.}\ \bibnamefont {Fisher}},\ }\bibfield  {title} {\bibinfo {title} {Infinite-randomness quantum {{Ising}} critical fixed points},\ }\href {https://doi.org/10.1103/PhysRevB.61.1160} {\bibfield  {journal} {\bibinfo  {journal} {Phys. Rev. B}\ }\textbf {\bibinfo {volume} {61}},\ \bibinfo {pages} {1160} (\bibinfo {year} {2000})}\BibitemShut {NoStop}%
\bibitem [{\citenamefont {Chandrashekar}\ and\ \citenamefont {Busch}(2014)}]{wave-localis_cb14}%
  \BibitemOpen
  \bibfield  {author} {\bibinfo {author} {\bibfnamefont {C.~M.}\ \bibnamefont {Chandrashekar}}\ and\ \bibinfo {author} {\bibfnamefont {T.}~\bibnamefont {Busch}},\ }\bibfield  {title} {\bibinfo {title} {Quantum percolation and transition point of a directed discrete-time quantum walk},\ }\href {https://doi.org/10.1038/srep06583} {\bibfield  {journal} {\bibinfo  {journal} {Sci. Rep.}\ }\textbf {\bibinfo {volume} {4}},\ \bibinfo {pages} {6583} (\bibinfo {year} {2014})}\BibitemShut {NoStop}%
\bibitem [{\citenamefont {Fostner}\ \emph {et~al.}(2014)\citenamefont {Fostner}, \citenamefont {Brown}, \citenamefont {Carr},\ and\ \citenamefont {Brown}}]{contin-percolation_fbcb14}%
  \BibitemOpen
  \bibfield  {author} {\bibinfo {author} {\bibfnamefont {S.}~\bibnamefont {Fostner}}, \bibinfo {author} {\bibfnamefont {R.}~\bibnamefont {Brown}}, \bibinfo {author} {\bibfnamefont {J.}~\bibnamefont {Carr}},\ and\ \bibinfo {author} {\bibfnamefont {S.~A.}\ \bibnamefont {Brown}},\ }\bibfield  {title} {\bibinfo {title} {Continuum percolation with tunneling},\ }\href {https://doi.org/10.1103/PhysRevB.89.075402} {\bibfield  {journal} {\bibinfo  {journal} {Phys. Rev. B}\ }\textbf {\bibinfo {volume} {89}},\ \bibinfo {pages} {075402} (\bibinfo {year} {2014})}\BibitemShut {NoStop}%
\bibitem [{\citenamefont {Skinner}\ \emph {et~al.}(2019)\citenamefont {Skinner}, \citenamefont {Ruhman},\ and\ \citenamefont {Nahum}}]{meas-induc_srn19}%
  \BibitemOpen
  \bibfield  {author} {\bibinfo {author} {\bibfnamefont {B.}~\bibnamefont {Skinner}}, \bibinfo {author} {\bibfnamefont {J.}~\bibnamefont {Ruhman}},\ and\ \bibinfo {author} {\bibfnamefont {A.}~\bibnamefont {Nahum}},\ }\bibfield  {title} {\bibinfo {title} {Measurement-{{Induced Phase Transitions}} in the {{Dynamics}} of {{Entanglement}}},\ }\href {https://doi.org/10.1103/PhysRevX.9.031009} {\bibfield  {journal} {\bibinfo  {journal} {Phys. Rev. X}\ }\textbf {\bibinfo {volume} {9}},\ \bibinfo {pages} {031009} (\bibinfo {year} {2019})}\BibitemShut {NoStop}%
\bibitem [{\citenamefont {Pant}\ \emph {et~al.}(2019{\natexlab{a}})\citenamefont {Pant}, \citenamefont {Towsley}, \citenamefont {Englund},\ and\ \citenamefont {Guha}}]{percol_pteg19}%
  \BibitemOpen
  \bibfield  {author} {\bibinfo {author} {\bibfnamefont {M.}~\bibnamefont {Pant}}, \bibinfo {author} {\bibfnamefont {D.}~\bibnamefont {Towsley}}, \bibinfo {author} {\bibfnamefont {D.}~\bibnamefont {Englund}},\ and\ \bibinfo {author} {\bibfnamefont {S.}~\bibnamefont {Guha}},\ }\bibfield  {title} {\bibinfo {title} {Percolation thresholds for photonic quantum computing},\ }\href {https://doi.org/10.1038/s41467-019-08948-x} {\bibfield  {journal} {\bibinfo  {journal} {Nat. Commun.}\ }\textbf {\bibinfo {volume} {10}},\ \bibinfo {pages} {1070} (\bibinfo {year} {2019}{\natexlab{a}})}\BibitemShut {NoStop}%
\bibitem [{\citenamefont {Gardner}\ \emph {et~al.}(1992)\citenamefont {Gardner}, \citenamefont {Turner}, \citenamefont {Dale},\ and\ \citenamefont {O'Neill}}]{gardner1992percolation}%
  \BibitemOpen
  \bibfield  {author} {\bibinfo {author} {\bibfnamefont {R.~H.}\ \bibnamefont {Gardner}}, \bibinfo {author} {\bibfnamefont {M.~G.}\ \bibnamefont {Turner}}, \bibinfo {author} {\bibfnamefont {V.~H.}\ \bibnamefont {Dale}},\ and\ \bibinfo {author} {\bibfnamefont {R.~V.}\ \bibnamefont {O'Neill}},\ }\bibfield  {title} {\bibinfo {title} {A percolation model of ecological flows},\ }in\ \href@noop {} {\emph {\bibinfo {booktitle} {Landscape Boundaries: Consequences for Biotic Diversity and Ecological Flows}}}\ (\bibinfo  {publisher} {{Springer}},\ \bibinfo {year} {1992})\ pp.\ \bibinfo {pages} {259--269}\BibitemShut {NoStop}%
\bibitem [{\citenamefont {Vigolo}\ \emph {et~al.}(2005)\citenamefont {Vigolo}, \citenamefont {Coulon}, \citenamefont {Maugey}, \citenamefont {Zakri},\ and\ \citenamefont {Poulin}}]{vigolo2005experimental}%
  \BibitemOpen
  \bibfield  {author} {\bibinfo {author} {\bibfnamefont {B.}~\bibnamefont {Vigolo}}, \bibinfo {author} {\bibfnamefont {C.}~\bibnamefont {Coulon}}, \bibinfo {author} {\bibfnamefont {M.}~\bibnamefont {Maugey}}, \bibinfo {author} {\bibfnamefont {C.}~\bibnamefont {Zakri}},\ and\ \bibinfo {author} {\bibfnamefont {P.}~\bibnamefont {Poulin}},\ }\bibfield  {title} {\bibinfo {title} {An experimental approach to the percolation of sticky nanotubes},\ }\href {https://doi.org/10.1126/science.1112835} {\bibfield  {journal} {\bibinfo  {journal} {Science}\ }\textbf {\bibinfo {volume} {309}},\ \bibinfo {pages} {920} (\bibinfo {year} {2005})}\BibitemShut {NoStop}%
\bibitem [{\citenamefont {Anekal}\ \emph {et~al.}(2006)\citenamefont {Anekal}, \citenamefont {Bahukudumbi},\ and\ \citenamefont {Bevan}}]{anekal2006dynamic}%
  \BibitemOpen
  \bibfield  {author} {\bibinfo {author} {\bibfnamefont {S.~G.}\ \bibnamefont {Anekal}}, \bibinfo {author} {\bibfnamefont {P.}~\bibnamefont {Bahukudumbi}},\ and\ \bibinfo {author} {\bibfnamefont {M.~A.}\ \bibnamefont {Bevan}},\ }\bibfield  {title} {\bibinfo {title} {Dynamic signature for the equilibrium percolation threshold of attractive colloidal fluids},\ }\href {https://doi.org/10.1103/PhysRevE.73.020403} {\bibfield  {journal} {\bibinfo  {journal} {Phys. Rev. E}\ }\textbf {\bibinfo {volume} {73}},\ \bibinfo {pages} {020403} (\bibinfo {year} {2006})}\BibitemShut {NoStop}%
\bibitem [{\citenamefont {Berkowitz}\ and\ \citenamefont {Balberg}(1993)}]{berkowitz1993percolation}%
  \BibitemOpen
  \bibfield  {author} {\bibinfo {author} {\bibfnamefont {B.}~\bibnamefont {Berkowitz}}\ and\ \bibinfo {author} {\bibfnamefont {I.}~\bibnamefont {Balberg}},\ }\bibfield  {title} {\bibinfo {title} {Percolation theory and its application to groundwater hydrology},\ }\href {https://doi.org/10.1029/92WR02707} {\bibfield  {journal} {\bibinfo  {journal} {Water Resour. Res.}\ }\textbf {\bibinfo {volume} {29}},\ \bibinfo {pages} {775} (\bibinfo {year} {1993})}\BibitemShut {NoStop}%
\bibitem [{\citenamefont {Berkowitz}\ and\ \citenamefont {Ewing}(1998)}]{berkowitz1998percolation}%
  \BibitemOpen
  \bibfield  {author} {\bibinfo {author} {\bibfnamefont {B.}~\bibnamefont {Berkowitz}}\ and\ \bibinfo {author} {\bibfnamefont {R.~P.}\ \bibnamefont {Ewing}},\ }\bibfield  {title} {\bibinfo {title} {Percolation theory and network modeling applications in soil physics},\ }\href {https://doi.org/10.1023/A:1006590500229} {\bibfield  {journal} {\bibinfo  {journal} {Surv. Geophys.}\ }\textbf {\bibinfo {volume} {19}},\ \bibinfo {pages} {23} (\bibinfo {year} {1998})}\BibitemShut {NoStop}%
\bibitem [{\citenamefont {Gaonac'h}\ \emph {et~al.}(2003)\citenamefont {Gaonac'h}, \citenamefont {Lovejoy},\ and\ \citenamefont {Schertzer}}]{gaonac2003percolating}%
  \BibitemOpen
  \bibfield  {author} {\bibinfo {author} {\bibfnamefont {H.}~\bibnamefont {Gaonac'h}}, \bibinfo {author} {\bibfnamefont {S.}~\bibnamefont {Lovejoy}},\ and\ \bibinfo {author} {\bibfnamefont {D.}~\bibnamefont {Schertzer}},\ }\bibfield  {title} {\bibinfo {title} {Percolating magmas and explosive volcanism},\ }\href {https://doi.org/10.1029/2002GL016022} {\bibfield  {journal} {\bibinfo  {journal} {Geophys. Res. Lett.}\ }\textbf {\bibinfo {volume} {30}},\ \bibinfo {pages} {1559} (\bibinfo {year} {2003})}\BibitemShut {NoStop}%
\bibitem [{\citenamefont {King}\ and\ \citenamefont {Masihi}(2019)}]{king2019percolation}%
  \BibitemOpen
  \bibfield  {author} {\bibinfo {author} {\bibfnamefont {P.~R.}\ \bibnamefont {King}}\ and\ \bibinfo {author} {\bibfnamefont {M.}~\bibnamefont {Masihi}},\ }\href@noop {} {\emph {\bibinfo {title} {Percolation Theory in Reservoir Engineering}}}\ (\bibinfo  {publisher} {{World Scientific}},\ \bibinfo {year} {2019})\BibitemShut {NoStop}%
\bibitem [{\citenamefont {Ghanbarian}\ and\ \citenamefont {Skaggs}(2022)}]{ghanbarian2022soil}%
  \BibitemOpen
  \bibfield  {author} {\bibinfo {author} {\bibfnamefont {B.}~\bibnamefont {Ghanbarian}}\ and\ \bibinfo {author} {\bibfnamefont {T.~H.}\ \bibnamefont {Skaggs}},\ }\bibfield  {title} {\bibinfo {title} {Soil water retention curve inflection point: {{Insight}} into soil structure from percolation theory},\ }\href {https://doi.org/10.1002/saj2.20360} {\bibfield  {journal} {\bibinfo  {journal} {Soil Sci. Soc. Am. J.}\ }\textbf {\bibinfo {volume} {86}},\ \bibinfo {pages} {338} (\bibinfo {year} {2022})}\BibitemShut {NoStop}%
\bibitem [{\citenamefont {Kacperski}\ and\ \citenamefont {Ho{\l}yst}(2000)}]{kacperski2000phase}%
  \BibitemOpen
  \bibfield  {author} {\bibinfo {author} {\bibfnamefont {K.}~\bibnamefont {Kacperski}}\ and\ \bibinfo {author} {\bibfnamefont {J.~A.}\ \bibnamefont {Ho{\l}yst}},\ }\bibfield  {title} {\bibinfo {title} {Phase transitions as a persistent feature of groups with leaders in models of opinion formation},\ }\href {https://doi.org/10.1016/S0378-4371(00)00398-8} {\bibfield  {journal} {\bibinfo  {journal} {Phys. Stat. Mech. Its Appl.}\ }\textbf {\bibinfo {volume} {287}},\ \bibinfo {pages} {631} (\bibinfo {year} {2000})}\BibitemShut {NoStop}%
\bibitem [{\citenamefont {Watts}(2002)}]{watts2002simple}%
  \BibitemOpen
  \bibfield  {author} {\bibinfo {author} {\bibfnamefont {D.~J.}\ \bibnamefont {Watts}},\ }\bibfield  {title} {\bibinfo {title} {A simple model of global cascades on random networks},\ }\href {https://doi.org/10.1073/pnas.082090499} {\bibfield  {journal} {\bibinfo  {journal} {Proc. Natl. Acad. Sci. U.S.A.}\ }\textbf {\bibinfo {volume} {99}},\ \bibinfo {pages} {5766} (\bibinfo {year} {2002})}\BibitemShut {NoStop}%
\bibitem [{\citenamefont {Centola}\ \emph {et~al.}(2007)\citenamefont {Centola}, \citenamefont {Egu{\'i}luz},\ and\ \citenamefont {Macy}}]{centola2007cascade}%
  \BibitemOpen
  \bibfield  {author} {\bibinfo {author} {\bibfnamefont {D.}~\bibnamefont {Centola}}, \bibinfo {author} {\bibfnamefont {V.~M.}\ \bibnamefont {Egu{\'i}luz}},\ and\ \bibinfo {author} {\bibfnamefont {M.~W.}\ \bibnamefont {Macy}},\ }\bibfield  {title} {\bibinfo {title} {Cascade dynamics of complex propagation},\ }\href {https://doi.org/10.1016/j.physa.2006.06.018} {\bibfield  {journal} {\bibinfo  {journal} {Phys. Stat. Mech. Its Appl.}\ }\textbf {\bibinfo {volume} {374}},\ \bibinfo {pages} {449} (\bibinfo {year} {2007})}\BibitemShut {NoStop}%
\bibitem [{\citenamefont {Iniguez}\ \emph {et~al.}(2009)\citenamefont {Iniguez}, \citenamefont {Kert{\'e}sz}, \citenamefont {Kaski},\ and\ \citenamefont {Barrio}}]{iniguez2009opinion}%
  \BibitemOpen
  \bibfield  {author} {\bibinfo {author} {\bibfnamefont {G.}~\bibnamefont {Iniguez}}, \bibinfo {author} {\bibfnamefont {J.}~\bibnamefont {Kert{\'e}sz}}, \bibinfo {author} {\bibfnamefont {K.~K.}\ \bibnamefont {Kaski}},\ and\ \bibinfo {author} {\bibfnamefont {R.~A.}\ \bibnamefont {Barrio}},\ }\bibfield  {title} {\bibinfo {title} {Opinion and community formation in coevolving networks},\ }\href {https://doi.org/10.1103/PhysRevE.80.066119} {\bibfield  {journal} {\bibinfo  {journal} {Phys. Rev. E}\ }\textbf {\bibinfo {volume} {80}},\ \bibinfo {pages} {066119} (\bibinfo {year} {2009})}\BibitemShut {NoStop}%
\bibitem [{\citenamefont {Echenique}\ \emph {et~al.}(2005)\citenamefont {Echenique}, \citenamefont {{G{\'o}mez-Gardenes}},\ and\ \citenamefont {Moreno}}]{echenique2005dynamics}%
  \BibitemOpen
  \bibfield  {author} {\bibinfo {author} {\bibfnamefont {P.}~\bibnamefont {Echenique}}, \bibinfo {author} {\bibfnamefont {J.}~\bibnamefont {{G{\'o}mez-Gardenes}}},\ and\ \bibinfo {author} {\bibfnamefont {Y.}~\bibnamefont {Moreno}},\ }\bibfield  {title} {\bibinfo {title} {Dynamics of jamming transitions in complex networks},\ }\href {https://doi.org/10.1209/epl/i2005-10080-8} {\bibfield  {journal} {\bibinfo  {journal} {Europhys. Lett.}\ }\textbf {\bibinfo {volume} {71}},\ \bibinfo {pages} {325} (\bibinfo {year} {2005})}\BibitemShut {NoStop}%
\bibitem [{\citenamefont {Vespignani}(2010)}]{vespignani2010fragility}%
  \BibitemOpen
  \bibfield  {author} {\bibinfo {author} {\bibfnamefont {A.}~\bibnamefont {Vespignani}},\ }\bibfield  {title} {\bibinfo {title} {The fragility of interdependency},\ }\href {https://doi.org/10.1038/464984a} {\bibfield  {journal} {\bibinfo  {journal} {Nature}\ }\textbf {\bibinfo {volume} {464}},\ \bibinfo {pages} {984} (\bibinfo {year} {2010})}\BibitemShut {NoStop}%
\bibitem [{\citenamefont {Li}\ \emph {et~al.}(2015)\citenamefont {Li}, \citenamefont {Fu}, \citenamefont {Wang}, \citenamefont {Lu}, \citenamefont {Berezin}, \citenamefont {Stanley},\ and\ \citenamefont {Havlin}}]{li2015percolation}%
  \BibitemOpen
  \bibfield  {author} {\bibinfo {author} {\bibfnamefont {D.}~\bibnamefont {Li}}, \bibinfo {author} {\bibfnamefont {B.}~\bibnamefont {Fu}}, \bibinfo {author} {\bibfnamefont {Y.}~\bibnamefont {Wang}}, \bibinfo {author} {\bibfnamefont {G.}~\bibnamefont {Lu}}, \bibinfo {author} {\bibfnamefont {Y.}~\bibnamefont {Berezin}}, \bibinfo {author} {\bibfnamefont {H.~E.}\ \bibnamefont {Stanley}},\ and\ \bibinfo {author} {\bibfnamefont {S.}~\bibnamefont {Havlin}},\ }\bibfield  {title} {\bibinfo {title} {Percolation transition in dynamical traffic network with evolving critical bottlenecks},\ }\href {https://doi.org/10.1073/pnas.1419185112} {\bibfield  {journal} {\bibinfo  {journal} {Proc. Natl. Acad. Sci. U.S.A.}\ }\textbf {\bibinfo {volume} {112}},\ \bibinfo {pages} {669} (\bibinfo {year} {2015})}\BibitemShut {NoStop}%
\bibitem [{\citenamefont {Del~Vicario}\ \emph {et~al.}(2016)\citenamefont {Del~Vicario}, \citenamefont {Bessi}, \citenamefont {Zollo}, \citenamefont {Petroni}, \citenamefont {Scala}, \citenamefont {Caldarelli}, \citenamefont {Stanley},\ and\ \citenamefont {Quattrociocchi}}]{del2016spreading}%
  \BibitemOpen
  \bibfield  {author} {\bibinfo {author} {\bibfnamefont {M.}~\bibnamefont {Del~Vicario}}, \bibinfo {author} {\bibfnamefont {A.}~\bibnamefont {Bessi}}, \bibinfo {author} {\bibfnamefont {F.}~\bibnamefont {Zollo}}, \bibinfo {author} {\bibfnamefont {F.}~\bibnamefont {Petroni}}, \bibinfo {author} {\bibfnamefont {A.}~\bibnamefont {Scala}}, \bibinfo {author} {\bibfnamefont {G.}~\bibnamefont {Caldarelli}}, \bibinfo {author} {\bibfnamefont {H.~E.}\ \bibnamefont {Stanley}},\ and\ \bibinfo {author} {\bibfnamefont {W.}~\bibnamefont {Quattrociocchi}},\ }\bibfield  {title} {\bibinfo {title} {The spreading of misinformation online},\ }\href {https://doi.org/10.1073/pnas.1517441113} {\bibfield  {journal} {\bibinfo  {journal} {Proc. Natl. Acad. Sci. U.S.A.}\ }\textbf {\bibinfo {volume} {113}},\ \bibinfo {pages} {554} (\bibinfo {year} {2016})}\BibitemShut {NoStop}%
\bibitem [{\citenamefont {Baggag}\ \emph {et~al.}(2018)\citenamefont {Baggag}, \citenamefont {Abbar}, \citenamefont {Zanouda},\ and\ \citenamefont {Srivastava}}]{baggag2018resilience}%
  \BibitemOpen
  \bibfield  {author} {\bibinfo {author} {\bibfnamefont {A.}~\bibnamefont {Baggag}}, \bibinfo {author} {\bibfnamefont {S.}~\bibnamefont {Abbar}}, \bibinfo {author} {\bibfnamefont {T.}~\bibnamefont {Zanouda}},\ and\ \bibinfo {author} {\bibfnamefont {J.}~\bibnamefont {Srivastava}},\ }\bibfield  {title} {\bibinfo {title} {Resilience analytics: Coverage and robustness in multi-modal transportation networks},\ }\href {https://doi.org/10.1140/epjds/s13688-018-0139-7} {\bibfield  {journal} {\bibinfo  {journal} {EPJ Data Sci.}\ }\textbf {\bibinfo {volume} {7}},\ \bibinfo {pages} {1} (\bibinfo {year} {2018})}\BibitemShut {NoStop}%
\bibitem [{\citenamefont {Barab{\'a}si}\ and\ \citenamefont {Albert}(1999)}]{Barabasi1999}%
  \BibitemOpen
  \bibfield  {author} {\bibinfo {author} {\bibfnamefont {A.-L.}\ \bibnamefont {Barab{\'a}si}}\ and\ \bibinfo {author} {\bibfnamefont {R.}~\bibnamefont {Albert}},\ }\bibfield  {title} {\bibinfo {title} {Emergence of scaling in random networks},\ }\href {https://doi.org/10.1126/science.286.5439.509} {\bibfield  {journal} {\bibinfo  {journal} {Science}\ }\textbf {\bibinfo {volume} {286}},\ \bibinfo {pages} {509} (\bibinfo {year} {1999})}\BibitemShut {NoStop}%
\bibitem [{\citenamefont {Zhou}\ \emph {et~al.}(2020)\citenamefont {Zhou}, \citenamefont {Meng},\ and\ \citenamefont {Stanley}}]{degree-degree-distance_zms20}%
  \BibitemOpen
  \bibfield  {author} {\bibinfo {author} {\bibfnamefont {B.}~\bibnamefont {Zhou}}, \bibinfo {author} {\bibfnamefont {X.}~\bibnamefont {Meng}},\ and\ \bibinfo {author} {\bibfnamefont {H.~E.}\ \bibnamefont {Stanley}},\ }\bibfield  {title} {\bibinfo {title} {Power-law distribution of degree\textendash degree distance: {{A}} better representation of the scale-free property of complex networks},\ }\href {https://doi.org/10.1073/pnas.1918901117} {\bibfield  {journal} {\bibinfo  {journal} {Proc. Natl. Acad. Sci. U.S.A.}\ }\textbf {\bibinfo {volume} {117}},\ \bibinfo {pages} {14812} (\bibinfo {year} {2020})}\BibitemShut {NoStop}%
\bibitem [{\citenamefont {Meng}\ and\ \citenamefont {Zhou}(2023)}]{scale-free_mz23}%
  \BibitemOpen
  \bibfield  {author} {\bibinfo {author} {\bibfnamefont {X.}~\bibnamefont {Meng}}\ and\ \bibinfo {author} {\bibfnamefont {B.}~\bibnamefont {Zhou}},\ }\bibfield  {title} {\bibinfo {title} {Scale-free networks beyond power-law degree distribution},\ }\href {https://doi.org/10.1016/j.chaos.2023.114173} {\bibfield  {journal} {\bibinfo  {journal} {Chaos Solitons Fractals}\ }\textbf {\bibinfo {volume} {176}},\ \bibinfo {pages} {114173} (\bibinfo {year} {2023})}\BibitemShut {NoStop}%
\bibitem [{\citenamefont {Bogun{\'a}}\ and\ \citenamefont {Serrano}(2005)}]{boguna2005generalized}%
  \BibitemOpen
  \bibfield  {author} {\bibinfo {author} {\bibfnamefont {M.}~\bibnamefont {Bogun{\'a}}}\ and\ \bibinfo {author} {\bibfnamefont {M.~{\'A}.}\ \bibnamefont {Serrano}},\ }\bibfield  {title} {\bibinfo {title} {Generalized percolation in random directed networks},\ }\href {https://doi.org/10.1103/PhysRevE.72.016106} {\bibfield  {journal} {\bibinfo  {journal} {Phys. Rev. E}\ }\textbf {\bibinfo {volume} {72}},\ \bibinfo {pages} {016106} (\bibinfo {year} {2005})}\BibitemShut {NoStop}%
\bibitem [{\citenamefont {{Badie-Modiri}}\ \emph {et~al.}(2022)\citenamefont {{Badie-Modiri}}, \citenamefont {Rizi}, \citenamefont {Karsai},\ and\ \citenamefont {Kivel{\"a}}}]{badie2022directed}%
  \BibitemOpen
  \bibfield  {author} {\bibinfo {author} {\bibfnamefont {A.}~\bibnamefont {{Badie-Modiri}}}, \bibinfo {author} {\bibfnamefont {A.~K.}\ \bibnamefont {Rizi}}, \bibinfo {author} {\bibfnamefont {M.}~\bibnamefont {Karsai}},\ and\ \bibinfo {author} {\bibfnamefont {M.}~\bibnamefont {Kivel{\"a}}},\ }\bibfield  {title} {\bibinfo {title} {Directed percolation in temporal networks},\ }\href {https://doi.org/10.1103/PhysRevResearch.4.L022047} {\bibfield  {journal} {\bibinfo  {journal} {Phys. Rev. Res.}\ }\textbf {\bibinfo {volume} {4}},\ \bibinfo {pages} {L022047} (\bibinfo {year} {2022})}\BibitemShut {NoStop}%
\bibitem [{\citenamefont {Cohen}\ \emph {et~al.}(2002)\citenamefont {Cohen}, \citenamefont {{ben-Avraham}},\ and\ \citenamefont {Havlin}}]{cohen2002percolation}%
  \BibitemOpen
  \bibfield  {author} {\bibinfo {author} {\bibfnamefont {R.}~\bibnamefont {Cohen}}, \bibinfo {author} {\bibfnamefont {D.}~\bibnamefont {{ben-Avraham}}},\ and\ \bibinfo {author} {\bibfnamefont {S.}~\bibnamefont {Havlin}},\ }\bibfield  {title} {\bibinfo {title} {Percolation critical exponents in scale-free networks},\ }\href {https://doi.org/10.1103/PhysRevE.66.036113} {\bibfield  {journal} {\bibinfo  {journal} {Phys. Rev. E}\ }\textbf {\bibinfo {volume} {66}},\ \bibinfo {pages} {036113} (\bibinfo {year} {2002})}\BibitemShut {NoStop}%
\bibitem [{\citenamefont {Gao}\ \emph {et~al.}(2011)\citenamefont {Gao}, \citenamefont {Buldyrev}, \citenamefont {Havlin},\ and\ \citenamefont {Stanley}}]{gao2011robustness}%
  \BibitemOpen
  \bibfield  {author} {\bibinfo {author} {\bibfnamefont {J.}~\bibnamefont {Gao}}, \bibinfo {author} {\bibfnamefont {S.~V.}\ \bibnamefont {Buldyrev}}, \bibinfo {author} {\bibfnamefont {S.}~\bibnamefont {Havlin}},\ and\ \bibinfo {author} {\bibfnamefont {H.~E.}\ \bibnamefont {Stanley}},\ }\bibfield  {title} {\bibinfo {title} {Robustness of a network of networks},\ }\href {https://doi.org/10.1103/PhysRevLett.107.195701} {\bibfield  {journal} {\bibinfo  {journal} {Phys. Rev. Lett.}\ }\textbf {\bibinfo {volume} {107}},\ \bibinfo {pages} {195701} (\bibinfo {year} {2011})}\BibitemShut {NoStop}%
\bibitem [{\citenamefont {Gao}\ \emph {et~al.}(2012{\natexlab{a}})\citenamefont {Gao}, \citenamefont {Buldyrev}, \citenamefont {Stanley},\ and\ \citenamefont {Havlin}}]{gao2012networks}%
  \BibitemOpen
  \bibfield  {author} {\bibinfo {author} {\bibfnamefont {J.}~\bibnamefont {Gao}}, \bibinfo {author} {\bibfnamefont {S.~V.}\ \bibnamefont {Buldyrev}}, \bibinfo {author} {\bibfnamefont {H.~E.}\ \bibnamefont {Stanley}},\ and\ \bibinfo {author} {\bibfnamefont {S.}~\bibnamefont {Havlin}},\ }\bibfield  {title} {\bibinfo {title} {Networks formed from interdependent networks},\ }\href {https://doi.org/10.1038/nphys2180} {\bibfield  {journal} {\bibinfo  {journal} {Nat. Phys.}\ }\textbf {\bibinfo {volume} {8}},\ \bibinfo {pages} {40} (\bibinfo {year} {2012}{\natexlab{a}})}\BibitemShut {NoStop}%
\bibitem [{\citenamefont {Kivel{\"a}}\ \emph {et~al.}(2014)\citenamefont {Kivel{\"a}}, \citenamefont {Arenas}, \citenamefont {Barthelemy}, \citenamefont {Gleeson}, \citenamefont {Moreno},\ and\ \citenamefont {Porter}}]{kivela2014multilayer}%
  \BibitemOpen
  \bibfield  {author} {\bibinfo {author} {\bibfnamefont {M.}~\bibnamefont {Kivel{\"a}}}, \bibinfo {author} {\bibfnamefont {A.}~\bibnamefont {Arenas}}, \bibinfo {author} {\bibfnamefont {M.}~\bibnamefont {Barthelemy}}, \bibinfo {author} {\bibfnamefont {J.~P.}\ \bibnamefont {Gleeson}}, \bibinfo {author} {\bibfnamefont {Y.}~\bibnamefont {Moreno}},\ and\ \bibinfo {author} {\bibfnamefont {M.~A.}\ \bibnamefont {Porter}},\ }\bibfield  {title} {\bibinfo {title} {Multilayer networks},\ }\href {https://doi.org/10.1093/comnet/cnu016} {\bibfield  {journal} {\bibinfo  {journal} {J. Complex Netw.}\ }\textbf {\bibinfo {volume} {2}},\ \bibinfo {pages} {203} (\bibinfo {year} {2014})}\BibitemShut {NoStop}%
\bibitem [{\citenamefont {Gao}\ \emph {et~al.}(2022)\citenamefont {Gao}, \citenamefont {Bashan}, \citenamefont {Shekhtman},\ and\ \citenamefont {Havlin}}]{gao2022introduction}%
  \BibitemOpen
  \bibfield  {author} {\bibinfo {author} {\bibfnamefont {J.}~\bibnamefont {Gao}}, \bibinfo {author} {\bibfnamefont {A.}~\bibnamefont {Bashan}}, \bibinfo {author} {\bibfnamefont {L.}~\bibnamefont {Shekhtman}},\ and\ \bibinfo {author} {\bibfnamefont {S.}~\bibnamefont {Havlin}},\ }\href@noop {} {\emph {\bibinfo {title} {Introduction to Networks of Networks}}}\ (\bibinfo  {publisher} {{IOP Publishing}},\ \bibinfo {year} {2022})\BibitemShut {NoStop}%
\bibitem [{\citenamefont {Buldyrev}\ \emph {et~al.}(2010)\citenamefont {Buldyrev}, \citenamefont {Parshani}, \citenamefont {Paul}, \citenamefont {Stanley},\ and\ \citenamefont {Havlin}}]{buldyrev2010catastrophic}%
  \BibitemOpen
  \bibfield  {author} {\bibinfo {author} {\bibfnamefont {S.~V.}\ \bibnamefont {Buldyrev}}, \bibinfo {author} {\bibfnamefont {R.}~\bibnamefont {Parshani}}, \bibinfo {author} {\bibfnamefont {G.}~\bibnamefont {Paul}}, \bibinfo {author} {\bibfnamefont {H.~E.}\ \bibnamefont {Stanley}},\ and\ \bibinfo {author} {\bibfnamefont {S.}~\bibnamefont {Havlin}},\ }\bibfield  {title} {\bibinfo {title} {Catastrophic cascade of failures in interdependent networks},\ }\href {https://doi.org/10.1038/nature08932} {\bibfield  {journal} {\bibinfo  {journal} {Nature}\ }\textbf {\bibinfo {volume} {464}},\ \bibinfo {pages} {1025} (\bibinfo {year} {2010})}\BibitemShut {NoStop}%
\bibitem [{\citenamefont {Parshani}\ \emph {et~al.}(2010{\natexlab{b}})\citenamefont {Parshani}, \citenamefont {Buldyrev},\ and\ \citenamefont {Havlin}}]{parshani2010interdependent}%
  \BibitemOpen
  \bibfield  {author} {\bibinfo {author} {\bibfnamefont {R.}~\bibnamefont {Parshani}}, \bibinfo {author} {\bibfnamefont {S.~V.}\ \bibnamefont {Buldyrev}},\ and\ \bibinfo {author} {\bibfnamefont {S.}~\bibnamefont {Havlin}},\ }\bibfield  {title} {\bibinfo {title} {Interdependent networks: {{Reducing}} the coupling strength leads to a change from a first to second order percolation transition},\ }\href {https://doi.org/10.1103/PhysRevLett.105.048701} {\bibfield  {journal} {\bibinfo  {journal} {Phys. Rev. Lett.}\ }\textbf {\bibinfo {volume} {105}},\ \bibinfo {pages} {048701} (\bibinfo {year} {2010}{\natexlab{b}})}\BibitemShut {NoStop}%
\bibitem [{\citenamefont {Huang}\ \emph {et~al.}(2011)\citenamefont {Huang}, \citenamefont {Gao}, \citenamefont {Buldyrev}, \citenamefont {Havlin},\ and\ \citenamefont {Stanley}}]{huang2011robustness}%
  \BibitemOpen
  \bibfield  {author} {\bibinfo {author} {\bibfnamefont {X.}~\bibnamefont {Huang}}, \bibinfo {author} {\bibfnamefont {J.}~\bibnamefont {Gao}}, \bibinfo {author} {\bibfnamefont {S.~V.}\ \bibnamefont {Buldyrev}}, \bibinfo {author} {\bibfnamefont {S.}~\bibnamefont {Havlin}},\ and\ \bibinfo {author} {\bibfnamefont {H.~E.}\ \bibnamefont {Stanley}},\ }\bibfield  {title} {\bibinfo {title} {Robustness of interdependent networks under targeted attack},\ }\href {https://doi.org/10.1103/PhysRevE.83.065101} {\bibfield  {journal} {\bibinfo  {journal} {Phys. Rev. E}\ }\textbf {\bibinfo {volume} {83}},\ \bibinfo {pages} {065101} (\bibinfo {year} {2011})}\BibitemShut {NoStop}%
\bibitem [{\citenamefont {Shao}\ \emph {et~al.}(2011)\citenamefont {Shao}, \citenamefont {Buldyrev}, \citenamefont {Havlin},\ and\ \citenamefont {Stanley}}]{shao2011cascade}%
  \BibitemOpen
  \bibfield  {author} {\bibinfo {author} {\bibfnamefont {J.}~\bibnamefont {Shao}}, \bibinfo {author} {\bibfnamefont {S.~V.}\ \bibnamefont {Buldyrev}}, \bibinfo {author} {\bibfnamefont {S.}~\bibnamefont {Havlin}},\ and\ \bibinfo {author} {\bibfnamefont {H.~E.}\ \bibnamefont {Stanley}},\ }\bibfield  {title} {\bibinfo {title} {Cascade of failures in coupled network systems with multiple support-dependence relations},\ }\href {https://doi.org/10.1103/PhysRevE.83.036116} {\bibfield  {journal} {\bibinfo  {journal} {Phys. Rev. E}\ }\textbf {\bibinfo {volume} {83}},\ \bibinfo {pages} {036116} (\bibinfo {year} {2011})}\BibitemShut {NoStop}%
\bibitem [{\citenamefont {Zhao}\ and\ \citenamefont {Bianconi}(2013)}]{zhao2013percolation}%
  \BibitemOpen
  \bibfield  {author} {\bibinfo {author} {\bibfnamefont {K.}~\bibnamefont {Zhao}}\ and\ \bibinfo {author} {\bibfnamefont {G.}~\bibnamefont {Bianconi}},\ }\bibfield  {title} {\bibinfo {title} {Percolation on interacting, antagonistic networks},\ }\href {https://doi.org/10.1088/1742-5468/2013/05/P05005} {\bibfield  {journal} {\bibinfo  {journal} {J. Stat. Mech. Theor. Exp.}\ }\textbf {\bibinfo {volume} {2013}},\ \bibinfo {pages} {P05005} (\bibinfo {year} {2013})}\BibitemShut {NoStop}%
\bibitem [{\citenamefont {Gao}\ \emph {et~al.}(2012{\natexlab{b}})\citenamefont {Gao}, \citenamefont {Buldyrev}, \citenamefont {Havlin},\ and\ \citenamefont {Stanley}}]{gao2012robustness}%
  \BibitemOpen
  \bibfield  {author} {\bibinfo {author} {\bibfnamefont {J.}~\bibnamefont {Gao}}, \bibinfo {author} {\bibfnamefont {S.~V.}\ \bibnamefont {Buldyrev}}, \bibinfo {author} {\bibfnamefont {S.}~\bibnamefont {Havlin}},\ and\ \bibinfo {author} {\bibfnamefont {H.~E.}\ \bibnamefont {Stanley}},\ }\bibfield  {title} {\bibinfo {title} {Robustness of a network formed by n interdependent networks with a one-to-one correspondence of dependent nodes},\ }\href {https://doi.org/10.1103/PhysRevE.85.066134} {\bibfield  {journal} {\bibinfo  {journal} {Phys. Rev. E}\ }\textbf {\bibinfo {volume} {85}},\ \bibinfo {pages} {066134} (\bibinfo {year} {2012}{\natexlab{b}})}\BibitemShut {NoStop}%
\bibitem [{\citenamefont {Dong}\ \emph {et~al.}(2012)\citenamefont {Dong}, \citenamefont {Gao}, \citenamefont {Tian}, \citenamefont {Du},\ and\ \citenamefont {He}}]{dong2012percolation}%
  \BibitemOpen
  \bibfield  {author} {\bibinfo {author} {\bibfnamefont {G.}~\bibnamefont {Dong}}, \bibinfo {author} {\bibfnamefont {J.}~\bibnamefont {Gao}}, \bibinfo {author} {\bibfnamefont {L.}~\bibnamefont {Tian}}, \bibinfo {author} {\bibfnamefont {R.}~\bibnamefont {Du}},\ and\ \bibinfo {author} {\bibfnamefont {Y.}~\bibnamefont {He}},\ }\bibfield  {title} {\bibinfo {title} {Percolation of partially interdependent networks under targeted attack},\ }\href {https://doi.org/10.1103/PhysRevE.85.016112} {\bibfield  {journal} {\bibinfo  {journal} {Phys. Rev. E}\ }\textbf {\bibinfo {volume} {85}},\ \bibinfo {pages} {016112} (\bibinfo {year} {2012})}\BibitemShut {NoStop}%
\bibitem [{\citenamefont {Dong}\ \emph {et~al.}(2013)\citenamefont {Dong}, \citenamefont {Gao}, \citenamefont {Du}, \citenamefont {Tian}, \citenamefont {Stanley},\ and\ \citenamefont {Havlin}}]{dong2013robustness}%
  \BibitemOpen
  \bibfield  {author} {\bibinfo {author} {\bibfnamefont {G.}~\bibnamefont {Dong}}, \bibinfo {author} {\bibfnamefont {J.}~\bibnamefont {Gao}}, \bibinfo {author} {\bibfnamefont {R.}~\bibnamefont {Du}}, \bibinfo {author} {\bibfnamefont {L.}~\bibnamefont {Tian}}, \bibinfo {author} {\bibfnamefont {H.~E.}\ \bibnamefont {Stanley}},\ and\ \bibinfo {author} {\bibfnamefont {S.}~\bibnamefont {Havlin}},\ }\bibfield  {title} {\bibinfo {title} {Robustness of network of networks under targeted attack},\ }\href {https://doi.org/10.1103/PhysRevE.87.052804} {\bibfield  {journal} {\bibinfo  {journal} {Phys. Rev. E}\ }\textbf {\bibinfo {volume} {87}},\ \bibinfo {pages} {052804} (\bibinfo {year} {2013})}\BibitemShut {NoStop}%
\bibitem [{\citenamefont {Baxter}\ \emph {et~al.}(2012)\citenamefont {Baxter}, \citenamefont {Dorogovtsev}, \citenamefont {Goltsev},\ and\ \citenamefont {Mendes}}]{baxter2012avalanche}%
  \BibitemOpen
  \bibfield  {author} {\bibinfo {author} {\bibfnamefont {G.~J.}\ \bibnamefont {Baxter}}, \bibinfo {author} {\bibfnamefont {S.~N.}\ \bibnamefont {Dorogovtsev}}, \bibinfo {author} {\bibfnamefont {A.~V.}\ \bibnamefont {Goltsev}},\ and\ \bibinfo {author} {\bibfnamefont {J.~F.~F.}\ \bibnamefont {Mendes}},\ }\bibfield  {title} {\bibinfo {title} {Avalanche collapse of interdependent networks},\ }\href {https://doi.org/10.1103/PhysRevLett.109.248701} {\bibfield  {journal} {\bibinfo  {journal} {Phys. Rev. Lett.}\ }\textbf {\bibinfo {volume} {109}},\ \bibinfo {pages} {248701} (\bibinfo {year} {2012})}\BibitemShut {NoStop}%
\bibitem [{\citenamefont {Liu}\ \emph {et~al.}(2020)\citenamefont {Liu}, \citenamefont {Maiorino}, \citenamefont {Halu}, \citenamefont {Glass}, \citenamefont {Prasad}, \citenamefont {Loscalzo}, \citenamefont {Gao},\ and\ \citenamefont {Sharma}}]{liu2020robustness}%
  \BibitemOpen
  \bibfield  {author} {\bibinfo {author} {\bibfnamefont {X.}~\bibnamefont {Liu}}, \bibinfo {author} {\bibfnamefont {E.}~\bibnamefont {Maiorino}}, \bibinfo {author} {\bibfnamefont {A.}~\bibnamefont {Halu}}, \bibinfo {author} {\bibfnamefont {K.}~\bibnamefont {Glass}}, \bibinfo {author} {\bibfnamefont {R.~B.}\ \bibnamefont {Prasad}}, \bibinfo {author} {\bibfnamefont {J.}~\bibnamefont {Loscalzo}}, \bibinfo {author} {\bibfnamefont {J.}~\bibnamefont {Gao}},\ and\ \bibinfo {author} {\bibfnamefont {A.}~\bibnamefont {Sharma}},\ }\bibfield  {title} {\bibinfo {title} {Robustness and lethality in multilayer biological molecular networks},\ }\href {https://doi.org/10.1038/s41467-020-19841-3} {\bibfield  {journal} {\bibinfo  {journal} {Nat. Commun.}\ }\textbf {\bibinfo {volume} {11}},\ \bibinfo {pages} {6043} (\bibinfo {year} {2020})}\BibitemShut {NoStop}%
\bibitem [{\citenamefont {D'Agostino}\ and\ \citenamefont {Scala}(2014)}]{d2014networks}%
  \BibitemOpen
  \bibfield  {author} {\bibinfo {author} {\bibfnamefont {G.}~\bibnamefont {D'Agostino}}\ and\ \bibinfo {author} {\bibfnamefont {A.}~\bibnamefont {Scala}},\ }\href@noop {} {\emph {\bibinfo {title} {Networks of Networks: The Last Frontier of Complexity}}},\ Vol.\ \bibinfo {volume} {340}\ (\bibinfo  {publisher} {{Springer}},\ \bibinfo {year} {2014})\BibitemShut {NoStop}%
\bibitem [{\citenamefont {Reis}\ \emph {et~al.}(2014)\citenamefont {Reis}, \citenamefont {Hu}, \citenamefont {Babino}, \citenamefont {Andrade~Jr}, \citenamefont {Canals}, \citenamefont {Sigman},\ and\ \citenamefont {Makse}}]{reis2014avoiding}%
  \BibitemOpen
  \bibfield  {author} {\bibinfo {author} {\bibfnamefont {S.~D.}\ \bibnamefont {Reis}}, \bibinfo {author} {\bibfnamefont {Y.}~\bibnamefont {Hu}}, \bibinfo {author} {\bibfnamefont {A.}~\bibnamefont {Babino}}, \bibinfo {author} {\bibfnamefont {J.~S.}\ \bibnamefont {Andrade~Jr}}, \bibinfo {author} {\bibfnamefont {S.}~\bibnamefont {Canals}}, \bibinfo {author} {\bibfnamefont {M.}~\bibnamefont {Sigman}},\ and\ \bibinfo {author} {\bibfnamefont {H.~A.}\ \bibnamefont {Makse}},\ }\bibfield  {title} {\bibinfo {title} {Avoiding catastrophic failure in correlated networks of networks},\ }\href {https://doi.org/10.1038/nphys3081} {\bibfield  {journal} {\bibinfo  {journal} {Nat. Phys.}\ }\textbf {\bibinfo {volume} {10}},\ \bibinfo {pages} {762} (\bibinfo {year} {2014})}\BibitemShut {NoStop}%
\bibitem [{\citenamefont {Boccaletti}\ \emph {et~al.}(2014)\citenamefont {Boccaletti}, \citenamefont {Bianconi}, \citenamefont {Criado}, \citenamefont {Del~Genio}, \citenamefont {{G{\'o}mez-Gardenes}}, \citenamefont {Romance}, \citenamefont {{Sendina-Nadal}}, \citenamefont {Wang},\ and\ \citenamefont {Zanin}}]{boccaletti2014structure}%
  \BibitemOpen
  \bibfield  {author} {\bibinfo {author} {\bibfnamefont {S.}~\bibnamefont {Boccaletti}}, \bibinfo {author} {\bibfnamefont {G.}~\bibnamefont {Bianconi}}, \bibinfo {author} {\bibfnamefont {R.}~\bibnamefont {Criado}}, \bibinfo {author} {\bibfnamefont {C.~I.}\ \bibnamefont {Del~Genio}}, \bibinfo {author} {\bibfnamefont {J.}~\bibnamefont {{G{\'o}mez-Gardenes}}}, \bibinfo {author} {\bibfnamefont {M.}~\bibnamefont {Romance}}, \bibinfo {author} {\bibfnamefont {I.}~\bibnamefont {{Sendina-Nadal}}}, \bibinfo {author} {\bibfnamefont {Z.}~\bibnamefont {Wang}},\ and\ \bibinfo {author} {\bibfnamefont {M.}~\bibnamefont {Zanin}},\ }\bibfield  {title} {\bibinfo {title} {The structure and dynamics of multilayer networks},\ }\href {https://doi.org/10.1016/j.physrep.2014.07.001} {\bibfield  {journal} {\bibinfo  {journal} {Phys. Rep.}\ }\textbf {\bibinfo {volume} {544}},\ \bibinfo {pages} {1} (\bibinfo {year} {2014})}\BibitemShut {NoStop}%
\bibitem [{\citenamefont {Bianconi}(2018)}]{bianconi2018multilayer}%
  \BibitemOpen
  \bibfield  {author} {\bibinfo {author} {\bibfnamefont {G.}~\bibnamefont {Bianconi}},\ }\href@noop {} {\emph {\bibinfo {title} {Multilayer Networks: Structure and Function}}}\ (\bibinfo  {publisher} {{Oxford University Press}},\ \bibinfo {year} {2018})\BibitemShut {NoStop}%
\bibitem [{\citenamefont {Gomez}\ \emph {et~al.}(2013)\citenamefont {Gomez}, \citenamefont {{Diaz-Guilera}}, \citenamefont {{Gomez-Gardenes}}, \citenamefont {{Perez-Vicente}}, \citenamefont {Moreno},\ and\ \citenamefont {Arenas}}]{gomez2013diffusion}%
  \BibitemOpen
  \bibfield  {author} {\bibinfo {author} {\bibfnamefont {S.}~\bibnamefont {Gomez}}, \bibinfo {author} {\bibfnamefont {A.}~\bibnamefont {{Diaz-Guilera}}}, \bibinfo {author} {\bibfnamefont {J.}~\bibnamefont {{Gomez-Gardenes}}}, \bibinfo {author} {\bibfnamefont {C.~J.}\ \bibnamefont {{Perez-Vicente}}}, \bibinfo {author} {\bibfnamefont {Y.}~\bibnamefont {Moreno}},\ and\ \bibinfo {author} {\bibfnamefont {A.}~\bibnamefont {Arenas}},\ }\bibfield  {title} {\bibinfo {title} {Diffusion dynamics on multiplex networks},\ }\href {https://doi.org/10.1103/PhysRevLett.110.028701} {\bibfield  {journal} {\bibinfo  {journal} {Phys. Rev. Lett.}\ }\textbf {\bibinfo {volume} {110}},\ \bibinfo {pages} {028701} (\bibinfo {year} {2013})}\BibitemShut {NoStop}%
\bibitem [{\citenamefont {Liu}\ \emph {et~al.}(2022)\citenamefont {Liu}, \citenamefont {Li}, \citenamefont {Ma}, \citenamefont {Szymanski}, \citenamefont {Stanley},\ and\ \citenamefont {Gao}}]{liu2022network}%
  \BibitemOpen
  \bibfield  {author} {\bibinfo {author} {\bibfnamefont {X.}~\bibnamefont {Liu}}, \bibinfo {author} {\bibfnamefont {D.}~\bibnamefont {Li}}, \bibinfo {author} {\bibfnamefont {M.}~\bibnamefont {Ma}}, \bibinfo {author} {\bibfnamefont {B.~K.}\ \bibnamefont {Szymanski}}, \bibinfo {author} {\bibfnamefont {H.~E.}\ \bibnamefont {Stanley}},\ and\ \bibinfo {author} {\bibfnamefont {J.}~\bibnamefont {Gao}},\ }\bibfield  {title} {\bibinfo {title} {Network resilience},\ }\href {https://doi.org/10.1016/j.physrep.2022.04.002} {\bibfield  {journal} {\bibinfo  {journal} {Phys. Rep.}\ }\textbf {\bibinfo {volume} {971}},\ \bibinfo {pages} {1} (\bibinfo {year} {2022})}\BibitemShut {NoStop}%
\bibitem [{\citenamefont {Bashan}\ \emph {et~al.}(2013)\citenamefont {Bashan}, \citenamefont {Berezin}, \citenamefont {Buldyrev},\ and\ \citenamefont {Havlin}}]{bashan2013extreme}%
  \BibitemOpen
  \bibfield  {author} {\bibinfo {author} {\bibfnamefont {A.}~\bibnamefont {Bashan}}, \bibinfo {author} {\bibfnamefont {Y.}~\bibnamefont {Berezin}}, \bibinfo {author} {\bibfnamefont {S.~V.}\ \bibnamefont {Buldyrev}},\ and\ \bibinfo {author} {\bibfnamefont {S.}~\bibnamefont {Havlin}},\ }\bibfield  {title} {\bibinfo {title} {The extreme vulnerability of interdependent spatially embedded networks},\ }\href {https://doi.org/10.1038/nphys2727} {\bibfield  {journal} {\bibinfo  {journal} {Nat. Phys.}\ }\textbf {\bibinfo {volume} {9}},\ \bibinfo {pages} {667} (\bibinfo {year} {2013})}\BibitemShut {NoStop}%
\bibitem [{\citenamefont {Lu}\ \emph {et~al.}(2014)\citenamefont {Lu}, \citenamefont {Yu}, \citenamefont {L{\"u}},\ and\ \citenamefont {Xue}}]{lu2014synchronization}%
  \BibitemOpen
  \bibfield  {author} {\bibinfo {author} {\bibfnamefont {R.}~\bibnamefont {Lu}}, \bibinfo {author} {\bibfnamefont {W.}~\bibnamefont {Yu}}, \bibinfo {author} {\bibfnamefont {J.}~\bibnamefont {L{\"u}}},\ and\ \bibinfo {author} {\bibfnamefont {A.}~\bibnamefont {Xue}},\ }\bibfield  {title} {\bibinfo {title} {Synchronization on complex networks of networks},\ }\href {https://doi.org/10.1109/TNNLS.2014.2305443} {\bibfield  {journal} {\bibinfo  {journal} {IEEE Trans. Neural Netw. Learn. Syst.}\ }\textbf {\bibinfo {volume} {25}},\ \bibinfo {pages} {2110} (\bibinfo {year} {2014})}\BibitemShut {NoStop}%
\bibitem [{\citenamefont {Ouyang}(2017)}]{ouyang2017mathematical}%
  \BibitemOpen
  \bibfield  {author} {\bibinfo {author} {\bibfnamefont {M.}~\bibnamefont {Ouyang}},\ }\bibfield  {title} {\bibinfo {title} {A mathematical framework to optimize resilience of interdependent critical infrastructure systems under spatially localized attacks},\ }\href {https://doi.org/10.1016/j.ejor.2017.04.022} {\bibfield  {journal} {\bibinfo  {journal} {Eur. J. Oper. Res.}\ }\textbf {\bibinfo {volume} {262}},\ \bibinfo {pages} {1072} (\bibinfo {year} {2017})}\BibitemShut {NoStop}%
\bibitem [{\citenamefont {Duan}\ \emph {et~al.}(2019)\citenamefont {Duan}, \citenamefont {Lv}, \citenamefont {Si}, \citenamefont {Wang}, \citenamefont {Li}, \citenamefont {Gao}, \citenamefont {Havlin}, \citenamefont {Stanley},\ and\ \citenamefont {Boccaletti}}]{duan2019universal}%
  \BibitemOpen
  \bibfield  {author} {\bibinfo {author} {\bibfnamefont {D.}~\bibnamefont {Duan}}, \bibinfo {author} {\bibfnamefont {C.}~\bibnamefont {Lv}}, \bibinfo {author} {\bibfnamefont {S.}~\bibnamefont {Si}}, \bibinfo {author} {\bibfnamefont {Z.}~\bibnamefont {Wang}}, \bibinfo {author} {\bibfnamefont {D.}~\bibnamefont {Li}}, \bibinfo {author} {\bibfnamefont {J.}~\bibnamefont {Gao}}, \bibinfo {author} {\bibfnamefont {S.}~\bibnamefont {Havlin}}, \bibinfo {author} {\bibfnamefont {H.~E.}\ \bibnamefont {Stanley}},\ and\ \bibinfo {author} {\bibfnamefont {S.}~\bibnamefont {Boccaletti}},\ }\bibfield  {title} {\bibinfo {title} {Universal behavior of cascading failures in interdependent networks},\ }\href {https://doi.org/10.1073/pnas.1904421116} {\bibfield  {journal} {\bibinfo  {journal} {Proc. Natl. Acad. Sci. U.S.A.}\ }\textbf {\bibinfo {volume} {116}},\ \bibinfo {pages} {22452} (\bibinfo {year} {2019})}\BibitemShut {NoStop}%
\bibitem [{\citenamefont {Cellai}\ \emph {et~al.}(2013)\citenamefont {Cellai}, \citenamefont {L{\'o}pez}, \citenamefont {Zhou}, \citenamefont {Gleeson},\ and\ \citenamefont {Bianconi}}]{cellai2013percolation}%
  \BibitemOpen
  \bibfield  {author} {\bibinfo {author} {\bibfnamefont {D.}~\bibnamefont {Cellai}}, \bibinfo {author} {\bibfnamefont {E.}~\bibnamefont {L{\'o}pez}}, \bibinfo {author} {\bibfnamefont {J.}~\bibnamefont {Zhou}}, \bibinfo {author} {\bibfnamefont {J.~P.}\ \bibnamefont {Gleeson}},\ and\ \bibinfo {author} {\bibfnamefont {G.}~\bibnamefont {Bianconi}},\ }\bibfield  {title} {\bibinfo {title} {Percolation in multiplex networks with overlap},\ }\href {https://doi.org/10.1103/PhysRevE.88.052811} {\bibfield  {journal} {\bibinfo  {journal} {Phys. Rev. E}\ }\textbf {\bibinfo {volume} {88}},\ \bibinfo {pages} {052811} (\bibinfo {year} {2013})}\BibitemShut {NoStop}%
\bibitem [{\citenamefont {Kenett}\ \emph {et~al.}(2015)\citenamefont {Kenett}, \citenamefont {Perc},\ and\ \citenamefont {Boccaletti}}]{kenett2015networks}%
  \BibitemOpen
  \bibfield  {author} {\bibinfo {author} {\bibfnamefont {D.~Y.}\ \bibnamefont {Kenett}}, \bibinfo {author} {\bibfnamefont {M.}~\bibnamefont {Perc}},\ and\ \bibinfo {author} {\bibfnamefont {S.}~\bibnamefont {Boccaletti}},\ }\bibfield  {title} {\bibinfo {title} {Networks of networks\textendash an introduction},\ }\href {https://doi.org/10.1016/j.chaos.2015.03.016} {\bibfield  {journal} {\bibinfo  {journal} {Chaos Solitons Fractals}\ }\textbf {\bibinfo {volume} {80}},\ \bibinfo {pages} {1} (\bibinfo {year} {2015})}\BibitemShut {NoStop}%
\bibitem [{\citenamefont {De~Domenico}(2023)}]{de2023more}%
  \BibitemOpen
  \bibfield  {author} {\bibinfo {author} {\bibfnamefont {M.}~\bibnamefont {De~Domenico}},\ }\bibfield  {title} {\bibinfo {title} {More is different in real-world multilayer networks},\ }\href@noop {} {\bibfield  {journal} {\bibinfo  {journal} {Nat. Phys.}\ }\textbf {\bibinfo {volume} {19}},\ \bibinfo {pages} {1247} (\bibinfo {year} {2023})}\BibitemShut {NoStop}%
\bibitem [{\citenamefont {Erd{\H o}s}\ and\ \citenamefont {R{\'e}nyi}(1959)}]{erdos-renyi_er59}%
  \BibitemOpen
  \bibfield  {author} {\bibinfo {author} {\bibfnamefont {P.}~\bibnamefont {Erd{\H o}s}}\ and\ \bibinfo {author} {\bibfnamefont {A.}~\bibnamefont {R{\'e}nyi}},\ }\bibfield  {title} {\bibinfo {title} {On random graphs},\ }\href@noop {} {\bibfield  {journal} {\bibinfo  {journal} {Publ. Math. Debr.}\ }\textbf {\bibinfo {volume} {6}},\ \bibinfo {pages} {290} (\bibinfo {year} {1959})}\BibitemShut {NoStop}%
\bibitem [{\citenamefont {Erd{\H o}s}\ and\ \citenamefont {R{\'e}nyi}(1960)}]{ER1960}%
  \BibitemOpen
  \bibfield  {author} {\bibinfo {author} {\bibfnamefont {P.}~\bibnamefont {Erd{\H o}s}}\ and\ \bibinfo {author} {\bibfnamefont {A.}~\bibnamefont {R{\'e}nyi}},\ }\bibfield  {title} {\bibinfo {title} {On the evolution of random graphs},\ }\href@noop {} {\bibfield  {journal} {\bibinfo  {journal} {Inst. Hung. Acad. Sci.}\ }\textbf {\bibinfo {volume} {5}},\ \bibinfo {pages} {17} (\bibinfo {year} {1960})}\BibitemShut {NoStop}%
\bibitem [{\citenamefont {Bollob{\'a}s}(1985)}]{Bollob1985}%
  \BibitemOpen
  \bibfield  {author} {\bibinfo {author} {\bibfnamefont {B.}~\bibnamefont {Bollob{\'a}s}},\ }\href@noop {} {\emph {\bibinfo {title} {Random Graphs}}}\ (\bibinfo  {publisher} {{Academic}},\ \bibinfo {address} {{London}},\ \bibinfo {year} {1985})\BibitemShut {NoStop}%
\bibitem [{\citenamefont {Bretto}(2013{\natexlab{b}})}]{bretto2013hypergraph}%
  \BibitemOpen
  \bibfield  {author} {\bibinfo {author} {\bibfnamefont {A.}~\bibnamefont {Bretto}},\ }\href@noop {} {\emph {\bibinfo {title} {Hypergraph Theory: An Introduction}}},\ Mathematical Engineering\ (\bibinfo  {publisher} {Springer International Publishing},\ \bibinfo {year} {2013})\BibitemShut {NoStop}%
\bibitem [{\citenamefont {Bianconi}(2021)}]{bianconi2021hyper}%
  \BibitemOpen
  \bibfield  {author} {\bibinfo {author} {\bibfnamefont {G.}~\bibnamefont {Bianconi}},\ }\href {https://doi.org/10.1017/9781108770996} {\emph {\bibinfo {title} {Higher-Order Networks}}},\ Elements in the Structure and Dynamics of Complex Networks\ (\bibinfo  {publisher} {Cambridge University Press},\ \bibinfo {year} {2021})\BibitemShut {NoStop}%
\bibitem [{\citenamefont {Bick}\ \emph {et~al.}(2023)\citenamefont {Bick}, \citenamefont {Gross}, \citenamefont {Harrington},\ and\ \citenamefont {Schaub}}]{bick2023hyper}%
  \BibitemOpen
  \bibfield  {author} {\bibinfo {author} {\bibfnamefont {C.}~\bibnamefont {Bick}}, \bibinfo {author} {\bibfnamefont {E.}~\bibnamefont {Gross}}, \bibinfo {author} {\bibfnamefont {H.}~\bibnamefont {Harrington}},\ and\ \bibinfo {author} {\bibfnamefont {M.}~\bibnamefont {Schaub}},\ }\bibfield  {title} {\bibinfo {title} {What are higher-order networks?},\ }\href {https://doi.org/10.1137/21M1414024} {\bibfield  {journal} {\bibinfo  {journal} {SIAM Review}\ }\textbf {\bibinfo {volume} {65}},\ \bibinfo {pages} {686} (\bibinfo {year} {2023})}\BibitemShut {NoStop}%
\bibitem [{\citenamefont {Klamt}\ \emph {et~al.}(2009)\citenamefont {Klamt}, \citenamefont {Haus},\ and\ \citenamefont {Theis}}]{klamt2009biology}%
  \BibitemOpen
  \bibfield  {author} {\bibinfo {author} {\bibfnamefont {S.}~\bibnamefont {Klamt}}, \bibinfo {author} {\bibfnamefont {U.}~\bibnamefont {Haus}},\ and\ \bibinfo {author} {\bibfnamefont {F.}~\bibnamefont {Theis}},\ }\bibfield  {title} {\bibinfo {title} {Hypergraphs and cellular networks},\ }\href {https://doi.org/10.1371/journal.pcbi.1000385} {\bibfield  {journal} {\bibinfo  {journal} {PLOS Computat. Biol.}\ }\textbf {\bibinfo {volume} {5}},\ \bibinfo {pages} {1} (\bibinfo {year} {2009})}\BibitemShut {NoStop}%
\bibitem [{\citenamefont {Jost}\ and\ \citenamefont {Mulas}(2019)}]{jost2019hyperL}%
  \BibitemOpen
  \bibfield  {author} {\bibinfo {author} {\bibfnamefont {J.}~\bibnamefont {Jost}}\ and\ \bibinfo {author} {\bibfnamefont {R.}~\bibnamefont {Mulas}},\ }\bibfield  {title} {\bibinfo {title} {Hypergraph laplace operators for chemical reaction networks},\ }\href {https://doi.org/10.1016/j.aim.2019.05.025} {\bibfield  {journal} {\bibinfo  {journal} {Adv. Math.}\ }\textbf {\bibinfo {volume} {351}},\ \bibinfo {pages} {870} (\bibinfo {year} {2019})}\BibitemShut {NoStop}%
\bibitem [{\citenamefont {Taramasco}\ \emph {et~al.}(2010)\citenamefont {Taramasco}, \citenamefont {J.},\ and\ \citenamefont {Roth}}]{taramasco2010collab}%
  \BibitemOpen
  \bibfield  {author} {\bibinfo {author} {\bibfnamefont {C.}~\bibnamefont {Taramasco}}, \bibinfo {author} {\bibfnamefont {C.}~\bibnamefont {J.}},\ and\ \bibinfo {author} {\bibfnamefont {C.}~\bibnamefont {Roth}},\ }\bibfield  {title} {\bibinfo {title} {Academic team formation as evolving hypergraphs},\ }\href {https://doi.org/10.1007/s11192-010-0226-4} {\bibfield  {journal} {\bibinfo  {journal} {Scientometrics}\ }\textbf {\bibinfo {volume} {85}},\ \bibinfo {pages} {721} (\bibinfo {year} {2010})}\BibitemShut {NoStop}%
\bibitem [{\citenamefont {Krumov}\ \emph {et~al.}(2011)\citenamefont {Krumov}, \citenamefont {Fretter}, \citenamefont {M\"{u}ller-Hannemann}, \citenamefont {Weihe},\ and\ \citenamefont {H\"{u}tt}}]{krumov2011collab}%
  \BibitemOpen
  \bibfield  {author} {\bibinfo {author} {\bibfnamefont {L.}~\bibnamefont {Krumov}}, \bibinfo {author} {\bibfnamefont {C.}~\bibnamefont {Fretter}}, \bibinfo {author} {\bibfnamefont {M.}~\bibnamefont {M\"{u}ller-Hannemann}}, \bibinfo {author} {\bibfnamefont {K.}~\bibnamefont {Weihe}},\ and\ \bibinfo {author} {\bibfnamefont {M.}~\bibnamefont {H\"{u}tt}},\ }\bibfield  {title} {\bibinfo {title} {Motifs in co-authorship networks and their relation to the impact of scientific publications},\ }\href {https://doi.org/10.1140/epjb/e2011-10746-5} {\bibfield  {journal} {\bibinfo  {journal} {Eur. Phys. J. B}\ }\textbf {\bibinfo {volume} {84}},\ \bibinfo {pages} {535} (\bibinfo {year} {2011})}\BibitemShut {NoStop}%
\bibitem [{\citenamefont {Neuh{\"a}user}\ \emph {et~al.}(2020)\citenamefont {Neuh{\"a}user}, \citenamefont {Mellor},\ and\ \citenamefont {Lambiotte}}]{neuhauser2020multibody}%
  \BibitemOpen
  \bibfield  {author} {\bibinfo {author} {\bibfnamefont {L.}~\bibnamefont {Neuh{\"a}user}}, \bibinfo {author} {\bibfnamefont {A.}~\bibnamefont {Mellor}},\ and\ \bibinfo {author} {\bibfnamefont {R.}~\bibnamefont {Lambiotte}},\ }\bibfield  {title} {\bibinfo {title} {Multibody interactions and nonlinear consensus dynamics on networked systems},\ }\href {https://doi.org/10.1103/PhysRevE.101.032310} {\bibfield  {journal} {\bibinfo  {journal} {Phys. Rev. E}\ }\textbf {\bibinfo {volume} {101}},\ \bibinfo {pages} {032310} (\bibinfo {year} {2020})}\BibitemShut {NoStop}%
\bibitem [{\citenamefont {Majhi}\ \emph {et~al.}(2022)\citenamefont {Majhi}, \citenamefont {Perc},\ and\ \citenamefont {Ghosh}}]{majhi2022higher}%
  \BibitemOpen
  \bibfield  {author} {\bibinfo {author} {\bibfnamefont {S.}~\bibnamefont {Majhi}}, \bibinfo {author} {\bibfnamefont {M.}~\bibnamefont {Perc}},\ and\ \bibinfo {author} {\bibfnamefont {D.}~\bibnamefont {Ghosh}},\ }\bibfield  {title} {\bibinfo {title} {Dynamics on higher-order networks: {{A}} review},\ }\href {https://doi.org/10.1098/rsif.2022.0043} {\bibfield  {journal} {\bibinfo  {journal} {J. Royal Soc. Interface}\ }\textbf {\bibinfo {volume} {19}},\ \bibinfo {pages} {20220043} (\bibinfo {year} {2022})}\BibitemShut {NoStop}%
\bibitem [{\citenamefont {Coutinho}\ \emph {et~al.}(2020)\citenamefont {Coutinho}, \citenamefont {Wu}, \citenamefont {Zhou},\ and\ \citenamefont {Liu}}]{coutinho2020percolhyper}%
  \BibitemOpen
  \bibfield  {author} {\bibinfo {author} {\bibfnamefont {B.}~\bibnamefont {Coutinho}}, \bibinfo {author} {\bibfnamefont {A.}~\bibnamefont {Wu}}, \bibinfo {author} {\bibfnamefont {H.}~\bibnamefont {Zhou}},\ and\ \bibinfo {author} {\bibfnamefont {Y.}~\bibnamefont {Liu}},\ }\bibfield  {title} {\bibinfo {title} {Covering problems and core percolations on hypergraphs},\ }\href {https://doi.org/10.1103/PhysRevLett.124.248301} {\bibfield  {journal} {\bibinfo  {journal} {Phys. Rev. Lett.}\ }\textbf {\bibinfo {volume} {124}},\ \bibinfo {pages} {248301} (\bibinfo {year} {2020})}\BibitemShut {NoStop}%
\bibitem [{\citenamefont {Sun}\ and\ \citenamefont {Bianconi}(2021)}]{sun2021percolhyper}%
  \BibitemOpen
  \bibfield  {author} {\bibinfo {author} {\bibfnamefont {H.}~\bibnamefont {Sun}}\ and\ \bibinfo {author} {\bibfnamefont {G.}~\bibnamefont {Bianconi}},\ }\bibfield  {title} {\bibinfo {title} {Higher-order percolation processes on multiplex hypergraphs},\ }\href {https://doi.org/10.1103/PhysRevE.104.034306} {\bibfield  {journal} {\bibinfo  {journal} {Phys. Rev. E}\ }\textbf {\bibinfo {volume} {104}},\ \bibinfo {pages} {034306} (\bibinfo {year} {2021})}\BibitemShut {NoStop}%
\bibitem [{\citenamefont {Sun}\ \emph {et~al.}(2023)\citenamefont {Sun}, \citenamefont {Radicchi}, \citenamefont {Kurths},\ and\ \citenamefont {Bianconi}}]{sun2023percolhyper}%
  \BibitemOpen
  \bibfield  {author} {\bibinfo {author} {\bibfnamefont {H.}~\bibnamefont {Sun}}, \bibinfo {author} {\bibfnamefont {F.}~\bibnamefont {Radicchi}}, \bibinfo {author} {\bibfnamefont {J.}~\bibnamefont {Kurths}},\ and\ \bibinfo {author} {\bibfnamefont {G.}~\bibnamefont {Bianconi}},\ }\bibfield  {title} {\bibinfo {title} {The dynamic nature of percolation on networks with triadic interactions},\ }\href {https://doi.org/10.1038/s41467-023-37019-5} {\bibfield  {journal} {\bibinfo  {journal} {Nat. Commun.}\ }\textbf {\bibinfo {volume} {14}},\ \bibinfo {pages} {1308} (\bibinfo {year} {2023})}\BibitemShut {NoStop}%
\bibitem [{\citenamefont {Lee}\ \emph {et~al.}(2023)\citenamefont {Lee}, \citenamefont {Goh}, \citenamefont {Lee},\ and\ \citenamefont {Kahng}}]{lee2023kqcore}%
  \BibitemOpen
  \bibfield  {author} {\bibinfo {author} {\bibfnamefont {J.}~\bibnamefont {Lee}}, \bibinfo {author} {\bibfnamefont {K.}~\bibnamefont {Goh}}, \bibinfo {author} {\bibfnamefont {D.}~\bibnamefont {Lee}},\ and\ \bibinfo {author} {\bibfnamefont {B.}~\bibnamefont {Kahng}},\ }\bibfield  {title} {\bibinfo {title} {(k,q)-core decomposition of hypergraphs},\ }\href {https://doi.org/10.1016/j.chaos.2023.113645} {\bibfield  {journal} {\bibinfo  {journal} {Chaos Solitons Fractals}\ }\textbf {\bibinfo {volume} {173}},\ \bibinfo {pages} {113645} (\bibinfo {year} {2023})}\BibitemShut {NoStop}%
\bibitem [{\citenamefont {Peng}\ \emph {et~al.}(2022)\citenamefont {Peng}, \citenamefont {Qian}, \citenamefont {Zhao}, \citenamefont {Zhong}, \citenamefont {Ling},\ and\ \citenamefont {Wang}}]{peng2022hyperattack}%
  \BibitemOpen
  \bibfield  {author} {\bibinfo {author} {\bibfnamefont {H.}~\bibnamefont {Peng}}, \bibinfo {author} {\bibfnamefont {C.}~\bibnamefont {Qian}}, \bibinfo {author} {\bibfnamefont {D.}~\bibnamefont {Zhao}}, \bibinfo {author} {\bibfnamefont {M.}~\bibnamefont {Zhong}}, \bibinfo {author} {\bibfnamefont {X.}~\bibnamefont {Ling}},\ and\ \bibinfo {author} {\bibfnamefont {W.}~\bibnamefont {Wang}},\ }\bibfield  {title} {\bibinfo {title} {Disintegrate hypergraph networks by attacking hyperedge},\ }\href {https://doi.org/10.1016/j.jksuci.2022.04.017} {\bibfield  {journal} {\bibinfo  {journal} {J. King Saud Univ. Comput. Inf. Sci.}\ }\textbf {\bibinfo {volume} {34}},\ \bibinfo {pages} {4679} (\bibinfo {year} {2022})}\BibitemShut {NoStop}%
\bibitem [{\citenamefont {Peng}\ \emph {et~al.}(2023)\citenamefont {Peng}, \citenamefont {Qian}, \citenamefont {Zhao}, \citenamefont {Zhong}, \citenamefont {Han}, \citenamefont {Li},\ and\ \citenamefont {Wang}}]{peng2023hypermp}%
  \BibitemOpen
  \bibfield  {author} {\bibinfo {author} {\bibfnamefont {H.}~\bibnamefont {Peng}}, \bibinfo {author} {\bibfnamefont {C.}~\bibnamefont {Qian}}, \bibinfo {author} {\bibfnamefont {D.}~\bibnamefont {Zhao}}, \bibinfo {author} {\bibfnamefont {M.}~\bibnamefont {Zhong}}, \bibinfo {author} {\bibfnamefont {J.}~\bibnamefont {Han}}, \bibinfo {author} {\bibfnamefont {R.}~\bibnamefont {Li}},\ and\ \bibinfo {author} {\bibfnamefont {W.}~\bibnamefont {Wang}},\ }\bibfield  {title} {\bibinfo {title} {Message passing approach to analyze the robustness of hypergraph},\ }\href@noop {} {\bibfield  {journal} {\bibinfo  {journal} {arXiv:2302.14594}\ } (\bibinfo {year} {2023})}\BibitemShut {NoStop}%
\bibitem [{\citenamefont {Bianconi}\ and\ \citenamefont {Dorogovtsev}(2023)}]{bianconi2023percoltheory}%
  \BibitemOpen
  \bibfield  {author} {\bibinfo {author} {\bibfnamefont {G.}~\bibnamefont {Bianconi}}\ and\ \bibinfo {author} {\bibfnamefont {S.}~\bibnamefont {Dorogovtsev}},\ }\bibfield  {title} {\bibinfo {title} {The theory of percolation on hypergraphs},\ }\href@noop {} {\bibfield  {journal} {\bibinfo  {journal} {arXiv:2305.12297}\ } (\bibinfo {year} {2023})}\BibitemShut {NoStop}%
\bibitem [{\citenamefont {Bose}\ \emph {et~al.}(1999)\citenamefont {Bose}, \citenamefont {Vedral},\ and\ \citenamefont {Knight}}]{QEP-series-rule_bvk99}%
  \BibitemOpen
  \bibfield  {author} {\bibinfo {author} {\bibfnamefont {S.}~\bibnamefont {Bose}}, \bibinfo {author} {\bibfnamefont {V.}~\bibnamefont {Vedral}},\ and\ \bibinfo {author} {\bibfnamefont {P.~L.}\ \bibnamefont {Knight}},\ }\bibfield  {title} {\bibinfo {title} {Purification via entanglement swapping and conserved entanglement},\ }\href {https://doi.org/10.1103/PhysRevA.60.194} {\bibfield  {journal} {\bibinfo  {journal} {Phys. Rev. A}\ }\textbf {\bibinfo {volume} {60}},\ \bibinfo {pages} {194} (\bibinfo {year} {1999})}\BibitemShut {NoStop}%
\bibitem [{\citenamefont {Perseguers}\ \emph {et~al.}(2008)\citenamefont {Perseguers}, \citenamefont {Cirac}, \citenamefont {Ac{\'i}n}, \citenamefont {Lewenstein},\ and\ \citenamefont {Wehr}}]{QEP-detail_pcalw08}%
  \BibitemOpen
  \bibfield  {author} {\bibinfo {author} {\bibfnamefont {S.}~\bibnamefont {Perseguers}}, \bibinfo {author} {\bibfnamefont {J.~I.}\ \bibnamefont {Cirac}}, \bibinfo {author} {\bibfnamefont {A.}~\bibnamefont {Ac{\'i}n}}, \bibinfo {author} {\bibfnamefont {M.}~\bibnamefont {Lewenstein}},\ and\ \bibinfo {author} {\bibfnamefont {J.}~\bibnamefont {Wehr}},\ }\bibfield  {title} {\bibinfo {title} {Entanglement distribution in pure-state quantum networks},\ }\href {https://doi.org/10.1103/PhysRevA.77.022308} {\bibfield  {journal} {\bibinfo  {journal} {Phys. Rev. A}\ }\textbf {\bibinfo {volume} {77}},\ \bibinfo {pages} {022308} (\bibinfo {year} {2008})}\BibitemShut {NoStop}%
\bibitem [{\citenamefont {Meng}\ \emph {et~al.}(2023)\citenamefont {Meng}, \citenamefont {Cui}, \citenamefont {Gao}, \citenamefont {Havlin},\ and\ \citenamefont {Ruckenstein}}]{det_mcghr23}%
  \BibitemOpen
  \bibfield  {author} {\bibinfo {author} {\bibfnamefont {X.}~\bibnamefont {Meng}}, \bibinfo {author} {\bibfnamefont {Y.}~\bibnamefont {Cui}}, \bibinfo {author} {\bibfnamefont {J.}~\bibnamefont {Gao}}, \bibinfo {author} {\bibfnamefont {S.}~\bibnamefont {Havlin}},\ and\ \bibinfo {author} {\bibfnamefont {A.~E.}\ \bibnamefont {Ruckenstein}},\ }\bibfield  {title} {\bibinfo {title} {Deterministic entanglement distribution on series-parallel quantum networks},\ }\href {https://doi.org/10.1103/PhysRevResearch.5.013225} {\bibfield  {journal} {\bibinfo  {journal} {Phys. Rev. Res.}\ }\textbf {\bibinfo {volume} {5}},\ \bibinfo {pages} {013225} (\bibinfo {year} {2023})}\BibitemShut {NoStop}%
\bibitem [{\citenamefont {Nielsen}(1999)}]{nielsen_n99}%
  \BibitemOpen
  \bibfield  {author} {\bibinfo {author} {\bibfnamefont {M.~A.}\ \bibnamefont {Nielsen}},\ }\bibfield  {title} {\bibinfo {title} {Conditions for a {{Class}} of {{Entanglement Transformations}}},\ }\href {https://doi.org/10.1103/PhysRevLett.83.436} {\bibfield  {journal} {\bibinfo  {journal} {Phys. Rev. Lett.}\ }\textbf {\bibinfo {volume} {83}},\ \bibinfo {pages} {436} (\bibinfo {year} {1999})}\BibitemShut {NoStop}%
\bibitem [{\citenamefont {{\.Z}ukowski}\ \emph {et~al.}(1993)\citenamefont {{\.Z}ukowski}, \citenamefont {Zeilinger}, \citenamefont {Horne},\ and\ \citenamefont {Ekert}}]{entangle-swap_zzhe93}%
  \BibitemOpen
  \bibfield  {author} {\bibinfo {author} {\bibfnamefont {M.}~\bibnamefont {{\.Z}ukowski}}, \bibinfo {author} {\bibfnamefont {A.}~\bibnamefont {Zeilinger}}, \bibinfo {author} {\bibfnamefont {M.~A.}\ \bibnamefont {Horne}},\ and\ \bibinfo {author} {\bibfnamefont {A.~K.}\ \bibnamefont {Ekert}},\ }\bibfield  {title} {\bibinfo {title} {``{{Event-Ready-Detectors}}'' {{Bell Experiment}} via {{Entanglement Swapping}}},\ }\href {https://doi.org/10.1103/PhysRevLett.71.4287} {\bibfield  {journal} {\bibinfo  {journal} {Phys. Rev. Lett.}\ }\textbf {\bibinfo {volume} {71}},\ \bibinfo {pages} {4287} (\bibinfo {year} {1993})}\BibitemShut {NoStop}%
\bibitem [{\citenamefont {Bennett}\ \emph {et~al.}(1996)\citenamefont {Bennett}, \citenamefont {Bernstein}, \citenamefont {Popescu},\ and\ \citenamefont {Schumacher}}]{entangle-conc_bbps96}%
  \BibitemOpen
  \bibfield  {author} {\bibinfo {author} {\bibfnamefont {C.~H.}\ \bibnamefont {Bennett}}, \bibinfo {author} {\bibfnamefont {H.~J.}\ \bibnamefont {Bernstein}}, \bibinfo {author} {\bibfnamefont {S.}~\bibnamefont {Popescu}},\ and\ \bibinfo {author} {\bibfnamefont {B.}~\bibnamefont {Schumacher}},\ }\bibfield  {title} {\bibinfo {title} {Concentrating partial entanglement by local operations},\ }\href {https://doi.org/10.1103/PhysRevA.53.2046} {\bibfield  {journal} {\bibinfo  {journal} {Phys. Rev. A}\ }\textbf {\bibinfo {volume} {53}},\ \bibinfo {pages} {2046} (\bibinfo {year} {1996})}\BibitemShut {NoStop}%
\bibitem [{\citenamefont {Vidal}(1999)}]{QEP-parallel-rule_v99}%
  \BibitemOpen
  \bibfield  {author} {\bibinfo {author} {\bibfnamefont {G.}~\bibnamefont {Vidal}},\ }\bibfield  {title} {\bibinfo {title} {Entanglement of {{Pure States}} for a {{Single Copy}}},\ }\href {https://doi.org/10.1103/PhysRevLett.83.1046} {\bibfield  {journal} {\bibinfo  {journal} {Phys. Rev. Lett.}\ }\textbf {\bibinfo {volume} {83}},\ \bibinfo {pages} {1046} (\bibinfo {year} {1999})}\BibitemShut {NoStop}%
\bibitem [{\citenamefont {Duffin}(1965)}]{series-parallel-netw_d65}%
  \BibitemOpen
  \bibfield  {author} {\bibinfo {author} {\bibfnamefont {R.~J.}\ \bibnamefont {Duffin}},\ }\bibfield  {title} {\bibinfo {title} {Topology of series-parallel networks},\ }\href {https://doi.org/10.1016/0022-247X(65)90125-3} {\bibfield  {journal} {\bibinfo  {journal} {J. Math. Anal. Appl.}\ }\textbf {\bibinfo {volume} {10}},\ \bibinfo {pages} {303} (\bibinfo {year} {1965})}\BibitemShut {NoStop}%
\bibitem [{\citenamefont {Kesten}\ \emph {et~al.}(1980)\citenamefont {Kesten} \emph {et~al.}}]{kesten1980critical}%
  \BibitemOpen
  \bibfield  {author} {\bibinfo {author} {\bibfnamefont {H.}~\bibnamefont {Kesten}} \emph {et~al.},\ }\bibfield  {title} {\bibinfo {title} {The critical probability of bond percolation on the square lattice equals 1/2},\ }\href {https://doi.org/10.1007/BF01197577} {\bibfield  {journal} {\bibinfo  {journal} {Commun. Math. Phys.}\ }\textbf {\bibinfo {volume} {74}},\ \bibinfo {pages} {41} (\bibinfo {year} {1980})}\BibitemShut {NoStop}%
\bibitem [{\citenamefont {Hill}\ and\ \citenamefont {Wootters}(1997)}]{concurrence_hw97}%
  \BibitemOpen
  \bibfield  {author} {\bibinfo {author} {\bibfnamefont {S.}~\bibnamefont {Hill}}\ and\ \bibinfo {author} {\bibfnamefont {W.~K.}\ \bibnamefont {Wootters}},\ }\bibfield  {title} {\bibinfo {title} {Entanglement of a {{Pair}} of {{Quantum Bits}}},\ }\href {https://doi.org/10.1103/PhysRevLett.78.5022} {\bibfield  {journal} {\bibinfo  {journal} {Phys. Rev. Lett.}\ }\textbf {\bibinfo {volume} {78}},\ \bibinfo {pages} {5022} (\bibinfo {year} {1997})}\BibitemShut {NoStop}%
\bibitem [{\citenamefont {Perseguers}\ \emph {et~al.}(2010)\citenamefont {Perseguers}, \citenamefont {Cavalcanti}, \citenamefont {Lapeyre~Jr.}, \citenamefont {Lewenstein},\ and\ \citenamefont {Ac{\'i}n}}]{QEP-GHZ_pclla10}%
  \BibitemOpen
  \bibfield  {author} {\bibinfo {author} {\bibfnamefont {S.}~\bibnamefont {Perseguers}}, \bibinfo {author} {\bibfnamefont {D.}~\bibnamefont {Cavalcanti}}, \bibinfo {author} {\bibfnamefont {G.~J.}\ \bibnamefont {Lapeyre~Jr.}}, \bibinfo {author} {\bibfnamefont {M.}~\bibnamefont {Lewenstein}},\ and\ \bibinfo {author} {\bibfnamefont {A.}~\bibnamefont {Ac{\'i}n}},\ }\bibfield  {title} {\bibinfo {title} {Multipartite entanglement percolation},\ }\href {https://doi.org/10.1103/PhysRevA.81.032327} {\bibfield  {journal} {\bibinfo  {journal} {Phys. Rev. A}\ }\textbf {\bibinfo {volume} {81}},\ \bibinfo {pages} {032327} (\bibinfo {year} {2010})}\BibitemShut {NoStop}%
\bibitem [{\citenamefont {Pant}\ \emph {et~al.}(2019{\natexlab{b}})\citenamefont {Pant}, \citenamefont {Krovi}, \citenamefont {Towsley}, \citenamefont {Tassiulas}, \citenamefont {Jiang}, \citenamefont {Basu}, \citenamefont {Englund},\ and\ \citenamefont {Guha}}]{q-netw-route_pkttjbeg19}%
  \BibitemOpen
  \bibfield  {author} {\bibinfo {author} {\bibfnamefont {M.}~\bibnamefont {Pant}}, \bibinfo {author} {\bibfnamefont {H.}~\bibnamefont {Krovi}}, \bibinfo {author} {\bibfnamefont {D.}~\bibnamefont {Towsley}}, \bibinfo {author} {\bibfnamefont {L.}~\bibnamefont {Tassiulas}}, \bibinfo {author} {\bibfnamefont {L.}~\bibnamefont {Jiang}}, \bibinfo {author} {\bibfnamefont {P.}~\bibnamefont {Basu}}, \bibinfo {author} {\bibfnamefont {D.}~\bibnamefont {Englund}},\ and\ \bibinfo {author} {\bibfnamefont {S.}~\bibnamefont {Guha}},\ }\bibfield  {title} {\bibinfo {title} {Routing entanglement in the quantum internet},\ }\href {https://doi.org/10.1038/s41534-019-0139-x} {\bibfield  {journal} {\bibinfo  {journal} {npj Quantum Inf.}\ }\textbf {\bibinfo {volume} {5}},\ \bibinfo {pages} {25} (\bibinfo {year} {2019}{\natexlab{b}})}\BibitemShut {NoStop}%
\bibitem [{\citenamefont {Kobayashi}\ \emph {et~al.}(2010)\citenamefont {Kobayashi}, \citenamefont {Le~Gall}, \citenamefont {Nishimura},\ and\ \citenamefont {R{\"o}tteler}}]{q-netw-code_klgnr10}%
  \BibitemOpen
  \bibfield  {author} {\bibinfo {author} {\bibfnamefont {H.}~\bibnamefont {Kobayashi}}, \bibinfo {author} {\bibfnamefont {F.}~\bibnamefont {Le~Gall}}, \bibinfo {author} {\bibfnamefont {H.}~\bibnamefont {Nishimura}},\ and\ \bibinfo {author} {\bibfnamefont {M.}~\bibnamefont {R{\"o}tteler}},\ }\bibfield  {title} {\bibinfo {title} {Perfect quantum network communication protocol based on classical network coding},\ }in\ \href {https://doi.org/10.1109/ISIT.2010.5513644} {\emph {\bibinfo {booktitle} {2010 {{IEEE International Symposium}} on {{Information Theory}}}}}\ (\bibinfo {year} {2010})\ pp.\ \bibinfo {pages} {2686--2690}\BibitemShut {NoStop}%
\bibitem [{\citenamefont {Cuquet}\ and\ \citenamefont {Calsamiglia}(2009)}]{QEP-q-swap_cc09}%
  \BibitemOpen
  \bibfield  {author} {\bibinfo {author} {\bibfnamefont {M.}~\bibnamefont {Cuquet}}\ and\ \bibinfo {author} {\bibfnamefont {J.}~\bibnamefont {Calsamiglia}},\ }\bibfield  {title} {\bibinfo {title} {Entanglement {{Percolation}} in {{Quantum Complex Networks}}},\ }\href {https://doi.org/10.1103/PhysRevLett.103.240503} {\bibfield  {journal} {\bibinfo  {journal} {Phys. Rev. Lett.}\ }\textbf {\bibinfo {volume} {103}},\ \bibinfo {pages} {240503} (\bibinfo {year} {2009})}\BibitemShut {NoStop}%
\bibitem [{\citenamefont {Lawler}(1976)}]{lawler1976matroids}%
  \BibitemOpen
  \bibfield  {author} {\bibinfo {author} {\bibfnamefont {E.}~\bibnamefont {Lawler}},\ }\href@noop {} {\emph {\bibinfo {title} {Combinatorial Optimization: {N}etworks and Matroids}}}\ (\bibinfo  {publisher} {Holtz, Rinehart and Winston},\ \bibinfo {year} {1976})\BibitemShut {NoStop}%
\bibitem [{\citenamefont {Monma}\ and\ \citenamefont {Sidney}(1979)}]{monma1979Algorithm}%
  \BibitemOpen
  \bibfield  {author} {\bibinfo {author} {\bibfnamefont {C.}~\bibnamefont {Monma}}\ and\ \bibinfo {author} {\bibfnamefont {J.}~\bibnamefont {Sidney}},\ }\bibfield  {title} {\bibinfo {title} {Sequencing with series-parallel precedence constraints},\ }\href {https://doi.org/10.1287/moor.4.3.215} {\bibfield  {journal} {\bibinfo  {journal} {Math. Oper. Res.}\ }\textbf {\bibinfo {volume} {4}},\ \bibinfo {pages} {215} (\bibinfo {year} {1979})}\BibitemShut {NoStop}%
\bibitem [{\citenamefont {Cohen}\ \emph {et~al.}(2000)\citenamefont {Cohen}, \citenamefont {Erez}, \citenamefont {{ben-Avraham}},\ and\ \citenamefont {Havlin}}]{netw-percolation_ceah00}%
  \BibitemOpen
  \bibfield  {author} {\bibinfo {author} {\bibfnamefont {R.}~\bibnamefont {Cohen}}, \bibinfo {author} {\bibfnamefont {K.}~\bibnamefont {Erez}}, \bibinfo {author} {\bibfnamefont {D.}~\bibnamefont {{ben-Avraham}}},\ and\ \bibinfo {author} {\bibfnamefont {S.}~\bibnamefont {Havlin}},\ }\bibfield  {title} {\bibinfo {title} {Resilience of the {{Internet}} to {{Random Breakdowns}}},\ }\href {https://doi.org/10.1103/PhysRevLett.85.4626} {\bibfield  {journal} {\bibinfo  {journal} {Phys. Rev. Lett.}\ }\textbf {\bibinfo {volume} {85}},\ \bibinfo {pages} {4626} (\bibinfo {year} {2000})}\BibitemShut {NoStop}%
\bibitem [{\citenamefont {Bern}\ \emph {et~al.}(1987)\citenamefont {Bern}, \citenamefont {Lawler},\ and\ \citenamefont {Wong}}]{bern1987subgraph}%
  \BibitemOpen
  \bibfield  {author} {\bibinfo {author} {\bibfnamefont {M.}~\bibnamefont {Bern}}, \bibinfo {author} {\bibfnamefont {E.}~\bibnamefont {Lawler}},\ and\ \bibinfo {author} {\bibfnamefont {A.}~\bibnamefont {Wong}},\ }\bibfield  {title} {\bibinfo {title} {Linear-time computation of optimal subgraphs of decomposable graphs},\ }\href {https://doi.org/https://doi.org/10.1016/0196-6774(87)90039-3} {\bibfield  {journal} {\bibinfo  {journal} {J. Algorithms}\ }\textbf {\bibinfo {volume} {8}},\ \bibinfo {pages} {216} (\bibinfo {year} {1987})}\BibitemShut {NoStop}%
\bibitem [{\citenamefont {Borie}\ \emph {et~al.}(1992)\citenamefont {Borie}, \citenamefont {Parker},\ and\ \citenamefont {Tovey}}]{borie1987subgraph}%
  \BibitemOpen
  \bibfield  {author} {\bibinfo {author} {\bibfnamefont {R.~B.}\ \bibnamefont {Borie}}, \bibinfo {author} {\bibfnamefont {R.~G.}\ \bibnamefont {Parker}},\ and\ \bibinfo {author} {\bibfnamefont {C.~A.}\ \bibnamefont {Tovey}},\ }\bibfield  {title} {\bibinfo {title} {Automatic generation of linear-time algorithms from predicate calculus descriptions of problems on recursively constructed graph families},\ }\href {https://doi.org/10.1007/BF01758777} {\bibfield  {journal} {\bibinfo  {journal} {Algorithmica}\ }\textbf {\bibinfo {volume} {7}},\ \bibinfo {pages} {555} (\bibinfo {year} {1992})}\BibitemShut {NoStop}%
\bibitem [{\citenamefont {Kikuno}\ \emph {et~al.}(1983)\citenamefont {Kikuno}, \citenamefont {Yoshida},\ and\ \citenamefont {Kakuda}}]{kikuno1983number}%
  \BibitemOpen
  \bibfield  {author} {\bibinfo {author} {\bibfnamefont {T.}~\bibnamefont {Kikuno}}, \bibinfo {author} {\bibfnamefont {N.}~\bibnamefont {Yoshida}},\ and\ \bibinfo {author} {\bibfnamefont {Y.}~\bibnamefont {Kakuda}},\ }\bibfield  {title} {\bibinfo {title} {A linear algorithm for the domination number of a series-parallel graph},\ }\href {https://doi.org/https://doi.org/10.1016/0166-218X(83)90003-3} {\bibfield  {journal} {\bibinfo  {journal} {Discret. Appl. Math.}\ }\textbf {\bibinfo {volume} {5}},\ \bibinfo {pages} {299} (\bibinfo {year} {1983})}\BibitemShut {NoStop}%
\bibitem [{\citenamefont {Takamizawa}\ \emph {et~al.}(1982)\citenamefont {Takamizawa}, \citenamefont {Nishizeki},\ and\ \citenamefont {Saito}}]{takamizawa1982linear}%
  \BibitemOpen
  \bibfield  {author} {\bibinfo {author} {\bibfnamefont {K.}~\bibnamefont {Takamizawa}}, \bibinfo {author} {\bibfnamefont {T.}~\bibnamefont {Nishizeki}},\ and\ \bibinfo {author} {\bibfnamefont {N.}~\bibnamefont {Saito}},\ }\bibfield  {title} {\bibinfo {title} {Linear-time computability of combinatorial problems on series-parallel graphs},\ }\href {https://doi.org/10.1145/322326.322328} {\bibfield  {journal} {\bibinfo  {journal} {J. ACM}\ }\textbf {\bibinfo {volume} {29}},\ \bibinfo {pages} {623–} (\bibinfo {year} {1982})}\BibitemShut {NoStop}%
\bibitem [{\citenamefont {Baffi}\ and\ \citenamefont {Petreschi}(1995)}]{baffi1995parallel}%
  \BibitemOpen
  \bibfield  {author} {\bibinfo {author} {\bibfnamefont {L.}~\bibnamefont {Baffi}}\ and\ \bibinfo {author} {\bibfnamefont {R.}~\bibnamefont {Petreschi}},\ }\bibfield  {title} {\bibinfo {title} {Parallel maximal matching on minimal vertex series parallel digraphs},\ }in\ \href@noop {} {\emph {\bibinfo {booktitle} {Algorithms, Concurrency and Knowledge}}},\ \bibinfo {editor} {edited by\ \bibinfo {editor} {\bibfnamefont {K.}~\bibnamefont {Kanchanasut}}\ and\ \bibinfo {editor} {\bibfnamefont {J.}~\bibnamefont {L{\'e}vy}}}\ (\bibinfo  {publisher} {Springer},\ \bibinfo {address} {Berlin},\ \bibinfo {year} {1995})\ pp.\ \bibinfo {pages} {34--47}\BibitemShut {NoStop}%
\bibitem [{\citenamefont {Bertolazzi}\ \emph {et~al.}(1992)\citenamefont {Bertolazzi}, \citenamefont {Cohen}, \citenamefont {Di~Battista}, \citenamefont {Tamassia},\ and\ \citenamefont {Tollis}}]{bertolazzi1992draw}%
  \BibitemOpen
  \bibfield  {author} {\bibinfo {author} {\bibfnamefont {P.}~\bibnamefont {Bertolazzi}}, \bibinfo {author} {\bibfnamefont {R.~F.}\ \bibnamefont {Cohen}}, \bibinfo {author} {\bibfnamefont {G.}~\bibnamefont {Di~Battista}}, \bibinfo {author} {\bibfnamefont {R.}~\bibnamefont {Tamassia}},\ and\ \bibinfo {author} {\bibfnamefont {I.~G.}\ \bibnamefont {Tollis}},\ }\bibfield  {title} {\bibinfo {title} {How to draw a series-parallel digraph},\ }in\ \href@noop {} {\emph {\bibinfo {booktitle} {Algorithm Theory --- SWAT '92}}},\ \bibinfo {editor} {edited by\ \bibinfo {editor} {\bibfnamefont {O.}~\bibnamefont {Nurmi}}\ and\ \bibinfo {editor} {\bibfnamefont {E.}~\bibnamefont {Ukkonen}}}\ (\bibinfo  {publisher} {Springer},\ \bibinfo {address} {Berlin, Germany},\ \bibinfo {year} {1992})\ pp.\ \bibinfo {pages} {272--283}\BibitemShut {NoStop}%
\bibitem [{\citenamefont {Rendl}(1986)}]{rendl1986assignment}%
  \BibitemOpen
  \bibfield  {author} {\bibinfo {author} {\bibfnamefont {F.}~\bibnamefont {Rendl}},\ }\bibfield  {title} {\bibinfo {title} {Quadratic assignment problems on series-parallel digraphs},\ }\href {https://doi.org/10.1007/BF01919179} {\bibfield  {journal} {\bibinfo  {journal} {Z. Oper. Res.}\ }\textbf {\bibinfo {volume} {30}},\ \bibinfo {pages} {A161–A17} (\bibinfo {year} {1986})}\BibitemShut {NoStop}%
\bibitem [{\citenamefont {Steiner}(1985)}]{steiner1985sp}%
  \BibitemOpen
  \bibfield  {author} {\bibinfo {author} {\bibfnamefont {G.}~\bibnamefont {Steiner}},\ }\bibfield  {title} {\bibinfo {title} {A compact labeling scheme for series-parallel graphs},\ }\href {https://doi.org/https://doi.org/10.1016/0166-218X(85)90079-4} {\bibfield  {journal} {\bibinfo  {journal} {Discret. Appl. Math.}\ }\textbf {\bibinfo {volume} {11}},\ \bibinfo {pages} {281} (\bibinfo {year} {1985})}\BibitemShut {NoStop}%
\bibitem [{\citenamefont {Bang-Jensen}\ and\ \citenamefont {Gutin}(2008)}]{bg1976digraphs}%
  \BibitemOpen
  \bibfield  {author} {\bibinfo {author} {\bibfnamefont {J.}~\bibnamefont {Bang-Jensen}}\ and\ \bibinfo {author} {\bibfnamefont {G.}~\bibnamefont {Gutin}},\ }\href {https://doi.org/10.1007/978-1-84800-998-1} {\emph {\bibinfo {title} {Digraphs: {T}heory, Algorithms and Applications}}}\ (\bibinfo  {publisher} {Springer},\ \bibinfo {year} {2008})\BibitemShut {NoStop}%
\bibitem [{\citenamefont {Valdes}\ \emph {et~al.}(1982)\citenamefont {Valdes}, \citenamefont {Tarjan},\ and\ \citenamefont {Lawler}}]{valdes1982sp}%
  \BibitemOpen
  \bibfield  {author} {\bibinfo {author} {\bibfnamefont {J.}~\bibnamefont {Valdes}}, \bibinfo {author} {\bibfnamefont {R.}~\bibnamefont {Tarjan}},\ and\ \bibinfo {author} {\bibfnamefont {E.}~\bibnamefont {Lawler}},\ }\bibfield  {title} {\bibinfo {title} {The recognition of series parallel digraphs},\ }\href {https://doi.org/10.1137/0211023} {\bibfield  {journal} {\bibinfo  {journal} {SIAM J. Comput.}\ }\textbf {\bibinfo {volume} {11}},\ \bibinfo {pages} {298} (\bibinfo {year} {1982})}\BibitemShut {NoStop}%
\bibitem [{\citenamefont {Bodlaender}\ and\ \citenamefont {van Antwerpen~de Fluiter}(2001)}]{bodlaender2001parallel}%
  \BibitemOpen
  \bibfield  {author} {\bibinfo {author} {\bibfnamefont {H.}~\bibnamefont {Bodlaender}}\ and\ \bibinfo {author} {\bibfnamefont {B.}~\bibnamefont {van Antwerpen~de Fluiter}},\ }\bibfield  {title} {\bibinfo {title} {Parallel algorithms for series parallel graphs and graphs with treewidth twos},\ }\href {https://doi.org/10.1007/s004530010070} {\bibfield  {journal} {\bibinfo  {journal} {Algorithmica}\ }\textbf {\bibinfo {volume} {29}},\ \bibinfo {pages} {534} (\bibinfo {year} {2001})}\BibitemShut {NoStop}%
\bibitem [{\citenamefont {Eppstein}(1992)}]{eppstein1992parallel}%
  \BibitemOpen
  \bibfield  {author} {\bibinfo {author} {\bibfnamefont {D.}~\bibnamefont {Eppstein}},\ }\bibfield  {title} {\bibinfo {title} {Parallel recognition of series-parallel graphs},\ }\href {https://doi.org/https://doi.org/10.1016/0890-5401(92)90041-D} {\bibfield  {journal} {\bibinfo  {journal} {Inform. and Comput.}\ }\textbf {\bibinfo {volume} {98}},\ \bibinfo {pages} {41} (\bibinfo {year} {1992})}\BibitemShut {NoStop}%
\bibitem [{\citenamefont {He}(1991)}]{he1991parallel}%
  \BibitemOpen
  \bibfield  {author} {\bibinfo {author} {\bibfnamefont {X.}~\bibnamefont {He}},\ }\bibfield  {title} {\bibinfo {title} {Efficient parallel algorithms for series parallel graphs},\ }\href {https://doi.org/https://doi.org/10.1016/0196-6774(91)90012-N} {\bibfield  {journal} {\bibinfo  {journal} {J. Algorithms}\ }\textbf {\bibinfo {volume} {12}},\ \bibinfo {pages} {409} (\bibinfo {year} {1991})}\BibitemShut {NoStop}%
\bibitem [{\citenamefont {He}\ and\ \citenamefont {Yesha}(1987)}]{he1987parallel}%
  \BibitemOpen
  \bibfield  {author} {\bibinfo {author} {\bibfnamefont {X.}~\bibnamefont {He}}\ and\ \bibinfo {author} {\bibfnamefont {Y.}~\bibnamefont {Yesha}},\ }\bibfield  {title} {\bibinfo {title} {Parallel recognition and decomposition of two terminal series parallel graphs},\ }\href {https://doi.org/https://doi.org/10.1016/0890-5401(87)90061-7} {\bibfield  {journal} {\bibinfo  {journal} {Inform. and Comput.}\ }\textbf {\bibinfo {volume} {75}},\ \bibinfo {pages} {15} (\bibinfo {year} {1987})}\BibitemShut {NoStop}%
\bibitem [{\citenamefont {Kron}(1939)}]{kron1939kron}%
  \BibitemOpen
  \bibfield  {author} {\bibinfo {author} {\bibfnamefont {G.}~\bibnamefont {Kron}},\ }\href@noop {} {\emph {\bibinfo {title} {Tensor Analysis of Networks}}}\ (\bibinfo  {publisher} {{Wiley}},\ \bibinfo {year} {1939})\BibitemShut {NoStop}%
\bibitem [{\citenamefont {Versfeld}(1970)}]{star-mesh_v70}%
  \BibitemOpen
  \bibfield  {author} {\bibinfo {author} {\bibfnamefont {L.}~\bibnamefont {Versfeld}},\ }\bibfield  {title} {\bibinfo {title} {Remarks on star-mesh transformation of electrical networks},\ }\href {https://doi.org/10.1049/el:19700417} {\bibfield  {journal} {\bibinfo  {journal} {Electron. Lett.}\ }\textbf {\bibinfo {volume} {6}},\ \bibinfo {pages} {597} (\bibinfo {year} {1970})}\BibitemShut {NoStop}%
\bibitem [{\citenamefont {Malik}\ \emph {et~al.}(2022)\citenamefont {Malik}, \citenamefont {Meng}, \citenamefont {Havlin}, \citenamefont {Korniss}, \citenamefont {Szymanski},\ and\ \citenamefont {Gao}}]{conpt_mmhksg22}%
  \BibitemOpen
  \bibfield  {author} {\bibinfo {author} {\bibfnamefont {O.}~\bibnamefont {Malik}}, \bibinfo {author} {\bibfnamefont {X.}~\bibnamefont {Meng}}, \bibinfo {author} {\bibfnamefont {S.}~\bibnamefont {Havlin}}, \bibinfo {author} {\bibfnamefont {G.}~\bibnamefont {Korniss}}, \bibinfo {author} {\bibfnamefont {B.~K.}\ \bibnamefont {Szymanski}},\ and\ \bibinfo {author} {\bibfnamefont {J.}~\bibnamefont {Gao}},\ }\bibfield  {title} {\bibinfo {title} {Concurrence percolation threshold of large-scale quantum networks},\ }\href {https://doi.org/10.1038/s42005-022-00958-4} {\bibfield  {journal} {\bibinfo  {journal} {Commun. Phys.}\ }\textbf {\bibinfo {volume} {5}},\ \bibinfo {pages} {1} (\bibinfo {year} {2022})}\BibitemShut {NoStop}%
\bibitem [{\citenamefont {Krattenthaler}(2015)}]{krattenthaler2015}%
  \BibitemOpen
  \bibfield  {author} {\bibinfo {author} {\bibfnamefont {C.}~\bibnamefont {Krattenthaler}},\ }\bibfield  {title} {\bibinfo {title} {Lattice path enumeration},\ }\href@noop {} {\bibfield  {journal} {\bibinfo  {journal} {arXiv:1503.05930}\ } (\bibinfo {year} {2015})}\BibitemShut {NoStop}%
\bibitem [{\citenamefont {Jensen}(2004)}]{Jensen_2004}%
  \BibitemOpen
  \bibfield  {author} {\bibinfo {author} {\bibfnamefont {I.}~\bibnamefont {Jensen}},\ }\bibfield  {title} {\bibinfo {title} {Enumeration of self-avoiding walks on the square lattice},\ }\href {https://doi.org/10.1088/0305-4470/37/21/002} {\bibfield  {journal} {\bibinfo  {journal} {J. Phys. A}\ }\textbf {\bibinfo {volume} {37}},\ \bibinfo {pages} {5503} (\bibinfo {year} {2004})}\BibitemShut {NoStop}%
\bibitem [{\citenamefont {Roberts}\ and\ \citenamefont {Kroese}(2007)}]{Roberts2007}%
  \BibitemOpen
  \bibfield  {author} {\bibinfo {author} {\bibfnamefont {B.}~\bibnamefont {Roberts}}\ and\ \bibinfo {author} {\bibfnamefont {D.~P.}\ \bibnamefont {Kroese}},\ }\bibfield  {title} {\bibinfo {title} {Estimating the number of s-t paths in a graph},\ }\href {https://doi.org/10.7155/jgaa.00142} {\bibfield  {journal} {\bibinfo  {journal} {J. Graph Algorithms Appl.}\ }\textbf {\bibinfo {volume} {11}},\ \bibinfo {pages} {195} (\bibinfo {year} {2007})}\BibitemShut {NoStop}%
\bibitem [{\citenamefont {Treinish}\ \emph {et~al.}(2023)\citenamefont {Treinish}, \citenamefont {Gambetta}, \citenamefont {Thomas}, \citenamefont {Nation}, \citenamefont {{qiskit-bot}}, \citenamefont {Kassebaum}, \citenamefont {Rodr{\'i}guez}, \citenamefont {Gonz{\'a}lez}, \citenamefont {Lishman}, \citenamefont {Hu}, \citenamefont {Bello}, \citenamefont {Garrison}, \citenamefont {Krsulich}, \citenamefont {Huang}, \citenamefont {Yu}, \citenamefont {Marques}, \citenamefont {Gacon}, \citenamefont {McKay}, \citenamefont {Arellano}, \citenamefont {Gomez}, \citenamefont {Capelluto}, \citenamefont {{Travis-S-IBM}}, \citenamefont {Mitchell}, \citenamefont {Panigrahi}, \citenamefont {{lerongil}}, \citenamefont {Rahman}, \citenamefont {Wood}, \citenamefont {Itoko}, \citenamefont {{Pozas-Kerstjens}},\ and\ \citenamefont {Wood}}]{qiskit_a21}%
  \BibitemOpen
  \bibfield  {author} {\bibinfo {author} {\bibfnamefont {M.}~\bibnamefont {Treinish}}, \bibinfo {author} {\bibfnamefont {J.}~\bibnamefont {Gambetta}}, \bibinfo {author} {\bibfnamefont {S.}~\bibnamefont {Thomas}}, \bibinfo {author} {\bibfnamefont {P.}~\bibnamefont {Nation}}, \bibinfo {author} {\bibnamefont {{qiskit-bot}}}, \bibinfo {author} {\bibfnamefont {P.}~\bibnamefont {Kassebaum}}, \bibinfo {author} {\bibfnamefont {D.~M.}\ \bibnamefont {Rodr{\'i}guez}}, \bibinfo {author} {\bibfnamefont {S.~d. l.~P.}\ \bibnamefont {Gonz{\'a}lez}}, \bibinfo {author} {\bibfnamefont {J.}~\bibnamefont {Lishman}}, \bibinfo {author} {\bibfnamefont {S.}~\bibnamefont {Hu}}, \bibinfo {author} {\bibfnamefont {L.}~\bibnamefont {Bello}}, \bibinfo {author} {\bibfnamefont {J.}~\bibnamefont {Garrison}}, \bibinfo {author} {\bibfnamefont {K.}~\bibnamefont {Krsulich}}, \bibinfo {author} {\bibfnamefont {J.}~\bibnamefont {Huang}}, \bibinfo {author} {\bibfnamefont {J.}~\bibnamefont {Yu}}, \bibinfo {author} {\bibfnamefont {M.}~\bibnamefont
  {Marques}}, \bibinfo {author} {\bibfnamefont {J.}~\bibnamefont {Gacon}}, \bibinfo {author} {\bibfnamefont {D.}~\bibnamefont {McKay}}, \bibinfo {author} {\bibfnamefont {E.}~\bibnamefont {Arellano}}, \bibinfo {author} {\bibfnamefont {J.}~\bibnamefont {Gomez}}, \bibinfo {author} {\bibfnamefont {L.}~\bibnamefont {Capelluto}}, \bibinfo {author} {\bibnamefont {{Travis-S-IBM}}}, \bibinfo {author} {\bibfnamefont {A.}~\bibnamefont {Mitchell}}, \bibinfo {author} {\bibfnamefont {A.}~\bibnamefont {Panigrahi}}, \bibinfo {author} {\bibnamefont {{lerongil}}}, \bibinfo {author} {\bibfnamefont {R.~I.}\ \bibnamefont {Rahman}}, \bibinfo {author} {\bibfnamefont {S.}~\bibnamefont {Wood}}, \bibinfo {author} {\bibfnamefont {T.}~\bibnamefont {Itoko}}, \bibinfo {author} {\bibfnamefont {A.}~\bibnamefont {{Pozas-Kerstjens}}},\ and\ \bibinfo {author} {\bibfnamefont {C.~J.}\ \bibnamefont {Wood}},\ }\href@noop {} {\bibinfo {title} {{{Qiskit}} 0.42.0}},\ \bibinfo {howpublished} {Zenodo} (\bibinfo {year} {2023})\BibitemShut {NoStop}%
\bibitem [{\citenamefont {{Calderon-Vargas}}\ \emph {et~al.}(2019)\citenamefont {{Calderon-Vargas}}, \citenamefont {Barron}, \citenamefont {Deng}, \citenamefont {Sigillito}, \citenamefont {Barnes},\ and\ \citenamefont {Economou}}]{q-cnot-fidel_cbdsbe19}%
  \BibitemOpen
  \bibfield  {author} {\bibinfo {author} {\bibfnamefont {F.~A.}\ \bibnamefont {{Calderon-Vargas}}}, \bibinfo {author} {\bibfnamefont {G.~S.}\ \bibnamefont {Barron}}, \bibinfo {author} {\bibfnamefont {X.-H.}\ \bibnamefont {Deng}}, \bibinfo {author} {\bibfnamefont {A.~J.}\ \bibnamefont {Sigillito}}, \bibinfo {author} {\bibfnamefont {E.}~\bibnamefont {Barnes}},\ and\ \bibinfo {author} {\bibfnamefont {S.~E.}\ \bibnamefont {Economou}},\ }\bibfield  {title} {\bibinfo {title} {Fast high-fidelity entangling gates for spin qubits in {{Si}} double quantum dots},\ }\href {https://doi.org/10.1103/PhysRevB.100.035304} {\bibfield  {journal} {\bibinfo  {journal} {Phys. Rev. B}\ }\textbf {\bibinfo {volume} {100}},\ \bibinfo {pages} {035304} (\bibinfo {year} {2019})}\BibitemShut {NoStop}%
\bibitem [{\citenamefont {Kandala}\ \emph {et~al.}(2021)\citenamefont {Kandala}, \citenamefont {Wei}, \citenamefont {Srinivasan}, \citenamefont {Magesan}, \citenamefont {Carnevale}, \citenamefont {Keefe}, \citenamefont {Klaus}, \citenamefont {Dial},\ and\ \citenamefont {McKay}}]{q-cnot-fidel_kwsmckkdm21}%
  \BibitemOpen
  \bibfield  {author} {\bibinfo {author} {\bibfnamefont {A.}~\bibnamefont {Kandala}}, \bibinfo {author} {\bibfnamefont {K.~X.}\ \bibnamefont {Wei}}, \bibinfo {author} {\bibfnamefont {S.}~\bibnamefont {Srinivasan}}, \bibinfo {author} {\bibfnamefont {E.}~\bibnamefont {Magesan}}, \bibinfo {author} {\bibfnamefont {S.}~\bibnamefont {Carnevale}}, \bibinfo {author} {\bibfnamefont {G.~A.}\ \bibnamefont {Keefe}}, \bibinfo {author} {\bibfnamefont {D.}~\bibnamefont {Klaus}}, \bibinfo {author} {\bibfnamefont {O.}~\bibnamefont {Dial}},\ and\ \bibinfo {author} {\bibfnamefont {D.~C.}\ \bibnamefont {McKay}},\ }\bibfield  {title} {\bibinfo {title} {Demonstration of a {{High-Fidelity}} {{CNOT}} {{Gate}} for {{Fixed-Frequency Transmons}} with {{Engineered}} {{ZZ}} {{Suppression}}},\ }\href {https://doi.org/10.1103/PhysRevLett.127.130501} {\bibfield  {journal} {\bibinfo  {journal} {Phys. Rev. Lett.}\ }\textbf {\bibinfo {volume} {127}},\ \bibinfo {pages} {130501} (\bibinfo {year} {2021})}\BibitemShut {NoStop}%
\bibitem [{\citenamefont {Bantysh}\ \emph {et~al.}(2021)\citenamefont {Bantysh}, \citenamefont {Chernyavskiy},\ and\ \citenamefont {Bogdanov}}]{q-tomogr_bcb21}%
  \BibitemOpen
  \bibfield  {author} {\bibinfo {author} {\bibfnamefont {B.~I.}\ \bibnamefont {Bantysh}}, \bibinfo {author} {\bibfnamefont {A.~Y.}\ \bibnamefont {Chernyavskiy}},\ and\ \bibinfo {author} {\bibfnamefont {Y.~I.}\ \bibnamefont {Bogdanov}},\ }\bibfield  {title} {\bibinfo {title} {Quantum tomography benchmarking},\ }\href {https://doi.org/10.1007/s11128-021-03285-9} {\bibfield  {journal} {\bibinfo  {journal} {Quantum Inf. Process.}\ }\textbf {\bibinfo {volume} {20}},\ \bibinfo {pages} {339} (\bibinfo {year} {2021})}\BibitemShut {NoStop}%
\bibitem [{\citenamefont {Li}\ \emph {et~al.}(2009)\citenamefont {Li}, \citenamefont {He},\ and\ \citenamefont {Jiang}}]{q-game_lhj09}%
  \BibitemOpen
  \bibfield  {author} {\bibinfo {author} {\bibfnamefont {Q.}~\bibnamefont {Li}}, \bibinfo {author} {\bibfnamefont {Y.}~\bibnamefont {He}},\ and\ \bibinfo {author} {\bibfnamefont {J.-p.}\ \bibnamefont {Jiang}},\ }\bibfield  {title} {\bibinfo {title} {A novel clustering algorithm based on quantum games},\ }\href {https://doi.org/10.1088/1751-8113/42/44/445303} {\bibfield  {journal} {\bibinfo  {journal} {J. Phys. A}\ }\textbf {\bibinfo {volume} {42}},\ \bibinfo {pages} {445303} (\bibinfo {year} {2009})}\BibitemShut {NoStop}%
\bibitem [{\citenamefont {Miszczak}\ \emph {et~al.}(2014)\citenamefont {Miszczak}, \citenamefont {Pawela},\ and\ \citenamefont {S{\l}adkowski}}]{q-game_mps14}%
  \BibitemOpen
  \bibfield  {author} {\bibinfo {author} {\bibfnamefont {J.~A.}\ \bibnamefont {Miszczak}}, \bibinfo {author} {\bibfnamefont {{\L}.}~\bibnamefont {Pawela}},\ and\ \bibinfo {author} {\bibfnamefont {J.}~\bibnamefont {S{\l}adkowski}},\ }\bibfield  {title} {\bibinfo {title} {General {{Model}} for an {{Entanglement-Enhanced Composed Quantum Game}} on a {{Two-Dimensional Lattice}}},\ }\href {https://doi.org/10.1142/S0219477514500126} {\bibfield  {journal} {\bibinfo  {journal} {Fluct. Noise Lett.}\ }\textbf {\bibinfo {volume} {13}},\ \bibinfo {pages} {1450012} (\bibinfo {year} {2014})}\BibitemShut {NoStop}%
\bibitem [{\citenamefont {Muhammad}\ \emph {et~al.}(2014)\citenamefont {Muhammad}, \citenamefont {Tavakoli}, \citenamefont {Kurant}, \citenamefont {Paw{\l}owski}, \citenamefont {{\.Z}ukowski},\ and\ \citenamefont {Bourennane}}]{q-game_mtkpb14}%
  \BibitemOpen
  \bibfield  {author} {\bibinfo {author} {\bibfnamefont {S.}~\bibnamefont {Muhammad}}, \bibinfo {author} {\bibfnamefont {A.}~\bibnamefont {Tavakoli}}, \bibinfo {author} {\bibfnamefont {M.}~\bibnamefont {Kurant}}, \bibinfo {author} {\bibfnamefont {M.}~\bibnamefont {Paw{\l}owski}}, \bibinfo {author} {\bibfnamefont {M.}~\bibnamefont {{\.Z}ukowski}},\ and\ \bibinfo {author} {\bibfnamefont {M.}~\bibnamefont {Bourennane}},\ }\bibfield  {title} {\bibinfo {title} {Quantum {{Bidding}} in {{Bridge}}},\ }\href {https://doi.org/10.1103/PhysRevX.4.021047} {\bibfield  {journal} {\bibinfo  {journal} {Phys. Rev. X}\ }\textbf {\bibinfo {volume} {4}},\ \bibinfo {pages} {021047} (\bibinfo {year} {2014})}\BibitemShut {NoStop}%
\bibitem [{\citenamefont {Pfleiderer}(2005)}]{first-order-q-phase-transit_p05}%
  \BibitemOpen
  \bibfield  {author} {\bibinfo {author} {\bibfnamefont {C.}~\bibnamefont {Pfleiderer}},\ }\bibfield  {title} {\bibinfo {title} {Why first order quantum phase transitions are interesting},\ }\href {https://doi.org/10.1088/0953-8984/17/11/031} {\bibfield  {journal} {\bibinfo  {journal} {J. Phys. Condens. Matter}\ }\textbf {\bibinfo {volume} {17}},\ \bibinfo {pages} {S987} (\bibinfo {year} {2005})}\BibitemShut {NoStop}%
\bibitem [{\citenamefont {{de Lima}}\ and\ \citenamefont {Gon{\c c}alves}(2008)}]{first-order-q-phase-transit_lg08}%
  \BibitemOpen
  \bibfield  {author} {\bibinfo {author} {\bibfnamefont {J.~P.}\ \bibnamefont {{de Lima}}}\ and\ \bibinfo {author} {\bibfnamefont {L.~L.}\ \bibnamefont {Gon{\c c}alves}},\ }\bibfield  {title} {\bibinfo {title} {Quantum transitions of the isotropic \${{XY}}\$ model with long-range interactions on the inhomogeneous periodic chain},\ }\href {https://doi.org/10.1103/PhysRevB.77.214424} {\bibfield  {journal} {\bibinfo  {journal} {Phys. Rev. B}\ }\textbf {\bibinfo {volume} {77}},\ \bibinfo {pages} {214424} (\bibinfo {year} {2008})}\BibitemShut {NoStop}%
\bibitem [{\citenamefont {Li}\ and\ \citenamefont {Lei}(2008)}]{first-order-q-phase-transit_ll08}%
  \BibitemOpen
  \bibfield  {author} {\bibinfo {author} {\bibfnamefont {J.}~\bibnamefont {Li}}\ and\ \bibinfo {author} {\bibfnamefont {S.}~\bibnamefont {Lei}},\ }\bibfield  {title} {\bibinfo {title} {A study of the spin-1/2 {{XZ}} chain with long-range interactions in the external magnetic field},\ }\href {https://doi.org/10.1016/j.jmmm.2008.06.025} {\bibfield  {journal} {\bibinfo  {journal} {J. Magn. Magn. Mater.}\ }\textbf {\bibinfo {volume} {320}},\ \bibinfo {pages} {2823} (\bibinfo {year} {2008})}\BibitemShut {NoStop}%
\end{thebibliography}%

\end{document}